\documentclass[showpacs,prd,preprintnumbers,nofootinbib,superscriptaddress,
]{revtex4-2}

\usepackage{amsmath,amsthm,amssymb}
\usepackage{mathrsfs} 
\usepackage{upgreek}
\usepackage[normalem]{ulem} 
\usepackage{graphicx}
\usepackage{latexsym}
\usepackage{url,hyperref}
\usepackage{textcomp}
\usepackage[dvipsnames]{xcolor} 
\usepackage{slashed}
\usepackage{epstopdf}
\usepackage{placeins}
\usepackage{array}
\usepackage{cancel}
\usepackage{verbatim}
\usepackage{caption}
\usepackage{subcaption}
\usepackage[splitrule,hang,flushmargin]{footmisc} 
\renewcommand{\footnoterule}{%
  \kern -3pt
  \hrule width \textwidth height 1pt
  \kern 2pt
}

\usepackage{chngcntr}
\usepackage{pdfpages}

\usepackage{setspace} 
\AtBeginDocument{\setstretch{1.125}}

\usepackage{outlines} 
\usepackage{tensor} 
\usepackage{csquotes} 

\usepackage{blindtext}
\usepackage{tcolorbox}

\hypersetup{colorlinks=true,linkcolor=blue,citecolor=magenta,filecolor=magenta,urlcolor=blue}

\numberwithin{equation}{section}
\renewcommand{\theequation}{\arabic{section}.\arabic{equation}}

\DeclareUnicodeCharacter{2212}{-} 

\usepackage{etoolbox}

\makeatletter
\patchcmd{\@outputpage@head}{\@ifx{\LS@rot\@undefined}{}{\LS@rot}}{}{}{}
\patchcmd{\frontmatter@abstract@produce} 
  {\vskip200\p@\@plus1fil
   \penalty-200\relax
   \vskip-200\p@\@plus-1fil}
  {}
  {}
  {}
\makeatother

\newcolumntype{C}{>{$}c<{$}} 
\newcolumntype{L}{>{$}l<{$}} 

\begin{document}

\title{A semi-analytic treatment of quasinormal excitation factors in the eikonal regime}

\author{Chun-Hung Chen}
\email{chun-hungc@nu.ac.th}
\affiliation{The Institute for Fundamental Study, \\
Naresuan University, Phitsanulok 65000, Thailand.}

\author{Hing-Tong Cho}
\email{htcho@mail.tku.edu.tw}
\affiliation{Department of Physics, Tamkang University,\\
Tamsui District, New Taipei City, Taiwan 25137.}

\author{Anna Chrysostomou}
\email{chrysostomou@ip2i.in2p3.fr}
\affiliation{Department of Physics, University of Johannesburg, \\
PO Box 524, Auckland Park 2006, South Africa.}
\affiliation{Institut de Physique des Deux Infinis de Lyon, Universit{\'e} de Lyon, UCBL, UMR 5822, CNRS/IN2P3,\\ 4 rue Enrico Fermi, 69622 Villeurbanne Cedex, France}

\author{Alan S. Cornell}
\email{acornell@uj.ac.za}
\affiliation{Department of Physics, University of Johannesburg, \\
PO Box 524, Auckland Park 2006, South Africa.}

\begin{abstract}
\par 

In this paper, we present an enhanced semi-analytic method for calculating quasinormal excitation factors in the eikonal regime, specifically for Schwarzschild black holes. To achieve improved accuracy in our quasinormal mode computations, we extend the Dolan and Ottewill inverse multipolar expansion technique and incorporate higher-order corrections from the WKB method of Iyer and Will. Our approach is carried out to a higher order than previous methods, thereby reducing the relative error, particularly for lower multipolar numbers. We validate our results by comparing them with those obtained using the Mano, Suzuki, and Takasugi method, demonstrating excellent agreement. A key advantage of our method is its ability to extract quasinormal excitation factors, which are crucial for accurately modeling gravitational wave signals from binary black hole mergers. This advancement provides a useful tool for future gravitational wave studies, enabling better quantification of quasinormal mode excitations and more precise identification of individual modes during black hole ringdowns.
\end{abstract}


\date{\today}
\maketitle

\section{Introduction}

\par The theoretical study of the evolution of black hole perturbations began with the seminal 1957 Schwarzschild stability analysis of Regge and Wheeler \cite{refRW}. As an intrinsically non-linear theory, general relativity (GR) does not readily allow for analytical approaches to such a problem. Under spherical symmetry and the assumption that the perturbations contribute negligibly to the space-time curvature, however, the evolution of the radial behaviour of the gravitational perturbations were shown to reduce to a wave equation of the form,
\begin{equation} \label{eq:SWEpsi}
\left[ \frac{d^2}{dr_{\star}^2} + \omega^2 - V(r) \right] {u}(r_{\star}) = 0   \;.
\end{equation}
\noindent Here, the tortoise coordinate $r_{\star} \in (-\infty,+\infty )$ is a bijection of the Schwarzschild circumferential radial coordinate $r$ that maps the black hole event horizon to negative infinity; $V(r)$ is the effective potential. This has been shown to hold for fields of all spin, with $V(r)$ encoding characteristic information about the perturbing test field (see reviews \cite{refBertiCardoso,refKonoplyaZhidenkoReview} for details). 

\par If we consider Eq. (\ref{eq:SWEpsi}) in the context of a scattering problem, where the black hole curvature performs the scattering and informs the out-going quasinormal mode boundary conditions \cite{Andersson2000_BHscattering,FuttermanHandlerMatzner1988_BHscattering}, the potential barrier is real and the eigenvalue $\omega$ is complex.  For a Schwarzschild black hole, $\omega$ depends on the angular momentum/multipolar number $\ell$ and a mode/overtone number $n$; for each $n$ there is an infinite number of multipoles \cite{Chandrasekhar1983}.

\par From the vantage point of a distant observer, it is well-established in the literature that the response of a black hole to an external perturbation can be separated into three phases: 
\begin{itemize}
    \item[$(i)$] a short outburst of gravitational radiations, where the emissions depend on the initial perturbing stimulus;
    \item[$(ii)$] a long period of damped oscillations dependent primarily on the black hole parameters, dominated by the exponentially-damped quasinormal modes (QNMs);
    \item[$(iii)$] a power-law tail that characterises the very late stage as the wave falls off with time \cite{Vishveshwara1970_scattering}.
\end{itemize}

\noindent QNMs and their corresponding quasinormal frequencies (QNFs) in particular offer tremendous insight into their black hole source: they convey key information about the parameters of the black hole whence they emerge \cite{Echeverria1989_BHpropertiesestimate}, the stability of the space-time in which they propagate \cite{Vishveshwara1970_stability}, and the possibility of violations of cosmic censorship \cite{Hintz2015_SCC}. Furthermore, the QNMs are closely related to the bound states of the inverted black hole potential \cite{refFerrMashh1,refFerrMashh2}. There are considerable theoretical implications for black hole QNMs, ranging from black hole thermodynamics to the gauge/gravity duality and braneworld scenarios \cite{refBertiCardoso}. Note that the QNM eigenvalue problem presented in Eq. (\ref{eq:SWEpsi}) has served as a platform for the development of a vast array of semi-classical and numerical techniques for approaching intrinsically dissipative systems \cite{refKonoplyaZhidenkoReview,refAIM,Cho:2011sf, Cornell:2022enn}.

\par While the response of a black hole to a perturbing field has been studied extensively, there are certain elements of the problem that require further scrutiny. Here, we shall focus on the degeneracy of the modes for each $n$, and the question of how to distinguish one mode from another. As discussed in Ref. \cite{Berti2005_BHspectroscopy}, identifying an individual mode within the superposition of QNMs observed is necessary for the development of waveform models and the testing of GR. Since the detectability of each mode depends on its relative excitation, a means by which to quantify the excitation of a QNM that is independent of the initial perturbing stimulus is vital.   

\par As the name suggests, quasinormal excitation factors (QNEFs) indicate how and by how much QNMs are excited \cite{refDolanOttewill2011}, {where we also summarize the key contribution of the QNEFs to the wave function in Appendix \ref{sec:scattering}}. Few papers are available on the subject, due in part to the complexity of the problem of determining the QNM wavefunction. A seminal attempt at the QNEF computation was carried out by Andersson in 1995 \cite{Andersson1995_SchWavefunction}. Following Nollert and Schmidt's construction of the QNMs as singularities of a Green's function \cite{NollertSchmidt1992_QNMsInhomogeneous}, Andersson applied the phase-integral method to the case of scalar QNMs in a Schwarzschild background. The final formulae he generated carried over naturally to the gravitational counterpart. It was only recently that the semi-analytic computation of QNEFs for the case of the Kerr black hole has been successfully carried out in Ref. \cite{Yang2014_KerrGreen} using a WKB analysis at leading order. Further important examples include the numerical treatments of the Kerr case in Refs. \cite{BertiCardoso2006_KerrExcitation} and \cite{Oshita2021_ExcitationOvertones}, where the latter explored Kerr QNEFs to the 20th overtone.


\par The calculation of a QNEF has received renewed interest following the successful detection of gravitational-waves (GWs) at the LIGO observatory \cite{refLIGO}. The now regular detection of GWs from binary black hole merger events by the LIGO-Virgo-KAGRA (LVK) collaboration \cite{refLIGO2018Run1_GWTC1,refLIGO2020Run2_GWTC2,refLIGOrecent_GWTC3} has allowed for a thorough examination of this relationship between QNFs and the parameters of post-merger black holes, enabling testing of GR in the strong regime \cite{LIGO2019_GWTC1-GRtest,LIGO2020_GWTC2-GRtest_pyRing3,LIGO2021_GWTC3-GRtest,Chrysostomou2023_EPJC}. However, it has also emphasised the importance of quantifying QNM excitations. Isolating the QNM-dominated phase of the post-merger gravitational radiation from the gravitational waveform is known to be highly non-trivial, due to the low signal-to-noise ratio (SNR) of the post-merger phase and the technical challenges in combining posterior probability densities from multiple events \cite{refLIGOguide,Baibhav2017_BHspectroscopyGW}. Quantifying QNM excitations is expected to become more complicated as GW detector sensitivity to higher overtones \cite{Oshita2021_ExcitationOvertones} and harmonics \cite{OtaChirenti2019_Harmonics} increases \cite{Baibhav2018_BHspectroscopyL}. 

\par As such, we shall consider a semi-analytical approach to the calculation of the QNEF. Specifically, we focus our attention on the scalar wave propagation study in a Schwarzschild background conducted by Dolan and Ottewill \cite{refDolanOttewill2011}. The semi-analytical method they employ is based on an inverse multipolar expansion method \cite{refDolanOttewill2009}, hereafter referred to as the \enquote{Dolan-Ottewill} technique. This method exploits the known relationship between the critical orbits of null geodesics and QNFs in the eikonal regime \cite{refGoebel1972,refCardosoLyapunov,Konoplya2023_BeyondEikonal} to construct a novel ansatz with which the QNM eigenvalue problem can be solved with relative ease, even at higher orders. This allows users to attain high levels of QNF accuracy with minimal computing power. Since the Dolan-Ottewill technique relies almost entirely on the nature of the space-time context, it can be applied to a variety of spherically-symmetric black holes \cite{refDolanOttewill2009,refFernandoCorrea} for perturbing fields of spin $s \in \{0,1/2,1,3/2,2\}$ \cite{refLiLinYang,refOurLargeL}. This includes extremal contexts where several tried-and-tested numerical methods are known to break down, such as black holes in asymptotically flat space-times whose charge-to-mass ratio tends to unity \cite{Berti2004}.

\par As a by-product, the Dolan-Ottewill technique produces an expression for the QNM wavefunction. The wavefunction manifests as a closed-form solution valid across the real line. With the aid of a WKB treatment near the peak of the potential, Dolan and Ottewill demonstrated in Ref. \cite{refDolanOttewill2011} that their technique can be used to compute QNEFs. We consider their method to be highly effective, but lacking in precision due to its cutoff at leading order. In this work, we therefore extend their computation to higher orders, where beginning in Section \ref{sec:Schwarz} we review the Dolan-Ottewill framework, elaborating on the original QNEF calculation outlined by the authors in Ref. \cite{refDolanOttewill2011}. In Section \ref{sec:QNEF}, we proceed with the calculation of the QNEF at higher orders. While the \enquote{exterior} solution for the wavefunction far from the global maximum of the potential can be computed in a relatively straightforward fashion, following the Dolan-Ottewill method, the \enquote{interior} solution represents a more involved calculation that borrows heavily from the Schutz-Iyer-Will formalism \cite{refBHWKB1}. Working within the eikonal approximation, we compute the interior and exterior wavefunctions separately. Then, through a systematic matching procedure, we demonstrate explicitly that at order $\mathcal{O}(L^{-2})$, we obtain a marked improvement on the leading-order QNEF values. 

\section{A review of the leading-order QNEF calculation \label{sec:Schwarz}}

\par Under geometric units ($c=G=1$), the Schwarzschild metric is given by
\begin{equation} \label{eq:metric}
    \mathrm{d}s^2= -f(r)\mathrm{d}t^2+f(r)^{-1}\mathrm{d}r^2+r^2 \mathrm{d}\Omega_2^2 \;,
\end{equation}
for $f(r)=1-2M/r$. Here, the Schwarzschild coordinates $(t,r,\theta,\phi)$ are defined on the regions $t \in (-\infty,+\infty)$, $r \in (r_+,+\infty)$, $\theta \in (0,\pi),$ and $\phi \in (0,2\pi)$. The event horizon is denoted by $r_+=2M$, where $M$ is the mass of the black hole, and $\mathrm{d}\Omega^2=\mathrm{d} \theta^2 + \sin^2 \theta \mathrm{d}\phi^2$ serves as the line element on the 2-sphere $\mathbb{S}^2$. From this point onwards, we shall set $M=1$.  


\par The QNFs of a Schwarzschild black hole within the large-$\ell$ regime can be expressed as
\begin{align}
\omega_{\ell n} & \equiv \omega_R - i \omega_I = L \Omega_R  - iN \Omega_I + \mathcal{O} (L^{-1}) \;, \label{eq:EikonalQuantities}
\end{align}
where $L \equiv \ell + 1/2$ and $N \equiv n + 1/2 \;.$ $\Omega_R$ refers to the orbital frequency of the $r_{orb}=3$ photon sphere\footnote{\enquote{Photon sphere} refers to the surface formed from the collection of null geodesics along which gravitationally-entrapped photons orbit a black hole. For a static black hole, photons orbit a black hole at this fixed radius; the orbit itself is unstable against radial perturbations.}. In the large-$\ell$ limit, this is equivalent to the Lyapunov exponent $\Omega_I$, which indicates the decay time scale of the gravitational perturbations. For the Schwarzschild space-time, $\Omega_R = \Omega_I = 1/\sqrt{27}$ \cite{refGoebel1972}. This relationship between eikonal QNFs and photon orbits has been observed in perturbing test fields of spin $s \in \{0,1/2,1,3/2\}$ within the region of a Schwarzschild black hole \cite{refCardosoLyapunov,refFerrMashh2,refPanJing,refShuShen,Konoplya2003} and a variety of different space-times, modulo some small augmentation (see for example Refs. \cite{refZhidenko2004,refOurLargeL}).

\par In Ref. \cite{refDolanOttewill2009}, Dolan and Ottewill presented a method by which to compute QNM frequencies and wavefunctions, invoking this relationship between photon orbits and the large-$\ell$ asymptotics of QNMs. The method centres on an ansatz for the wavefunction that is inspired by the trajectory of a distant photon along a null geodesic that ends on the photon sphere, locked in orbit about a spherically-symmetric black hole. The ansatz can be written concisely as
\begin{equation} \label{eq:DOansatz}
u_{\ell \omega}(r) = e^{i \omega z(r_{\star})}v(r) \;, \quad z(r_{\star}) = \int^{r_{\star}} \rho (r) dr_{\star}  \;, \quad \rho(r) = b_c k_c (r) \;.
\end{equation}
The integrand is a function of the impact parameter $b$ and $k^2(r)=1/b^2-f(r)/r^2$, with $r_{\star} = \displaystyle \int dr/f(r)$ serving as the usual tortoise coordinate. The subscript $c$ denotes a quantity evaluated at the critical radius $r_c= 2 f(r)/ \partial_r f(r) \Big|_{r=r_c}$ at which $k^2(r,b_c)$ has repeated roots and satisfies the condition $k^2 (r_c,b_c) = \partial_r k^2 (r_c,b_c) = 0$ (see Appendix \ref{sub:DO} for details). The QNF is expressed as a linear expansion in inverse mutlipolar numbers parameterised as $L=\ell+1/2$, that is
\begin{equation} \label{eq:QNFseries}
\omega_{\ell n} = \sum_{k=-1} 
{\overline{\omega}_{k} }L^{-k} \;,
\end{equation}
{where recall that $n$ is the overtone number. The coefficients $\overline{\omega}_{k}$ are functions of $n$ (see Eq. (\ref{eq:QNF(2)})).}

\noindent Similarly, the function $v(r)$ is a series expansion in $L^{-k}$,
\begin{equation} \label{eq:v(r)n}
v_{\ell n} (r) = \left[\left(1 - \frac{3}{r}\right)^n + \sum_{i=1}^{n} \sum_{j=1}^{\infty} \alpha_{_{ijn}} L^{-j} \left(1-\frac{3}{r} \right)^{n-i}  \right] \times  \exp \bigg \{ \sum_{k=0} {S_{k n}} (r)L^{-k} \bigg \} \;.
\end{equation}
Eqs. (\ref{eq:DOansatz})-(\ref{eq:v(r)n}) may then be substituted into Eq. (\ref{eq:SWEpsi}) to obtain
\begin{equation}  \label{eq:ODE2}
f(r) \frac{d}{dr} \left[ f(r) \frac{dv_{\ell n}(r)}{dr} \right] + 2i \omega_{\ell n} \rho(r)\frac{dv_{\ell n}(r)}{dr} +  \left[  i \omega_{\ell n}  f(r) \frac{d \rho}{dr} +  \left[ 1 - \rho(r)^2 \right] \;\omega_{\ell n}^2  - V(r) \right]v_{\ell n}(r) = 0 \;.
\end{equation}

\par The primary objective of the Dolan-Ottewill multipolar expansion method is then to solve for the coefficients of Eq. (\ref{eq:QNFseries}) using Eq. (\ref{eq:ODE2}) in an iterative fashion, order by order. To compute ${\overline{\omega}_{k+1}}$ for ${k\geq 0}$, we require an expression for ${dS_{k n}/dr}$. Through solving for the quantities of ${\overline{\omega}_{k}}$ and ${dS_{k n}/dr}$, the QNF can be generated as a function of $L$. For lightly-damped modes, the QNF spectrum for spin-0 fields in a Schwarzschild black hole space-time is given by
\begin{equation}
    \omega_{n \ell} \approx \frac{1}{\sqrt{27}} \left[L - iN + \left[ \frac{29}{432}-\frac{5N^{{2}}}{36} \right] \frac{1}{L}- iN \left[ \frac{313}{15552} +\frac{235N^2}{3888}\right] \frac{1}{L^2} + ... \right] \;.
    \label{eq:QNF(2)}
\end{equation}

\par Similarly, the radial component of the wavefunction in Eq. (\ref{eq:v(r)n}) can be constructed by integrating the explicit expressions for ${dS_{k n}/dr}$. As shown in Ref. \cite{refDolanOttewill2011}, these can be used to construct the wavefunction and to study the properties of large-$\ell$ asymptotics in the context of QNM wave propagation at leading order. We shall see in Section \ref{sec:QNEF} how this enables us to express the wavefunction, as well as the QNEF for higher-orders in $L$. Note that throughout the text, we shall refer to 
{{the \enquote{lowest-order} QNF as $\omega^{(-1)}_{\ell n} = {\overline{\omega}_{-1} L} =L/\sqrt{27}$ and the \enquote{leading-order} QNF as $\omega^{(0)}_{\ell n} = {\overline{\omega}_{-1 }L + \overline{\omega}_{0 } }= (L - iN)/\sqrt{27}$.}} 
{Furthermore, we make use of the superscript ``$(k)$" to represent the ``order" of each function or variable.} 

\par Recall that the QNEF indicates how (and by how much) the QNMs are excited by some given initial data. As already mentioned, to compute this QNEF expression, Dolan and Ottewill extended their QNF analysis of Ref. \cite{refDolanOttewill2009}. However, a hallmark of that method is a breakdown at the critical orbit $r=r_c=3$ that cannot be circumvented in the  case of the wavefunction computation. As such, the solution space is divided into an \enquote{interior} region near $r=3$ and \enquote{exterior} regions defined by $r> (3+ {\epsilon '})$ and $r < (3-{\epsilon '})$ for some infinitesimally-small ${\epsilon '}$. While the exterior solution is based on the photon-sphere arguments of the QNF calculation made in Refs. \cite{refDolanOttewill2009,refDolanOttewill2011}, the interior solution follows the low-order WKB method of Refs. \cite{refBHWKB0,refBHWKB0.5}; specifically, the interior solution reduces to the asymptotic form of the parabolic cylinder function $\psi_1 \sim D_{{a}} ((-1+i)z)$. Here, we shall closely follow Appendix A of Ref. \cite{refDolanOttewill2011}, in order to review and clarify the method for the computation of the QNEF at leading order. 

\par The procedure begins with the introduction of a perturbation factor into the QNF series expansion Eq. (\ref{eq:QNFseries}),
\begin{equation} \label{eq:omega0perturbed}
\widetilde{\omega}_{\ell n} = \omega_{\ell n} + \epsilon   \;.
\end{equation}
\noindent {Throughout, we indicate variables or functions subjected to a perturbation with an overhead tilde ``$ \sim $".} {We distinguish between terms independent of $\epsilon$ and coupled to $\epsilon$ as being of order $\mathcal{O}(\epsilon^0)$ and of order $\mathcal{O}(\epsilon)$, respectively. Note also that $\epsilon$ is not related to $\epsilon^{\prime}$.}

\setlength{\parindent}{1em}
\subsection{Interior solution: using the parabolic cylinder function \label{sssec:intDO}}

\par The first step is an algebraic manipulation of Eq. (\ref{eq:SWEpsi}). We introduce $\ell = L -1/2$ in the interior function ansatz $u_{\ell \omega}(r) = f^{-1/2}\psi$. This yields
\begin{equation}
 f^{3/2} \psi^{\prime \prime} + \frac{2
 }{r^3}f^{1/2}\psi + \frac{1
 }{r^4} f^{-1/2}\psi +  \left[ \omega^2 - f \left[ \frac{L^2-1/4}{r^2} \right] - f\frac{2
 }{r^3} \right](f^{-1/2}\psi)   =  0 \;, 
\end{equation}
which we multiply by $f^{-3/2}$ to obtain
\begin{equation} \label{eq:SWEpsiU}
\frac{d^2\psi}{dr^2} +U(r) \psi = 0 \;, \hspace{0.65cm} U(r) = f^{-2} \left[ \omega^2 - f \left( \frac{L^2-1/4}{r^2} \right) + \frac{1
}{r^4} \right] \;.
\end{equation}

\par Consider the change of variables,
\begin{equation} \label{eq:rtoz}
r= 3+ \sqrt{\frac{3}{L}} z \;,
\end{equation}
and the first two terms of Eq. (\ref{eq:QNFseries}) evaluated at the leading-order QNF,
\begin{equation}
(\omega^{(0)}_{{\ell n}})^2 \approx ( \overline{\omega}{_{-1}}L + \overline{\omega}{_{0}})^2= \frac{1}{27}L^2 + \frac{2 \overline{\omega}{_{0}} L}{\sqrt{27}} + \overline{\omega}{_{0}}^2 \;.
\end{equation}
Substituting these into Eq. ({\ref{eq:SWEpsiU}}) and extracting the dominant $r$-term, \textit{viz.} $f^{-2}\omega^2$, produces the potential
\begin{equation} \label{eq:SWEpsiUnew}
U(r) = \left( 2\sqrt{3} \overline{\omega}{_{0}}+ \frac{z^2}{3} \right)L + \mathcal{O} (L^{1/2}) 
\end{equation}
at leading order ($\mathcal{O}(L^1)$). Eq. (\ref{eq:SWEpsiU}) can then be written as
\begin{equation}
\frac{d^2 \psi}{dz^2} + \left(2 \sqrt{27} \overline{\omega}{_{0}} +z^2 \right)\psi = 0 \;;
\end{equation}
upon the replacement of $\overline{\omega}{_{0}}$ with the perturbed leading-order Eq. (\ref{eq:omega0perturbed}), we obtain
\begin{equation} \label{eq:InteriorODE}
\frac{d^2 \psi}{dz^2} + (-2iN + 2\sqrt{27}\epsilon +z^2)\psi = 0 \;.
\end{equation}
Note that an equivalent expression was determined in Ref. \cite{Yang2014_KerrGreen} using arguments based on the WKB results of Ref. \cite{refBHWKB1}. Eq. (\ref{eq:InteriorODE}) leads us to two independent solutions that can be expressed in terms of the parabolic cylinder functions \cite{refBHWKB0,refBHWKB0.5,refmathbible}, 
\begin{align}
\psi_1 & =  D_{n+ \eta} \left[ z(-1+i) \right] \; , \label{eq:psi1Schwarz} \\
\psi_2 & =  D_{n+\eta} \left[ z(+1 - i) \right] \label{eq:psi2Schwarz} \;.
\end{align}
Here, $n$ is the overtone number and $\eta = i\epsilon \sqrt{27}$. The asymptotic behaviour of Eq. (\ref{eq:psi1Schwarz}) is given by
\begin{align} \label{eq:psi1asympSchwarz}
\psi_1 \sim
\begin{cases}
\displaystyle 2^{(n+\eta)/2}e^{-\;i \pi (n+\eta)/4} \vert z \vert^{(n+\eta)} e^{+iz^2/2} \;, & z \rightarrow -\infty 
\\
\displaystyle 2^{(n+\eta)/2} e^{+3i \pi (n+\eta)/4} \vert z \vert^{(n+\eta)} e^{+iz^2/2} \;   & 
\\
\displaystyle \hspace{0.3cm} -\frac{(2 \pi)^{1/2}}{\Gamma (-(n+\eta))} \frac{e^{+i \pi (n+\eta)} e^{-iz^2/2} }{ e^{3i \pi (n+\eta +1)/4} \; 2^{(n+\eta +1)/2} \; \vert z \vert^{(n+\eta +1)}}  \;, & z \rightarrow +\infty \;. 
\end{cases}
\end{align}
The appropriate solution for this physical context is $\psi_1$: as $z \rightarrow \infty$, the two terms of $\psi_1$ correspond properly to out-going and in-going waves; $\psi_2$ on the other hand has both out-going and in-going parts for $z \rightarrow - \infty$ (see Appendix \ref{sec:scattering} for details). To satisfy the requirement that waves are purely in-going at the horizon, $\psi_2$ cannot be used so we employ $\psi_1$ only. 

\par We then take the lowest-order approximation in $\epsilon$ (through $\eta=i\epsilon\sqrt{27} \rightarrow 0$), such that Eq. (\ref{eq:psi1asympSchwarz}) becomes
\begin{align} \label{eq:psi1asympSchwarzApprox}
\psi_1 \sim
\begin{cases}
\displaystyle 2^{n/2}e^{-\;i \pi n/4} \vert z \vert^{n} e^{+iz^2/2} \;, & z \rightarrow -\infty \; , \\
\displaystyle 2^{n/2} e^{+3i \pi n/4} \vert z \vert^{n} e^{+iz^2/2}   +\eta \frac{(2 \pi)^{1/2} \Gamma(n+1) e^{-iz^2/2} }{ e^{3i \pi (n+1)/4} \; 2^{(n +1)/2} \; \vert z \vert^{(n+1)}}  \;, & z \rightarrow +\infty \;. 
\end{cases}
\end{align}

\subsection{Exterior solution: using the Dolan-Ottewill ansatz \label{sssec:extDO}}

\par We utilise the Dolan-Ottewill ansatz introduced in Eq. (\ref{eq:DOansatz}),
\begin{equation} \label{eq:Upm}
u^{\pm} (r) = \exp \Bigg \{ \pm i \omega \int^r_3 \left(1 + \frac{6}{r} \right)^{1/2} \left(1 - \frac{3}{r} \right) dr_{\star} \Bigg \} v^{\pm}(r) \;,
\end{equation} 
where the ``$\pm$" superscripts indicate that we take into account {two types of boundary conditions in the exterior region. The ``$+$" sign represents the purely in-going wave near the event horizon and purely out-going wave at spatial infinity. These are the QNM boundary conditions; as such, only $u^+$ corresponds to QNMs. On the other hand, the ``$-$" sign represents the purely out-going wave near the event horizon and the purely in-going wave at spatial infinity. Note carefully, however, that under the perturbation of the QNFs as in Eq. (\ref{eq:omega0perturbed}), the first order correction of the in-going coefficient in the matching region also contributes to the study of QNEFs. As such $u^{-}$ is required to provide a linear combination of in- and out-going waves in the matching region. With this setup, the subsequent wave equation {is given by}
\begin{equation} 
f \frac{d^2 v^{\pm}(r)}{dr^2} + \left[\frac{2}{r^2} \pm 2i \omega \left( 1 + \frac{6}{r} \right)^{1/2} \left(1 - \frac{3}{r} \right) \right] \frac{dv^{\pm}(r)}{dr} + \bigg[ \frac{27 \omega^2 - L^2}{r^2} \pm \frac{27 i \omega}{r^3} \left(1 + \frac{6}{r} \right)^{-1/2} + \frac{1}{4r^2} - \frac{2 }{r^3} \bigg] v^{\pm}(r) = 0 \;. \label{eq:SWEdo}
\end{equation}
Near the critical orbit $r=r_c=3$, we can make the approximation by taking $z = r(1-3/r) \sqrt{L/3}$ from Eq. (\ref{eq:rtoz}),
\begin{equation} \label{eq:expL0}
\exp \Bigg \{ \pm i \omega \int^{r_{\star}}_{3} \left( 1 + \frac{6}{r} \right)^{-1/2} \left( 1 - \frac{3}{r} \right) dr_{\star} \Bigg \} \approx  \exp \{ \pm i z^2/2 + \mathcal{O} (L^{-1}) \}  \;.
\end{equation}
With this, we have the exponential component of $u^{\pm}$. At leading order, $v^{\pm}(r)$ of Eq. (\ref{eq:SWEdo}) reduces to
\begin{equation}
\label{eq:barspm0n}
    v^{{\pm}(0)} (r) \approx \exp \big \{ {\widetilde{S}^{\pm}_{0n}} (r) L^0 \big \} \;.
\end{equation}
\noindent For each side of the potential barrier, the perturbed ${\widetilde{S}^{\pm}_{0 n}} (r)$ functions may be written as:
\begin{align}
\widetilde{S}^{+}_{{0 n }} & =  S_{{0 n}} (r) + \eta Z_{0}(r) \; , \label{eq:S+onpert}\\
\widetilde{S}^{-}_{{0 n}} & =  S_{{0 n}} (r) - \left[ 2N + \eta \right] Z_{0}(r) \label{eq:S-onpert} \;.
\end{align}
To determine $S_{{0 n}} (r)$, we substitute the leading-order expression for $v_{{\ell n}}(r)$ into Eq. (\ref{eq:ODE2}) and solve for $dS_{{0 n}}/dr$. We then integrate over $r$. Recall that to determine an explicit expression for $dS_{{0 n}}/dr$, we need to introduce $\omega$ to order $\mathcal{O} (L^{-k})$. As such, for $dS_{{0 n}}/dr$, we evaluate the function at $\omega_{\ell n} =(L -iN)/\sqrt{27}$. Note that this leading-order expression is independent of the spin of the perturbing field. 

\par Similarly, {for $\widetilde{S}^{\pm}_{{0 n }}$ we} 
insert $\omega \rightarrow \widetilde{\omega}_{{\ell n }} = (L-i/2)/\sqrt{27} + \epsilon$ {and Eq. (\ref{eq:barspm0n})} into Eq. (\ref{eq:SWEdo}). After the integration, the term linear in $\epsilon$ (i.e. the $\mathcal{O} \left(\epsilon \right)$ term) in Eq. (\ref{eq:S+onpert}) can be realised as $\epsilon dZ_{0}/dr$. }
We then integrate this term with respect to $r$ and impose that $Z_{0} (r) \rightarrow 0$ as $r \rightarrow 2$. As such,
\begin{align}
\frac{d Z_{0}(r)}{dr}  &= \frac{\sqrt{27}}{r^2(1-\frac{3}{r})\sqrt{1+\frac{6}{r}}}, \\
\Rightarrow Z_{0}(r) & =  \ln |r-3| - \ln ( 3 + 2r + \sqrt{3r(r+6))} + \ln \xi, \; \label{eq:Z0}
\end{align} 
where $\xi=\left(2+\sqrt{3} \right)^2$. Note that $Z_{0}(r)$ is also independent of the spin of the field and the overtone number. {We emphasise that solving for $\widetilde{S}^{-}_{0n}$, following the same procedure, yields $S_{0n}(r)$ and $[2N + \eta]Z_{0}(r)$; $S_{0n}(r)$ contributes terms of order $\mathcal{O}(\epsilon^0)$ while $[2N + \eta]Z_{0}(r)$ contributes terms of both $\mathcal{O}(\epsilon^0)$ and $\mathcal{O}(\epsilon)$.} Using $\eta = \sqrt{27}i \epsilon$ and $y=\sqrt{1+6/r}$, {the explicit expressions for Eqs. (\ref{eq:S+onpert}) and (\ref{eq:S-onpert}) may produced as}
\begin{align}
\widetilde{S}^{+}_{0 n} & = \frac{1}{2} \ln \bigg \{ \frac{2}{y} \bigg \} + 2 N \ln \bigg \{ \frac{2 + \sqrt{3}}{y + \sqrt{3}} \bigg \}  + \eta \ln \bigg \{ \frac{r-3}{3 + 2r + \sqrt{3}yr} \xi \bigg \} \; , \\
\widetilde{S}^{-}_{0 n} & = \frac{1}{2} \ln \bigg \{ \frac{2}{y} \bigg \} + 2 N \ln \bigg \{ \frac{2 + \sqrt{3}}{y + \sqrt{3}} \bigg \}  - \left[ 2N + \eta \right] \ln \bigg \{ \frac{r-3}{3 + 2r + \sqrt{3}yr} \xi \bigg \} \; .
\end{align}
We then apply the change of variables from Eq. (\ref{eq:rtoz}) 
 into $S{_{0n}(r)}$ and $Z{_{0}(r)}$, and impose $z/\sqrt{L} \rightarrow 0$ (for $|z| \gg 1$). This yields
\begin{align}
S{_{0n}} (z) & \approx  \ln \Bigg \{ \left( \frac{4}{3} \right)^{1/4} \left( \frac{\xi}{12} \right)^N \Bigg \} \;,\label{eq:approxS0n} \\
Z{_{0}} (z) & \approx  \ln \left( \frac{\xi |z| }{2 \sqrt{27L}} \right) \label{eq:approxZ0n} \;.
\end{align}
\noindent In Eq. (\ref{eq:approxZ0n}), we include the $\xi$  missing from Eq. (A25) of the original work (c.f. the wavefunction of Eq. (A26) in Ref. \cite{refDolanOttewill2011}, where $\xi$ is included once again).
Finally, we substitute the above approximations into Eq. (\ref{eq:Upm}) to obtain the asymptotics of $u^{\pm}$,
\begin{align}
u^{+ (0)} &\approx  e^{+iz^2/2} \times \left(1-\frac{3}{r} \right)^n \exp \{ S{_{0 n}}(z)  + \eta Z{_{0}}(z) \} \nonumber \\
& \approx  \sqrt{2} e^{+iz^2/2} \left(\frac{z}{L^{1/2}} \right)^n \left( \frac{\xi}{4 \sqrt{27}}\right)^N \left(\frac{\xi |z|}{2 \sqrt{27L}} \right)^{\eta} \;;
 \label{eq:Uplus} \\
u^{- (0)} &\approx e^{-iz^2/2} \times \left(1-\frac{3}{r} \right)^n \exp \{ S{_{0n}}(z) - [2N + \eta] Z{_{0}}(z) \} \nonumber \\
& \approx \sqrt{2} e^{-iz^2/2} \left(\frac{z}{L^{1/2}} \right)^{-(n+1)} \left( \frac{\xi}{\sqrt{27}}\right)^{-N} \left(\frac{\xi |z|}{2 \sqrt{27L}} \right)^{-\eta} \;. \label{eq:Uminus}
\end{align}

\subsection{Matching procedure \label{sss:matchy}}

\par We now need to match the solutions for $r \sim 3$ $(u^{(0)}_{r=3})$\footnote{Recall that for the interior function, $u_{\ell \omega} = f^{-1/2}\psi$. At leading order, Dolan and Ottewill approximate this as $u_{\ell \omega} \sim \psi$. At higher orders, we shall see that this prefactor plays a more significant role.}, $r>(3+ {\epsilon'})$ $(u^{(0)}_{r>3})$, and $r < (3 - {\epsilon'})$ $(u^{(0)}_{r<3})$. For the region near $r=r_c=3$, we have already justified that the $\psi_1$ is the appropriate choice, such that 
\begin{equation}
u^{(0)}_{r=3} \approx \psi^{(0)}_1 = D_{n + \eta} (z(-1+i)) \;.
\end{equation} 
Then for the asymptotic regions,
\begin{align}
u^{(0)}_{r<3} & =  C^{(0)}_{in}u^{+ (0)} \;, \\
u^{(0)}_{r>3} & =  B^{(0)}_{out}u^{+ (0)} + B^{(0)}_{in}u^{- \; (0)} \;,
\end{align}
for $z \rightarrow - \infty \; (r < 3)$ and $z \rightarrow + \infty \; (r>3)$, respectively. 
For the former, we determine $C^{(0)}_{in}$ by matching $\psi_1$ for $z \rightarrow -\infty$ of Eq. (\ref{eq:psi1asympSchwarz}) to Eq. (\ref{eq:Uplus}). For the latter, we determine $B^{(0)}_{out}$ by matching the first term of Eq. (\ref{eq:psi1asympSchwarz}) for $z \rightarrow +\infty$ with Eq. (\ref{eq:Uplus}) and $B_{in}^{(0)}$ by matching the second term of Eq. (\ref{eq:psi1asympSchwarz}) for $z \rightarrow +\infty$ with Eq. (\ref{eq:Uminus}). {Note that these asymptotic limits  for $z$ are set for ``mathematical convenience"; the exact matching region is described by a sufficiently large $|z|$ with a finite (sufficiently small)  $|z|/\sqrt{L}$. This is because $\epsilon'\sim z/\sqrt{L}$, as presented in Eq. (\ref{eq:rtoz}). With this in place, the matching region shall satisfy both interior and exterior waveforms. We elaborate upon this in Section \ref{subsec:matching}.} Explicitly, for $C^{(0)}_{in}$,
\begin{align}
\psi^{(0)}_{z \rightarrow - \infty} & =  C^{(0)}_{in} u^{+ (0)} \nonumber \\
2^{n + \eta} e^{-i \pi (n+\eta)/4} e^{+iz^2/2} & =  C^{(0)}_{in} \sqrt{2} e^{+iz^2/2} \left(\frac{z}{L^{1/2}} \right)^n \left( \frac{\xi}{4 \sqrt{27}}\right)^N \left(\frac{\xi |z|}{2 \sqrt{27L}} \right)^{\eta} \; , \nonumber \\
\mathrm{and} \; \mathrm{by} \; \mathrm{imposing} \; \eta \rightarrow 0  \;\; \;\; \;\;
C^{(0)}_{in} & =  2^{2n} 2^{(n+1)/2} L^{n/2} \left( \frac{\xi}{ \sqrt{27}}\right)^{-N} {e^{-i \pi n/4} }\;.
\label{eq:cin0}
\end{align}
Similarly, for $B^{(0)}_{out}$, we use the $e^{+iz^2/2}$ term of Eq. (\ref{eq:psi1asympSchwarz}) to obtain
\begin{align}
\psi^{(0)}_{z \rightarrow + \infty}  & =  B^{(0)}_{out} u^{+ (0)} \nonumber \\
2^{(n+\eta)/2} e^{+3i \pi (n+\eta)/4} \vert z \vert^{(n+\eta)} e^{+iz^2/2}  & =  B^{(0)}_{out} \sqrt{2} e^{+iz^2/2} \left(\frac{z}{L^{1/2}} \right)^n \left( \frac{\xi}{4 \sqrt{27}}\right)^N \left(\frac{\xi |z|}{2 \sqrt{27L}} \right)^{\eta} \; , \nonumber \\
\mathrm{and} \; \mathrm{by} \; \mathrm{imposing} \; \eta \rightarrow 0 \;\; \;\; \;\; B^{(0)}_{out} & =  (-1)^n C^{(0)}_{in} \;.
\label{eq:bout0}
\end{align}  
Note how we have exploited $e^{+3i \pi n/4}=e^{ i \pi n-i \pi n/4}$, where $(e^{i \pi})^n=(-1)^n$, to express $B^{(0)}_{out}$ as a function of $C^{(0)}_{in}$. 

\par Finally, for $B^{(0)}_{in}$ we use the $e^{-iz^2/2}$ term of Eq. (\ref{eq:psi1asympSchwarz}) and Eq. (\ref{eq:Uminus}). 
\begin{align}
\psi^{(0)}_{z \rightarrow + \infty} & =  B^{(0)}_{in} u^{- (0)} \nonumber \\
-\frac{(2 \pi)^{1/2}}{\Gamma (-(n+\eta))} \frac{e^{+i \pi (n+\eta)} e^{-iz^2/2} }{ e^{3i \pi (n+\eta +1)/4} \; 2^{(n+\eta +1)/2} \; \vert z \vert^{(n+\eta +1)}}
  & = B^{(0)}_{in} \sqrt{2} e^{-iz^2/2} \left(\frac{z}{L^{1/2}} \right)^{-(n+1)} \left( \frac{\xi}{\sqrt{27}}\right)^{-N}\left(\frac{\xi |z|}{2 \sqrt{27L}} \right)^{-\eta} \nonumber \\
\mathrm{up~ to ~order} \; \mathcal{O}(\epsilon):  \;\; \;\; \;\; \eta \frac{(2 \pi)^{1/2} \Gamma(n+1) e^{-iz^2/2} }{ e^{3i \pi (n+1)/4} \; 2^{(n +1)/2} \; \vert z \vert^{(n+1)}} 
 & =  B_{in}^{(0)} \sqrt{2} e^{-iz^2/2} \left(\frac{z}{L^{1/2}} \right)^{-(n+1)} \left( \frac{\xi}{\sqrt{27}}\right)^{-N} \; .
\end{align}  
Note in particular the use of the property $\left(\Gamma (-(n+\eta))\right)^{-1} = + \eta \; \Gamma(n+1)$. Through a number of algebraic manipulations to accommodate the introduction of the expression for $C^{(0)}_{in}$, we find that
\begin{equation}
\label{eq:bin0}
B^{(0)}_{in} = \epsilon \Gamma(n+1) 2^{-n} L^{-n} \left( \frac{27 \pi}{iL} \right)^{1/2} e^{-i \pi n /2} \left(\frac{\xi}{2 \sqrt{27}} \right)^{2N} C^{(0)}_{in}\;,
\end{equation} 
\noindent where we have corrected a minor typographical error by including the $C^{(0)}_{in}$ omitted from Eq. (A34) of the original work, Ref. \cite{refDolanOttewill2011}.

\subsection{in-going and out-going coefficients \label{sss:coef}}

\par To compute the in-going and out-going coefficients, $A^{\pm (0)}_{\ell n}$, we require the contribution from the \enquote{phase factors} \cite{refDolanOttewill2011}: 
\begin{align}
\alpha_1^{(0)} & =  \exp \Bigg \{ +i \omega \int^{r=2}_{r=3} \left(1 + \frac{6}{r} \right)^{1/2} \left(1 - \frac{3}{r} \right) \frac{dr}{f} \Bigg \} \exp \{ + i \omega r_{\star} \}                                                   \nonumber \\
& =\exp \{ i \omega [6 - \sqrt{27} + 8 \ln 2 - 3 \ln \xi ] \} \; , \\
\beta_1^{(0)} & =  \exp \Bigg \{ +i \omega \int^{r=\infty}_{r=3} \left(1 + \frac{6}{r} \right)^{1/2} \left(1 - \frac{3}{r} \right) \frac{dr}{f} \Bigg \} \exp \{ - i \omega r_{\star} \}                                                   \nonumber \\
& =\exp \{ i \omega [3 - \sqrt{27} + 4 \ln 2 - 3 \ln \xi ] \}  \; , \\
\gamma_1^{(0)} & =  \exp \Bigg \{ -i \omega \int^{r=\infty}_{r=3} \left(1 + \frac{6}{r} \right)^{1/2} \left(1 - \frac{3}{r} \right) \frac{dr}{f} \Bigg \} \exp \{ + i \omega r_{\star} \}                                                    = 1/\beta_1^{(0)} \;,
\end{align}
\begin{align}
\alpha_2^{(0)} & =  \lim_{r \rightarrow 2} e^{{\widetilde{S}^{+}_{0n}}} = 1 \;, \\
\beta_2^{(0)} & =  \lim_{r \rightarrow \infty} e^{{\widetilde{S}^{+}_{0n}}} = 2^{1/2} \left( \sqrt{\xi}/2 \right)^N \;, \\
\gamma_2^{(0)} & =  \lim_{r \rightarrow \infty} e^{{\widetilde{S}^{-}_{0n}}} = 2^{1/2} \left( 2 \sqrt{\xi} \right)^{-N} \;.
\end{align}

\par At leading order and evaluated at the perturbed QNF of Eq. ({\ref{eq:omega0perturbed}}),
\begin{align}
A^{+ (0)}_{\ell n}\Big \vert_{\omega \rightarrow \widetilde{\omega}_{\ell n}^{(0)}} & = 
\frac{\beta_1^{(0)} \beta_2^{(0)}}{\alpha_1^{(0)} \alpha_2^{(0)}} \frac{B^{(0)}_{out}}{C^{(0)}_{in}} \nonumber \\
& \approx   2^{1/2} (-1)^n \left( \frac{\sqrt{\xi}}{2} \right)^N \exp \{- i {\omega_{\ell n}^{(0)}} [3 + 4 \ln 2] + \mathcal{O} (\epsilon)\} \;, \label{eq:AplusLO}
\end{align}
since $\exp \{ 3 \pi i n /4 + \pi i n/4 \} = (e^{i \pi})^n = (-1)^n$. Then, evaluating at the perturbed QNF of Eq. (\ref{eq:omega0perturbed}),
\begin{align}
A^{- (0)}_{\ell n} \Big \vert_{\omega \rightarrow \widetilde{\omega}_{\ell n}^{(0)}} & =  \frac{\gamma_1^{(0)} \gamma_2^{(0)}}{\alpha_1^{(0)} \alpha_2^{(0)}} \frac{B^{(0)}_{in}}{C^{(0)}_{in}} \nonumber \\
& \approx  \epsilon \Gamma (n+1) 2^{-n} L^{-n} \left( \frac{27 \pi }{i L} \right)^{1/2} \left( \frac{\xi}{2 \sqrt{27}} \right)^{2N} 2^{1/2} \left(2 \sqrt{ \xi } \right)^{-N} {e^{-i n \pi/2}}\nonumber \\
& \quad \quad \times \exp \{ -i {\omega_{\ell n}^{(0)}} [9 - 2\sqrt{27} + 12 \ln 2 - 6 \ln \xi ] + \mathcal{O} (\epsilon)\} \;. \label{eq:AminusLO}
\end{align}
\noindent From Eq. (\ref{eq:AminusLO}), it is clear that the in-going coefficient at order {$\mathcal{O}(\epsilon^{0})$} vanishes; it is from the perturbation in the QNF {(see Eq. (\ref{eq:omega0perturbed})}) that we obtain a non-zero contribution from $A^{- (0)}_{\ell n}.$ 
Eq. (\ref{eq:AminusLO}) can be viewed as a Taylor-expansion of $A_{\ell n}^{-(0)}$ with respect to $\tilde{\omega}_{\ell n}^{(0)}$ around the QNF $\omega_{\ell n}^{(0)}$. We can demonstrate this explicitly by writing the full perturbed expression for $A^{-(k)}_{\ell n}$,
\begin{equation}
A^{-(k)}_{\ell n}|_{\omega\rightarrow\widetilde{\omega}_{\ell n}^{(k)}}=0+\epsilon\left(\frac{\partial A^{-(k)}_{\ell n}}{\partial \omega}\right)\Bigg \vert_{\substack{\omega \rightarrow \omega_{\ell n}^{(k)}}} +\mathcal{O} (\epsilon^{2}).
\end{equation}
In this case, Eq. (\ref{eq:AminusLO}) is naturally of order $\mathcal{O}(\epsilon)$ and higher, where the coefficient of $\epsilon$ shall be the first partial derivative term of $A^{-(k)}_{\ell n}$ evaluated at $\omega_{\ell n}^{(k)}$. That is, for the leading order herewith studied,
\begin{align}
\left ( \frac{\partial A^{-(0) }_{\ell n}}{\partial \omega}  \right) \Bigg \vert_{\omega\rightarrow\omega_{\ell n}^{(0)}}&= \Gamma (n+1) 2^{-n} L^{-n} \left( \frac{27 \pi }{i L} \right)^{1/2} \left( \frac{\xi}{2 \sqrt{27}} \right)^{2N} 2^{1/2} \left(2 \sqrt{ \xi } \right)^{-N} e^{-i n \pi/2}\nonumber \\
& \quad \quad \times \exp \{ -i \omega_{\ell n}^{(0)} [9 - 2\sqrt{27} + 12 \ln 2 - 6 \ln \xi ] \} \;.
\end{align}
The $\mathcal{O} (\epsilon^{2})$ term will not contribute to the evaluation of the QNEFs. This is also the justification for maintaining $A^{-(k)}_{\ell n}$ only up to order $\mathcal{O}(\epsilon)$ for our higher-order $L^{-k}$ results.

\subsection{The leading-order quasinormal excitation factor \label{sss:QNEF}}

\begin{figure}[t]
\centering
\includegraphics[width=0.6\textwidth]{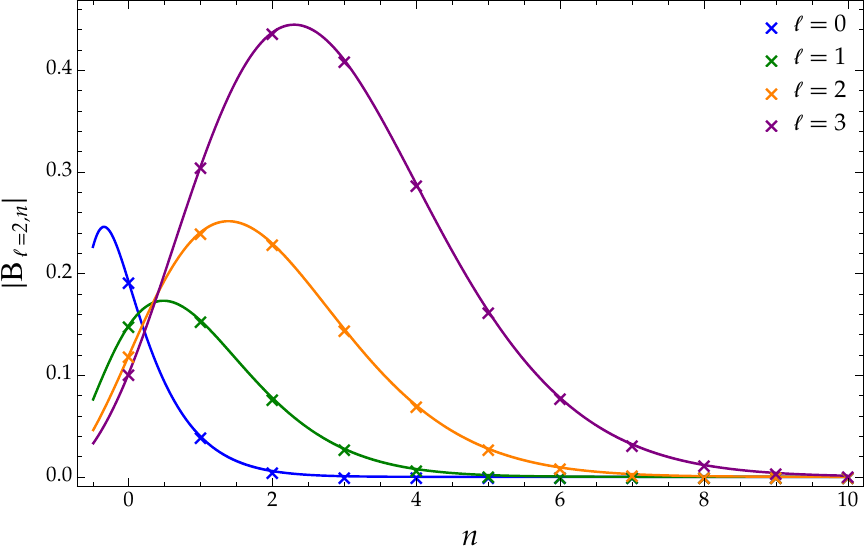}
\vskip 0.3cm
\caption{\textit{The magnitude of the $\ell=0,1,2,3$ QNEFs for increasing values of $n$ at leading order, following Eq. (\ref{eq:QNEFLO_Fig4}).}}
\label{fig:QNEFvsOvertones}
\end{figure}

\par The QNEF at order $L^k$ is defined as
\begin{equation} \label{eq:QNEF}
\mathcal{B}^{(k)}_{\ell n} \equiv 
\frac{A^{+ (k)}_{\ell n}}{2 {\omega}} \left( \frac{ \partial A^{- (k)}_{\ell n}}{\partial {\omega}} \right)^{-1} \Bigg \vert_{\substack{{\omega}\rightarrow\omega^{(k)}_{\ell n}}} 
\end{equation}
Using the expressions for Eqs. (\ref{eq:AplusLO}) and (\ref{eq:AminusLO}), we find that the leading-order QNEF for general $n$ becomes
\begin{equation}
\mathcal{B}_{\ell n}^{(0)} = \frac{(-i L)^{n-1/2}}{n!(\sqrt{27}\omega^{(0)}_{\ell n}/L)}\frac{\exp \{ 2 i \omega^{(0)}_{\ell n}\zeta \}}{\sqrt{8\pi}}\left(\frac{216}{\xi}\right)^{n+1/2} \label{eq:QNEFLO} \;,
\end{equation}
\noindent where we make use of the constant defined in Ref. \cite{refDolanOttewill2011},
\begin{equation}
\zeta=3-\sqrt{27}+4\ln{2}-6\ln{\left(2+\sqrt{3}\right)} \label{eq:zita} \;,
\end{equation}
\noindent and the property $\Gamma(n+1) = n!$. This is precisely Eq. (A34) of Ref. \cite{refDolanOttewill2011}, where we have corrected a minor typographical error in the denominator. Note, however, that the QNEF result quoted in the main text of Ref. \cite{refDolanOttewill2011}, Eq. (31), can be obtained by substituting $\omega^{(0)}_{\ell n} \approx (L-iN)/\sqrt{27}$ into the $\exp \{ 2 i \omega^{(0)}_{\ell n}\zeta \}$ of Eq. (\ref{eq:QNEFLO}), but only with the lowest-order $\omega^{(-1)}_{\ell n} \approx L/\sqrt{27}$ term being in the denominator. For the associated Fig. 4 of Ref. \cite{refDolanOttewill2011}, the values plotted correspond to Eq. (\ref{eq:QNEFLO}) evaluated at $\omega^{(0)}_{\ell n} \approx (L-iN)/\sqrt{27}$ for $n=0$, $viz.$
\begin{equation}
\mathcal{B}^{(0)}_{\ell  n} = \left(\frac{i}{L} \right)^{1/2} \frac{B \; e^{2 i \zeta L/\sqrt{27}}}{1-\frac{i \left(n+ \frac{1}{2} \right)}{L}}   \frac{\left(-i \kappa L \right)^n}{n!} \Bigg \vert_{n=0} = \left(\frac{i}{L} \right)^{1/2} \frac{B \; e^{2 i \zeta L/\sqrt{27}}}{1-\frac{i}{2L}} \;, \quad  B = \left( \frac{27}{\xi \pi} \right)^{1/2}e^{\zeta /\sqrt{27}} \;,\;\; \kappa = \frac{216 e^{2 \zeta /\sqrt{27}}}{\xi} \label{eq:QNEFLO_Fig4} \;, 
\end{equation}
where we have made use of the constants $B$ and $\kappa$ introduced in Ref. \cite{refDolanOttewill2011}. We can simplify Eq. (\ref{eq:QNEFLO_Fig4}) further as
\begin{equation}
\mathcal{B}_{\ell 0}^{(0)}=\left(\frac{\exp \{ 2 i \omega_{\ell 0}^{(0)} \zeta \}}{\omega_{\ell 0}^{(0)}}\right)\left(\frac{\sqrt{L}}{\left(2+\sqrt{3}\right)\sqrt{-i\pi}}\right),\label{eq:QNEFloPenultimate}
\end{equation}
\noindent where $\omega_{\ell 0}^{(0)}= \overline{\omega}_{-1}L + \overline{\omega}_0$ refers to the QNF for $n=0$ at leading order $L^{(0)}$.

\par For the QNEF corresponding to the first four harmonics for increasing overtone numbers, see Fig. \ref{fig:QNEFvsOvertones}. For the fundamental mode, Eq. (\ref{eq:QNEFLO_Fig4}) is in agreement with our leading-order result, plotted in Fig. \ref{fig:QNEFNNLO}.

 \par In Fig. \ref{fig:QNEFvsOvertones}, we plot the results for the leading-order QNEF for the first several harmonics and overtones. There, the QNEF can be seen to increase with $n$, bar the slight fall for $n=1$ (where if we interpolate between the discrete $n$ points, we observe that the peaks of the $\ell = 0$ and $\ell=2$ curves are almost equivalent). We note with interest that the contribution of the QNEF is at a maximum when $\ell=n$ for each QNEF. 

\section{Quasinormal wavefunctions and excitation factors \label{sec:QNEF}}

\par In Section \ref{sec:Schwarz}, we provided a review of the leading-order calculation of the QNEF using the Dolan-Ottewill method. In this section, we demonstrate a detailed calculation of our higher-order QNEF computation for the $n=0$ case. We show explicit results up to $\mathcal{O} (L^{-2})$. To elevate the interior solution to higher orders, we follow the strategy of Ref. \cite{refBHWKB1} in which a solution of the form $\psi \sim (g^{\prime}(z))^{-1/2} D_{a}[g(z)]$ is used to achieve a third-order expression, {where $D_{a}[g(z)]$ is the parabolic cylinder function}. For the exterior solution, higher-order expressions can be achieved if we use the Dolan-Ottewill method to compute expressions for the first few $S_{k0}(r)$ functions of Eq. (\ref{eq:v(r)n}). The final step of the QNEF calculation then primarily involves the appropriate matching of the solutions.

\subsection{Preliminaries \label{subsec:preamble}}

\par As before, we begin with the effective QNM potential for a scalar test field for $f(r)=1-2/r$,
\begin{equation} \label{eq:Uraw}
\frac{d^2\psi}{dr^2} + U(r) \psi = 0 \;, \quad \quad U(r) = \frac{1}{f(r)^2} \left[ \omega^2 - f(r) \left( \frac{L^2 - \frac{1}{4}}{r^2} \right) + \frac{1}{r^4} \right] \;,
\end{equation}
where the perturbed QNF is given by
\begin{align} \label{eq:omega2perturbed}
\widetilde{\omega}_{\ell n} &= \omega_{\ell n} + \epsilon \nonumber \\
& \approx  \overline{\omega}_{-1}L + \overline{\omega}_{0} + \overline{\omega}_{1} L^{-1} + \overline{\omega}_{2} L^{-2} + \epsilon + \mathcal{O} \left(L^{-3} \right)\;.
\end{align}

\noindent Here, we shall retain terms from the QNF series expansion of Eq. (\ref{eq:QNFseries}) up to order $\mathcal{O}(L^{-2})$. Following the standard Schultz-Iyer-Will technique \cite{refBHWKB0,refBHWKB0.5,refBHWKB1}, we expand the barrier potential $U(r)$ in a Taylor series about the maximum $r = r_p$,
\begin{equation} \label{eq:UTaylorsVersion}
U(r) \approx U(r_p) + \frac{1}{2} U^{\prime \prime}(r_p)(r - r_p)^2 + \frac{1}{6} U^{(3)} (r - r_p)^3 + ... \; .
\end{equation}
If we solve for $r_p = r_0 + r_1 L^{-1} + r_2 L^{-2} + ...$ using $U^{\prime}(r_p)=0$ in the large-$\ell$ limit, we find that at order $\mathcal{O}(L^{-2})$,
\begin{equation} \label{eq:rp}
r_p = 3 + 12 \sqrt{3} \overline{\omega} _0 L^{-1} + \left( \frac{11}{6} + 342 \overline{\omega} _0^2+12 \sqrt{3} \overline{\omega} _1 \right) L^{-2} + \mathcal{O} \left(L^{-3} \right) \;.
\end{equation}

\par Upon substituting Eq. (\ref{eq:rp}) into Eq. (\ref{eq:Uraw}),
\begin{equation}
U(r_p) = 2 \sqrt{3} L \overline{\omega} _0 + \left( \frac{7}{36} - 39 \overline{\omega} _0^2 + 2 \sqrt{3} \overline{\omega} _1 \right) + \frac{2 \left(504 \sqrt{3} \overline{\omega} _0^3-22 \sqrt{3} \overline{\omega} _0-351 \overline{\omega} _1 \overline{\omega} _0 + 9 \sqrt{3} \overline{\omega}_2\right)}{9} \frac{1}{L} +...
\end{equation}
for large-$\ell$ at higher orders. We continue in this way for each term of Eq. (\ref{eq:UTaylorsVersion}),
\begin{align}
\frac{1}{2} U^{\prime \prime} (r_p) & = \frac{L^2}{9} - \frac{16 \overline{\omega}_0}{\sqrt{3}}L + \left(-\frac{233}{324}  + \frac{888 \overline{\omega} _0^2}{3}- \frac{16 \overline{\omega}_1}{\sqrt{3}} \right)  +... \; , \label{eq:dUs} \\
\frac{1}{6} U^{(3)} (r_p) & = -\frac{20 L^2}{81} + \frac{1160 L \overline{\omega} _0}{27 \sqrt{3}} +... \; , \nonumber \\
\frac{1}{24} U^{(4)} (r_p) & = \frac{95 L^2}{243}-\frac{6544 L \overline{\omega} _0}{81 \sqrt{3}}  +... \; , \nonumber
\end{align} 
parameterising $r_p$ as
\begin{equation}
r - r_p = \frac{z}{\alpha} \;. \label{eq:zalpha}
\end{equation}
At leading order, $r_p=3$ and $\alpha = \sqrt{L/3}$ (c.f. Eq. (\ref{eq:rtoz}) of Section \ref{sec:Schwarz}). We determine $\alpha$ by substituting Eqs. (\ref{eq:dUs}) and (\ref{eq:zalpha}) into Eq. (\ref{eq:UTaylorsVersion}). However, the introduction of Eq. (\ref{eq:zalpha}) into the ordinary differential equation implies that the new equation to solve is
\begin{equation}
\frac{d^2}{d(r_p + z/\alpha)^2} \rightarrow \alpha^2 \frac{d^2}{dz^2} \quad \Rightarrow \quad \frac{d^2\psi}{dz^2} + \frac{1}{\alpha^2}U(z) \psi = 0 \;.
\end{equation}
This brings us to
\begin{align}
\frac{1}{\alpha^2} U(z) & \approx  \left[ 2 \sqrt{3} L \overline{\omega} _0 + \left( \frac{7}{36} - 39 \overline{\omega} _0^2 + 2 \sqrt{3} \overline{\omega} _1 \right) +... \right]\frac{1}{\alpha^2} + \left[ \frac{L^2}{9} - \frac{16 \overline{\omega}_0}{\sqrt{3}}L +... \right] \frac{z^2}{\alpha^4} \nonumber \\
& \hspace{1cm} + \left[-\frac{20 L^2}{81} + \frac{1160 L \overline{\omega} _0}{27 \sqrt{3}} +... \right] \frac{z^3}{\alpha^5} + \left[ \frac{95 L^2}{243}-\frac{6544 L \overline{\omega} _0}{81 \sqrt{3}} + ... \right] \frac{z^4}{\alpha^6} + ... \; .
\end{align}

\par The coefficient of the quadratic term in $z$ is expected to be 1 \cite{refBHWKB0,refBHWKB0.5,refBHWKB1}, {which allows us to solve for $\alpha$. Therefore,} 
we can rewrite Eq. (\ref{eq:zalpha}) as
\begin{equation} \label{eq:rtozNLO}
r(z) - \alpha^{-1} z= 3 + 12 \sqrt{3} \overline{\omega} _0 L^{-1} + \left( \frac{11}{6} + 342 \overline{\omega} _0^2+12 \sqrt{3} \overline{\omega} _1 \right) L^{-2} + \mathcal{O} \left(L^{-3}\right)  \; ,
\end{equation}
\noindent where 
\begin{equation}
    {\alpha^{-1}} = \frac{\sqrt{3}}{\sqrt{L}}+\frac{36 \overline{\omega} _0}{L^{3/2}}+\left[ \frac{233}{48 \sqrt{3}}+414 \sqrt{3} \overline{\omega} _0^2+36 \overline{\omega} _1 \right] \frac{1}{L^{5/2}} +\mathcal{O} \left(L^{-\frac{7}{2}}\right)
    \;.
\end{equation}
\begin{figure}[t]
    \centering
    \includegraphics[width=0.6\linewidth]{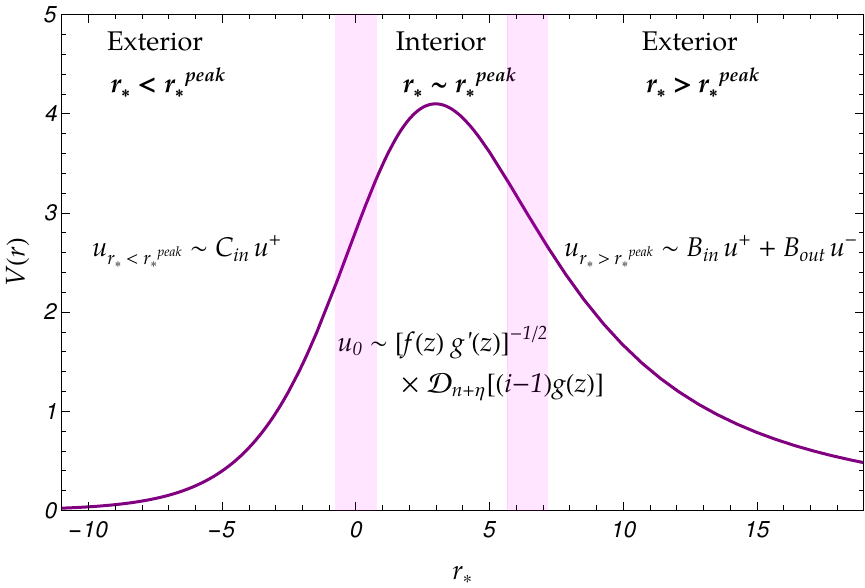}
    \caption{\textit{Diagrammatic representation of the solutions, where asymptotics are matched across the shaded regions. At leading-order, $g(z) \sim z$. 
    }}
    \label{fig:DOscattering}
\end{figure}

\subsection{Interior solution: using the parabolic cylinder function \label{sec:OurInterior}}

\par While the spin of the perturbing field is irrelevant at leading order, deviations emerge at higher orders of $L^{-k}$. Here, we consider $s=0$ perturbations, which allows us to use Eq. (\ref{eq:SWEpsi}) as before in order to obtain a new effective potential (c.f. Eq. (\ref{eq:SWEpsiUnew})). Following Refs. \cite{refBHWKB0,refBHWKB0.5,refBHWKB1}, we elevate the leading-order expressions to next-to-next-to-leading order by substituting a power series expansion in $g(z)$ (c.f. Eq. (3.16) of Ref. \cite{refBHWKB1}) in place of $z$ within the ordinary differential equation. Thereafter, we solve the resultant ordinary differential equation (c.f. Eq. (3.14) of Ref. \cite{refBHWKB1}) using functions of the parabolic cylinder functions of the form $\psi \sim (g^{\prime}(z))^{-1/2} D_{a}[g(z)]$.

\par We begin by substituting Eqs (\ref{eq:zalpha}) and (\ref{eq:rtozNLO}) into Eq. (\ref{eq:UTaylorsVersion}), such that the radial equation becomes
\begin{align}
\frac{d^2 \psi}{dz^2} & + \Bigg ( \left[ 6 \sqrt{3} \overline{\omega}_0 + \left( \frac{7}{12} + 315 \overline{\omega}_0^2 + 6 \sqrt{3} \overline{\omega}_1 \right) \frac{1}{L} + ... \right] + z^2  \nonumber \\
&+ \left[ -\frac{20\sqrt{3}}{9}\frac{1}{L^{1/2}} -\frac{40 \overline{\omega} _0}{3} \frac{1}{L^{3/2}}+... \right]z^3 + \left[\frac{95}{9} \frac{1}{L} +\frac{296\sqrt{3} \overline{\omega} _0}{9} \frac{1}{L^2} + ...\right]z^4 + ... \Bigg ) \psi = 0 
\end{align}
\begin{eqnarray}
\Rightarrow  \quad \frac{d^2 \psi}{dz^2} + Q(z) \psi &=& 0 \;. \label{eq:ODEQ}
\end{eqnarray}

\par Analogous to Eq. (3.15) of Ref. \cite{refBHWKB1} and Eq. (\ref{eq:SWEpsiUnew}), we can rewrite the potential as
\begin{equation} \label{eq:NLOpotQ}
Q(z) =  \sum_{n=0}^{\infty} \left(\frac{1}{L} \right)^{n/2} U_{n/2} (z)  \;,
\end{equation} 
and solve for each term, $viz.$

\begin{align}
U_0 & = 6 \sqrt{3} \overline{\omega}_0 + z^2 \;,\label{eq:U0} \\ 
U_{1/2} & = -\frac{20\sqrt{3} z^3}{9} \;, \\
U_1 & =\frac{7}{12} +  315 \overline{\omega} _0^2+6 \sqrt{3} \overline{\omega} _1+\frac{95 z^4}{9}  \;,  \\
U_{3/2} & =  -\frac{40\overline{\omega} _0 z^3}{3} - \frac{392 \sqrt{3} z^{5}}{27} \;, \\
U_2 & =   5088\sqrt{3}\overline{\omega} _0^3+\frac{75 \sqrt{3} \overline{\omega} _0}{4}+630 \overline{\omega} _0 \overline{\omega} _1 + 6\sqrt{3} \overline{\omega}_2+ \frac{296 \sqrt{3} \overline{\omega} _0 z^4}{9} +\frac{1499 z^{6}}{27} \;.
\end{align}
Following Eq. (3.14) of Ref. \cite{refBHWKB1} and Eq. (\ref{eq:psi1Schwarz}), we set the solution of Eq. (\ref{eq:ODEQ}) as
\begin{equation} \label{eq:psi1}
\psi = \left( \frac{dg}{dz} \right)^{-1/2} \mathcal{D}_a ((-1+i) g(z)) \;,
\end{equation}
where $\mathcal{D}_a ((-1+i) g(z))$ satisfies the parabolic cylinder function,
\begin{equation}
\left(\frac{d^2}{dg^2}+\lambda+g^{2}(z)\right)\mathcal{D}_a ((-1+i) g(z)) = 0 \;,
\end{equation}
and $\lambda$ is an unsolved variable that is constant in $z$, and $a=-(1-i\lambda)/2$. With Eq. (\ref{eq:psi1}), we can rewrite Eq. (\ref{eq:ODEQ}) in terms of $g(z)$,
\begin{equation}
\left( \frac{dg}{dz} \right)^2 \left[\lambda+ g(z)^2 \right] + \frac{1}{2} \left(\frac{dg}{dz} \right)^{-1} \left( \frac{d^3g}{dz^3} \right) - \frac{3}{4} \left( \frac{dg}{dz} \right)^{-2} \left( \frac{d^2g}{dz^2} \right)^2 - \sum_{n=0}^{\infty} \left(\frac{1}{L} \right)^{n/2} U_n (z) = 0 \;.
\end{equation}
Solving the above equation by taking
\begin{equation}
g(z) =  z+\sum_{n=1}^{\infty} \left( \frac{1}{L} \right)^{n/2} A_{n/2}(z) \; ,
\end{equation}
and 
\begin{equation}
\lambda=\sum_{k=0}^{\infty} \left( \frac{1}{L} \right)^{k} \lambda_{k} ~~ , ~~ 
A_{n/2}(z)\sim \sum_{k=0}^{\infty} p_{nk} z^k \;,
\end{equation}
we can solve for $\lambda$ and $A_{n/2} (z)$ iteratively by considering each coefficient of $L^{-n/2}$. For example, at the order of $L^{0}$ and $L^{-1/2}$, we generate
\begin{align}
L^{0}:  &\quad \lambda_{0}-6 \sqrt{3}\overline{\omega}_{0}=0 \;, \\
L^{-1/2}:  &\quad \frac{1}{18} \left(9 A_{1/2}^{(3)}(z)+36 z^2 A_{1/2}'(z)+72( \lambda_0-3\sqrt{3}\overline{\omega}_{0}) A_{1/2}'(z)+36 z A_{1/2}(z)+40 \sqrt{3} z^3\right) = 0 \;, 
\end{align}
and solve for each coefficient of $z$. Following this procedure, we find the result up to and including $\mathcal{O}(L^{-2})$:  
\begin{align}
\lambda &= \sum_{n=0}^{4} \left(\frac{1}{L} \right)^{n/2} U_{n/2} (0) -\frac{\Lambda}{L}-\frac{\Omega}{L^{2}} \;,\label{lambdahigherorder}\\
A_{1/2} &= \frac{40 \overline{\omega}_0 }{9 \sqrt{3}}-\frac{10 z^2}{9 \sqrt{3}} \;;\\
A_1 & = \frac{1265 z^3}{1944}+\frac{2605 \overline{\omega}_0 z}{216 \sqrt{3}} \;, \quad \quad \Lambda =\frac{155}{216} +\frac{645 \overline{\omega}_0^2}{2}\; , \\
A_{3/2} &= \frac{254179 \overline{\omega} _0^2}{1215 \sqrt{3}}+\frac{40 \overline{\omega} _1}{3}-\frac{58987 z^4}{43740 \sqrt{3}}-\frac{185179 \overline{\omega} _0 z^2}{43740}+\frac{6781}{3645 \sqrt{3}} \; ,  \\
A_2 &= \frac{7702759 z^5}{7558272}+\frac{4699901 \overline{\omega} _0 z^3}{629856 \sqrt{3}}+\left(\frac{108646309 \overline{\omega} _0^2}{839808}+\frac{2605 \overline{\omega} _1}{216 \sqrt{3}}+\frac{443665}{1679616}\right) z \;, \nonumber\\
\Omega &= \frac{121727 \overline{\omega} _0^3}{8 \sqrt{3}}+645 \overline{\omega} _1 \overline{\omega} _0+\frac{48623 \overline{\omega} _0}{864 \sqrt{3}} \;.
\end{align}
Furthermore, the index of the parabolic cylinder function $D_{a}((-1+i)g(z))$ is given by $a=-1/2 +i \lambda/2$. With the perturbation in Eq. (\ref{eq:omega2perturbed}) and the result in Eq. (\ref{lambdahigherorder}), we may write $ a=n+ i \epsilon \sqrt{27} + \mathcal{O}\left( \frac{1}{L} \right)$.

\par We now have all the necessary components to write the interior wavefunction. For the parabolic cylinder function, we use Eq. (\ref{eq:psi1Schwarz}) and Eq. (3.22) of Ref. \cite{refBHWKB1}, with $z \rightarrow g(z)$. For $\eta = i \epsilon \sqrt{27}$, this produces solutions of the form
\begin{equation}
\psi \sim (g^{\prime}(z))^{-1/2} D_{n + \eta + \mathcal{O} \left( \frac{1}{L} \right)} [ (-1 + i) g(z)] \label{eq:IYDO} \;.
\end{equation}  
\noindent Since we are concerned only with the fundamental QNM in this calculation, we shall set $n=0$ from this point.

\par Recall also that the interior function in the Dolan-Ottewill framework is given by the ansatz $u_0~=~f^{-1/2} \psi$. With the introduction of the $r \rightarrow z$ replacement of Eq. (\ref{eq:rtozNLO}), the asymptotic behaviour of this prefactor is 
\begin{align}
    f(z)^{-1/2} & \sim \left(1 - \frac{2}{r_p + \alpha^{-1} z }\right)^{-1/2} 
    \nonumber \\
    & \sim  \sqrt{3}\left[1 -\frac{z}{\sqrt{3 L}} + \left(\frac{2i}{3} +  \frac{5 z^2}{6}  \right) \frac{1}{L} + \left( -\frac{13 z^3}{6 \sqrt{3} }-\frac{4 i z}{3 \sqrt{3} } \right)\frac{1}{L^{3/2}} + \left( \frac{139 z^4}{72}+i z^2-\frac{115}{162} \right) \frac{1}{L^2} +...\right]\;.
\label{eq:prefactor}
\end{align} 
\noindent For the sake of brevity, we shall present only the asymptotic forms here and impose that $n=0$ for the overtone. In the eikonal limit,
\begin{align}
    \psi_{z \rightarrow - \infty} & \sim e^{+i g(z)^2/2} \left( \frac{dg}{dz}\right)^{-1/2}  \label{eq:int-posinf-asymp} \\
    \psi_{z \rightarrow + \infty} &\sim e^{+i g(z)^2/2}  \left( \frac{dg}{dz}\right)^{-1/2} - e^{-i g(z)^2/2}  \left( \frac{dg}{dz}\right)^{-1/2} \frac{\sqrt{2 \pi} e^{-3i \pi/4} 2^{-1/2}}{\Gamma \big \{ -n - \eta + \mathcal{O} \left( \frac{1}{L} \right) \big \}}  \frac{1}{g(z)} \left[ 1 + \frac{i}{2} g(z)^{-2} - \frac{3}{4} g(z)^{-4} + ... \right] \label{eq:int-neginf-asymp} \;,
\end{align}
where the asymptotics of each component are given by: 
\begin{align}
   e^{\pm i g(z)^2/2}  &\sim  \exp \Bigg \{\pm \frac{i z^2}{2} \pm \frac{(20 z - 10 i z^3) }{3 \sqrt{27}}\frac{1}{L^{1/2}} \pm \left[ - \frac{200 i}{243} - \frac{595 z^2}{3888} +\frac{185 i z^4}{216} \right] \frac{1}{L}  \mp \bigg[ \frac{839 i z^5}{405 \sqrt{3}}+\frac{527 z^3}{131220 \sqrt{3}}
   \nonumber  \\
   & \hspace{1.5cm} +\frac{93269 i z}{65610 \sqrt{3}} \bigg] \frac{1}{L^{3/2}}  \pm \left[\frac{60547 i z^6}{34992}+\frac{853499 z^4}{7558272} -\frac{27982081 i z^2}{45349632}+\frac{8837}{177147} \right] \frac{1}{L^2} +... \Bigg \} \;.\\
 \left( \frac{dg}{dz}\right)^{-1/2} &\sim 1 + \frac{10z}{9 \sqrt{3 L}}+\left[-\frac{155 z^2}{432}+\frac{2605 i}{7776}\right] \frac{1}{L}+\left[\frac{51323 z^3}{87480 \sqrt{3}}+\frac{215767 i z}{524880 \sqrt{3}}\right]\frac{1}{L^{3/2}}
   \nonumber\\
 & \hspace{1.5cm} +\left[-\frac{11897483 z^4}{30233088}-\frac{1100411 i z^2}{10077696}+\frac{100425053}{362797056}\right]\frac{1}{L^{2}}+...~.
\end{align}

\begin{align}
&\frac{1}{g(z)} \left[ 1 + \frac{i}{2g(z)^{2}}  - \frac{3}{4 g(z)^{4}}  + ... \right] \nonumber \\
& = \frac{1}{z} + \frac{i}{2 z^3} -\frac{3}{4 z^5} + \left[ \frac{10}{9 \sqrt{3}} +\frac{35 i}{9 \sqrt{3} z^2} -\frac{5 \sqrt{3}}{2 z^4} -\frac{25 i}{3 \sqrt{3} z^6} \right]\frac{1}{L^{1/2}} + \left[ -\frac{155 z}{648} +\frac{5005 i}{1944 z} -\frac{9505}{972 z^3} -\frac{44875 i}{1728 z^5} + \frac{500}{27 z^7}  \right] \frac{1}{L} \nonumber \\
&\hspace{1cm} + \left[ \frac{70000 i}{729 \sqrt{3} z^8}+\frac{762343}{3888 \sqrt{3} z^6}-\frac{32791703 i}{349920 \sqrt{3} z^4}+\frac{15737 z^2}{43740 \sqrt{3}}-\frac{583061}{32805 \sqrt{3} z^2}+\frac{79477 i}{131220 \sqrt{3}} \right] \frac{1}{L^{3/2}} \nonumber \\
& \hspace{1cm} + \left[-\frac{2800000}{19683 z^9}+\frac{30926495 i}{78732 z^7}+\frac{32239603847}{120932352 z^5}-\frac{1743973 z^3}{7558272}-\frac{9730444495 i}{181398528 z^3}-\frac{329953 i z}{15116544}-\frac{152696747}{45349632 z} \right] \frac{1}{L^2} + ...\;.
\end{align}

\noindent Finally, applying the $\Gamma \{ -n - \eta + \mathcal{O} \left( L^{-1} \right) \}$ approximation in the $n = 0$ limit yields,
\begin{equation}
\frac{1}{\Gamma \big \{ -n - \eta + \mathcal{O} \left( \frac{1}{L} \right) \big \}} \sqrt{2 \pi} e^{-3i \pi/4} 2^{-1/2} \sim \left(-i \sqrt{27} \epsilon \right)  \left[1+\frac{5 i}{36 L}-\frac{659}{7776 L^2} +...\right]\sqrt{\pi} e^{-3i \pi/4} \;.
\end{equation}

\subsection{Exterior solution: using the Dolan-Ottewill ansatz}

\par For the exterior solution, we closely follow the method described in Section \ref{sssec:extDO} for
\begin{equation}
\label{eq.3.54}
 u^{\pm} (r) = \exp \Bigg \{ \pm i \omega \int^r_3 \left(1 + \frac{6}{r} \right)^{1/2} \left(1 - \frac{3}{r} \right) dr_{\star} \Bigg \} v^{\pm}(r) \;.   
\end{equation}
To obtain the exponential component of the wavefunction, we perform a power series expansion on the integrand around $r = 3$, and then integrate directly with respect to the tortoise coordinate. We then perform the substitution of $r \rightarrow r_p + \alpha^{-1} z$, {$\omega \rightarrow \omega_{\ell 0}^{(2)}\approx  \overline{\omega}_{-1}L + \overline{\omega}_{0} + \overline{\omega}_{1} L^{-1} + \overline{\omega}_{2} L^{-2}$, and $\overline{\omega}_{k}$ give in Eq. (\ref{eq:QNF(2)})}, which gives
\begin{align}
\label{eq.3.55}
   \pm &i \omega \int^r_3 \left(1 + \frac{6}{r} \right)^{1/2} \left(1 - \frac{3}{r} \right) dr_{\star}  \nonumber \\ 
   &\sim \pm \Big [
   \frac{i z^2}{2} 
   + \left[ \frac{2 z}{\sqrt{3}}-\frac{10 i z^3}{9 \sqrt{3}} \right]\frac{1}{L^{1/2}} +\left[-\frac{2 i}{3} + \frac{z^2}{36} + \frac{185 i z^4}{216} +\right]\frac{1}{L} - \left[ \frac{95 i z}{54 \sqrt{3}} +\frac{10 z^3}{27 \sqrt{3}} +\frac{839 i z^5}{405 \sqrt{3}} \right] \frac{1}{L^{3/2}}  \nonumber \\
   & \hspace{1cm} + \left[ -\frac{4}{27}-\frac{1135 i z^2}{1944}+\frac{1457 z^4}{3888}+\frac{60547 i z^6}{34992} \right]\frac{1}{L^2}
   + \mathcal{O}(L^{-5/2}) 
    \Big] \;.
\end{align}

\par To construct the higher-order expressions for the function $v^{\pm (k)}$, we require the higher-order $\widetilde{S}_{kn} (r)$ functions (see Eqs. (\ref{eq:v(r)n}) { and (\ref{eq:SWEdo})}). For $n=0$, we can express this as
\begin{equation}
v^{\pm (k)} (r) = \exp \{ \widetilde{S}^{\pm}_{00}(r)  + \widetilde{S}^{\pm}_{10} (r) L^{-1} + ... + \widetilde{S}^{\pm}_{k0} (r) L^{-k}  \} \;. \label{eq:vNLO}
\end{equation}

To extract these terms, we follow the same iterative Dolan-Ottewill QNF calculation. Recall that to determine each components of the QNF series expansion $\overline{\omega}_{k}$, we require $S_{kn} (r)$ and its derivatives with respect to $r$. {Recall also that $v^{\pm}(r)$ and $\widetilde{S}^{\pm}_{kn}(r)$ are the functions after introducing the QNF perturbation (see Eq. (\ref{eq:omega2perturbed}), together with Eq. (\ref{eq:QNF(2)})) into Eq. (\ref{eq:SWEdo}), such that we solve $\widetilde{S}^{\pm}_{k0}(r)$ order by order.} 
Their explicit forms are provided in Appendix \ref{app:ext}, up to and including $k=2$. {To determine the asymptotics of these perturbed $\widetilde{S}^{\pm}_{k0}$ functions, we evaluate the functions at the perturbed QNF $\widetilde{\omega}^{(k)}_{\ell 0}$, perform the substitution $r \rightarrow r_p + \alpha^{-1} z$, and subject the result to a power series expansion about $L \rightarrow \infty$ to second order in $L$. }Using these asymptotic forms of the components, {together with Eqs. (\ref{eq.3.55}) and (\ref{eq:vNLO}), the explicit forms of Eq. (\ref{eq.3.54}), $u^{\pm(k=2)}$, may} produce
at the order of $\mathcal{O}(L^{-2})$,
\begin{align}
    u^{+ (2)} &\approx \sqrt{2} e^{+iz^2/2}  \left(\frac{\xi }{4 \sqrt{27}} \right)^{1/2} \times \exp \Bigg \{  \left[ \frac{7 z}{3 \sqrt{3} }-\frac{10 i z^3}{9 \sqrt{3}}\right] \frac{1}{\sqrt{L}} + \left[-\frac{367 i}{324} +\frac{67 i}{144 \sqrt{3}}-\frac{11 z^2}{216}+\frac{185 i z^4}{216} \right] \frac{1}{L} \nonumber \\
   & \hspace{1cm} + \left[ -\frac{166 i z}{81 \sqrt{3}}-\frac{151 z^3}{486 \sqrt{3}}-\frac{839 i z^5}{405 \sqrt{3}} \right] \frac{1}{L^{3/2}} + \left[ -\frac{11687}{93312}-\frac{47}{2592 \sqrt{3}}-\frac{9289 i z^2}{23328}+\frac{5581 z^4}{15552}+\frac{60547 i z^6}{34992}\right] \frac{1}{L^2}\Bigg \} \; , \label{eq:UplusNLO}
\end{align}
and
\begin{align}
    u^{- (2)} & \approx \sqrt{2} e^{-iz^2/2} \left(\frac{z}{L^{1/2}} \right)^{-1} \left( \frac{\xi}{\sqrt{27}}\right)^{-1/2} \exp \Bigg \{ -\frac{5}{8 z^4}+\frac{i}{2 z^2} + \left[ -\frac{5 i}{\sqrt{3} z^5}-\frac{5}{\sqrt{3} z^3}+\frac{10 i}{3 \sqrt{3} z}-\frac{z}{\sqrt{3}}+\frac{10 i z^3}{9 \sqrt{3}}\right] \frac{1}{\sqrt{L}} \nonumber \\
    & \hspace{1cm}+ \left[ \frac{679
   i}{324}-\frac{67 i}{144 \sqrt{3}}+\frac{25}{3 z^6}-\frac{13 i}{z^4}-\frac{265}{54 z^2}-\frac{41 z^2}{216}-\frac{185 i z^4}{216}\right] \frac{1}{L} + \Bigg[ \frac{100 i}{3 \sqrt{3}
   z^7}+\frac{8243}{108 \sqrt{3} z^5}-\frac{1991 i}{54 \sqrt{3} z^3}-\frac{55}{9 \sqrt{3} z} \nonumber \\
   &\hspace{1.5cm} +\frac{32 i z}{81 \sqrt{3}}+\frac{77 z^3}{162 \sqrt{3}}+\frac{839 i z^5}{405\sqrt{3}} \Bigg ] \frac{1}{L^{3/2}}  + \Bigg [\frac{6031}{93312}+\frac{47}{2592 \sqrt{3}}-\frac{350}{9 z^8} +\frac{19615 i}{162 z^6}+\frac{73271}{864 z^4} -\frac{56915 i}{3888
   z^2} \nonumber \\
   & \hspace{2cm} +\frac{23945 i z^2}{23328}-\frac{18679 z^4}{46656}-\frac{60547 i z^6}{34992}\Bigg]\frac{1}{L^2}  \Bigg \} \label{eq:UminusNLO} \; .
\end{align}

\subsection{Matching procedure \label{subsec:matching}}

\par Within the matching regions, the expression
\begin{equation}
    r-3 \propto\frac{z}{\sqrt{L}}+...~,
\end{equation}
governs the behaviour of the in-going and out-going solutions, following the form of Eq. (\ref{eq:rtoz}) { and Eq. (\ref{eq:zalpha})}. Specifically, $r-3$ is small and $L$ is large. In other words, when we apply a power series expansion to the exponential function, terms of the order $\mathcal{O}(z^aL^{-b/2})$ for $a \leq |b|$ are affected. In so doing, exponential terms of appropriate order in $z/\sqrt{L}$ reduce to polynomial form. For the interior solutions, $z \rightarrow \pm \infty$, we match these against the exterior solutions. 

\par We find that we can obtain constant (i.e. $z$-independent) terms for $C_{in}$, $B_{in}$, and $B_{out}$ as a consequence of the cancellation of higher-order terms during the matching procedure. For terms with powers of $z$ in the denominator that do not cancel, we have determined their contribution to be negligible. This is evidenced when we solve for these constant coefficients at increasingly higher orders: terms that retain powers of $z$ in the denominator at order $k$ cancel at orders $k+1$ and higher during the matching procedure; only constant and $L$-dependent terms remain unmodified as we progress to higher orders. Finally, we note that each of these reduce to their leading-order form for $L \rightarrow \infty$ (c.f. Section \ref{sss:matchy}).

\subsubsection{To obtain $C_{in} \sim u^{r\to -\infty}_0 / u^+$}

   \par Through the perturbative power series expansion, the exterior solution in the region $r < r_p$ becomes 
   \begin{align}
       u^{+ (2)} &\sim \left(\frac{\xi }{2 \sqrt{27}} \right)^{1/2} \exp \Bigg \{ +\frac{i z^2}{2}-\frac{10 i z^3}{9 \sqrt{3}} \frac{1}{\sqrt{L}} +\frac{185 i z^4}{216} \frac{1}{L} -\frac{839 i z^5}{405 \sqrt{3}} \frac{1}{L^{3/2}} +\frac{60547 i
   z^6}{34992} \frac{1}{L^2} \Bigg \} \nonumber \\
   & \quad \times \Bigg [ 1+\frac{7 z}{3 \sqrt{3}} \frac{1}{\sqrt{L}} +\left[-\frac{367 i}{324} +\frac{67 i}{144 \sqrt{3}}+\frac{185 z^2}{216} \right] \frac{1}{L}+\left[ \frac{469 i z}{1296}-\frac{4561 i z}{972
   \sqrt{3}}+\frac{179 z^3}{648 \sqrt{3}} \right] \frac{1}{L^{3/2}} \nonumber \\
   & \quad \quad + \left[-\frac{2696959}{3359232}+\frac{23743}{46656 \sqrt{3}}-\frac{103657 i z^2}{34992}+\frac{12395 i z^2}{31104 \sqrt{3}}+\frac{6517 z^4}{31104} \right] \frac{1}{L^2} \Bigg ] \;.
   \end{align}
For the interior solution, we retain only the $z \rightarrow - \infty$ asymptotic forms presented in Section \ref{sec:OurInterior}. Recall that the complete solution is given by
\begin{equation}
    u_0^{z \rightarrow - \infty} \sim  f^{-1/2}e^{+i g(z)^2/2} \left( \frac{dg}{dz}\right)^{-1/2}  \;.
    \label{eq:3.63}
\end{equation}
Following the same linearisation for terms of order $\mathcal{O}(z^aL^{-b/2})$  for $a \leq |b|$, we find that 
   \begin{align}
    u_0^{z \rightarrow - \infty} &\sim  \exp \Bigg \{ +\frac{i z^2}{2}-\frac{10 i z^3}{9 \sqrt{3}} \frac{1}{\sqrt{L}} +\frac{185 i z^4}{216} \frac{1}{L} -\frac{839 i z^5}{405 \sqrt{3}} \frac{1}{L^{3/2}} +\frac{60547 i
   z^6}{34992} \frac{1}{L^2} \Bigg \} \nonumber \\
   & \quad \times \sqrt{3}\Bigg [ 1+\frac{7 z}{3 \sqrt{3}} \frac{1}{\sqrt{L}} +\left[\frac{463 i}{2592}+\frac{185 z^2}{216} \right] \frac{1}{L}+\left[ -\frac{12695 i z}{7776 \sqrt{3}}+\frac{179 z^3}{648 \sqrt{3}} \right] \frac{1}{L^{3/2}} \nonumber \\
   & \quad \quad + \left[-\frac{1623193}{13436928} -\frac{1029697 i z^2}{559872} +\frac{6517 z^4}{31104} \right] \frac{1}{L^2} \Bigg ] +... \;.
   \end{align}

\par Terms that are higher-order in $z$ cancel, leaving us with
\begin{align}
C^{(2)}_{in} \sim \sqrt{2}\left(\frac{\xi }{3 \sqrt{27}} \right)^{-1/2} \left[ 1 + \left[ \frac{1133 i}{864}-\frac{67 i}{144 \sqrt{3}} \right ] \frac{1}{L} + \left(\frac{78167}{124416 \sqrt{3}}-\frac{1307101}{1492992} \right) \frac{1}{L^2}   \right] \;.
\end{align}

\subsubsection{To obtain $B_{out} \sim u_{0}^{r\to +\infty} / u^+$}

\noindent To obtain $B^{(2)}_{out},$ we must match the asymptotics of $f^{-1/2} \psi^{(2)}_{z \rightarrow + \infty} (e^{+iz^2/2})$ over $u^{+ (2)}$, {where $\psi^{(2)}_{z \rightarrow + \infty} (e^{+iz^2/2})$ is the first term of Eq. (\ref{eq:int-neginf-asymp}). Up to and including $\mathcal{O}(L^{-2})$}, this gives 
\begin{align}
B^{(2)}_{out} \sim \sqrt{2}\left(\frac{\xi }{3 \sqrt{27}} \right)^{-1/2} \left[ 1 + \left[ \frac{1133 i}{864}-\frac{67 i}{144 \sqrt{3}} \right ] \frac{1}{L} + \left(\frac{78167}{124416 \sqrt{3}}-\frac{1307101}{1492992} \right) \frac{1}{L^2}  \right] =C^{(2)}_{in}\; .
\end{align}

\subsubsection{To obtain $B_{in} \sim u_0^{r\to +\infty} / u^-$ }

\par To obtain $B^{(2)}_{in},$ we must match the asymptotics of {$f^{-1/2} \psi^{(2)}_{z \rightarrow + \infty} (e^{-iz^2/2})$ over $u^{- (2)}$,  where $\psi^{(2)}_{z \rightarrow + \infty} (e^{-iz^2/2})$ is the second term of Eq. (\ref{eq:int-neginf-asymp}) at $\mathcal{O}(L^{-2})$}.  Through the perturbative power series expansion, the exterior solution in the region $r > r_p$ becomes 
   \begin{align}
       u^{- (2)} & \sim  \left(\frac{z}{L^{1/2}} \right)^{-1} \left( \frac{\xi}{2\sqrt{27}}\right)^{-1/2} \exp \Bigg \{ -\frac{i z^2}{2}-\frac{10 i z^3}{9 \sqrt{3}} \frac{1}{\sqrt{L}} +\frac{185 i z^4}{216} \frac{1}{L} +\frac{839 i z^5}{405 \sqrt{3}} \frac{1}{L^{3/2}} -\frac{60547 i
   z^6}{34992} \frac{1}{L^2} \Bigg \} \nonumber \\
   & \qquad \times \Bigg [ 1 -\frac{i}{3 z^6}-\frac{3}{4 z^4}+\frac{i}{2 z^2} + \left[ 
   \mathcal{O} \left(\frac{1}{z^7} \right)-\frac{29 i}{3 \sqrt{3} z^5}-\frac{71}{12 \sqrt{3} z^3}+\frac{17 i}{6 \sqrt{3}
   z}-\frac{z}{\sqrt{3}} \right] \frac{1}{\sqrt{L}} \nonumber \\
   & \qquad \quad + \Bigg [  
   \mathcal{O} \left(\frac{1}{z^{6}} \right)
 -\frac{26135 i}{1296 z^4}+\frac{67 i}{192 \sqrt{3} z^4}-\frac{14431}{2592 z^2}+\frac{67}{288 \sqrt{3}
   z^2}-\frac{5 z^2}{216}  
   + \frac{1261 i}{1296}-\frac{67 i}{144 \sqrt{3}} \Bigg ] \frac{1}{L} \nonumber \\
   & \qquad \qquad + \Bigg [ 
   \mathcal{O} \left(\frac{1}{z^{5}} \right) 
   +\frac{4757 i}{5184 z^3}-\frac{190799 i}{3888 \sqrt{3} z^3} +\frac{1139}{2592 z} 
   -\frac{45023}{7776 \sqrt{3} z}+\frac{67 i z}{432}-\frac{1909 i z}{1296 \sqrt{3}}+\frac{395 z^3}{648 \sqrt{3}} \Bigg ] \frac{1}{L^{3/2}} \nonumber \\
    & \qquad \qquad \quad + \Bigg [ 
   \mathcal{O} \left(\frac{1}{z^{4}} \right) 
   -\frac{103604815 i}{6718464 z^2}+\frac{970261 i}{373248 \sqrt{3} z^2}+\frac{694049 i
   z^2}{559872}+\frac{335 i z^2}{31104 \sqrt{3}}-\frac{52981 z^4}{93312} + \frac{118373}{209952} \nonumber \\
   & \qquad \qquad \qquad +\frac{87871}{186624 \sqrt{3}} \Bigg ] \frac{1}{L^2} \;.
   \end{align}
Since terms satisfying the condition $\mathcal{O}(z^aL^{-b/2})$ for $a \leq |b|$ are linearised through the power series expansion, we are left with terms in inverse powers of $z$. The contribution of these terms can be neglected as discussed above.

\par As before, for the interior solution, we retain only the $z \rightarrow + \infty$ asymptotic forms presented in Section \ref{sec:OurInterior}. Here, however, we are concerned only with the $\exp \{ - ig(z)^2/2 \}$ component. Recall that the complete solution is given by
\begin{equation}
    u_0^{z \rightarrow + \infty} \sim  -f^{-1/2}  e^{-i g(z)^2/2}  \left( \frac{dg}{dz}\right)^{-1/2} \frac{\sqrt{2 \pi}}{\Gamma \big \{-n - \eta + \mathcal{O} \left( \frac{1}{L} \right) \big \}} e^{-3i \pi/4} 2^{-1/2} \frac{1}{g(z)} \left[ 1 + \frac{i}{2} g(z)^{-2} - \frac{3}{4} g(z)^{-4} + ... \right]  \;.
    \label{eq:3.68}
\end{equation}
Following the same linearisation for terms of order $\mathcal{O}(z^aL^{-b/2})$  for $a \leq |b|$, we find that 
   \begin{align}
    u_0^{z \rightarrow + \infty} &\sim \frac{9}{z} \sqrt{\frac{\pi}{i}} \epsilon \exp \Bigg \{ -\frac{i z^2}{2}+\frac{10 i z^3}{9 \sqrt{3}} \frac{1}{\sqrt{L}} -\frac{185 i z^4}{216} \frac{1}{L} +\frac{839 i z^5}{405 \sqrt{3}} \frac{1}{L^{3/2}} -\frac{60547 i
   z^6}{34992} \frac{1}{L^2} \Bigg \} \nonumber \\
   & \quad \times \Bigg [ 1 -\frac{3}{4 z^4}+\frac{i}{2 z^2} + \left[ -\frac{25 i}{3 \sqrt{3} z^5}-\frac{71}{12 \sqrt{3} z^3}+\frac{17 i}{6 \sqrt{3} z}-\frac{z}{\sqrt{3}} \right] \frac{1}{\sqrt{L}} \nonumber \\
   & \quad +\left[\frac{5963 i}{2592}+\frac{500}{27 z^6}-\frac{74573 i}{3456 z^4}-\frac{32303}{5184 z^2}-\frac{5 z^2}{216} \right] \frac{1}{L}+\Bigg [ \frac{70000 i}{729 \sqrt{3} z^7}+\frac{4043783}{23328 \sqrt{3} z^5} \nonumber \\
   & \quad \quad -\frac{5544353 i}{93312 \sqrt{3} z^3}-\frac{148543}{15552 \sqrt{3} z}-\frac{7259 i z}{2592
   \sqrt{3}}+\frac{395 z^3}{648 \sqrt{3}} \Bigg] \frac{1}{L^{3/2}} + \Bigg [-\frac{23383681}{13436928} \nonumber \\
   & \quad \quad \quad -\frac{2800000}{19683 z^8}+\frac{56938865 i}{157464 z^6}+\frac{30705895193}{161243136 z^4}-\frac{629568725 i}{26873856 z^2}+\frac{169211 i
   z^2}{139968}-\frac{52981 z^4}{93312} \Bigg] \frac{1}{L^2} \Bigg ] +...\;.
   \end{align}
Upon performing the matching procedure, we then obtain 
\begin{align}
    B^{(2)}_{in}&\sim \epsilon\left(\frac{27\pi\xi}{i 2\sqrt{3} L}\right)^{1/2}\bigg[1 + \mathcal{O} \left( \frac{1}{z^6} \right) + \mathcal{O} \left( \frac{1}{z^5} \right)\frac{1}{L^{1/2}}+\left[\frac{1147 i}{864}+\frac{67 i}{144 \sqrt{3}} + \mathcal{O} \left( \frac{1}{z^4} \right)\right] \frac{1}{L} + \mathcal{O} \left( \frac{1}{z^3} \right) \frac{1}{L^{3/2}} \nonumber \\
    & \qquad + \left[ -\frac{1619197}{1492992}-\frac{79105}{124416 \sqrt{3}} + \mathcal{O} \left( \frac{1}{z^2} \right)\right] \frac{1}{L^2}  \bigg] \;.\label{Bin-2}
\end{align}
\noindent We find constant coefficients only for the terms with integer powers of $(1/L)$, but not the ones with half-integer powers. Though we have retained coefficients with inverse powers of $z$, we expect them to be cancelled by higher order contributions, while the constant coefficients as shown in Eq. (\ref{Bin-2}) will remain unmodified. To illustrate this explicitly, let us consider the above procedure only to order $\mathcal{O}(L^{-1})$, 
\begin{align}
    B^{(1)}_{in}&\sim \epsilon\left(\frac{27\pi\xi}{i 2\sqrt{3} L}\right)^{1/2}\bigg[1 + \mathcal{O} \left( \frac{1}{z^4} \right) + \mathcal{O} \left( \frac{1}{z^3} \right)\frac{1}{L^{1/2}}+\left[\frac{1147 i}{864}+\frac{67 i}{144 \sqrt{3}} + \mathcal{O} \left( \frac{1}{z^2} \right)\right] \frac{1}{L} \bigg] \;.\label{Bin-1}
\end{align}
\noindent If we compare Eq. (\ref{Bin-2}) and Eq. (\ref{Bin-1}), we observe that despite the changes in the functions of $z$, the constant coefficients of $L^{-1}$ remain the same. 
\par We note also that when we compare our higher-order matching variables, $C^{(2)}_{in}$, $B^{(2)}_{out}$, $B^{(2)}_{in}$, with the $n=0$ the leading-order results of Eqs. (\ref{eq:cin0}), (\ref{eq:bout0}), and (\ref{eq:bin0}), respectively, there is a $\sqrt{3}$ discrepancy for the leading coefficients. This is due to our necessary inclusion of the prefactor $f^{-1/2}$ in Eqs. (\ref{eq:3.63}) and (\ref{eq:3.68}). For the leading order, the prefactor is just a constant $\sqrt{3}$ and will fully cancel out in the calculation of the $A^{\pm(0)}_{\ell n} $ coefficients. As such, the leading order results will not change when excluding this prefactor. However, for the higher-order studies, the prefactor provides a polynomial of $\mathcal{O}(z^{a}L^{-b/2})$ as in Eq. (\ref{eq:prefactor}), which contributes to the cancellation of $z$-dependent terms in the higher-order matching.

\subsection{The quasinormal excitation factor}

\par Finally, we can bring together the components of the calculation in order to construct the QNEF. To define the QNEF, we use Eq. (\ref{eq:QNEF}) for {$k=2$} and $n=0$,
 \begin{equation}
\mathcal{B}^{(2)}_{\ell 0} \equiv 
\frac{A^{+ (2)}_{\ell 0}}{2 \omega} \left( \frac{ \partial A^{- (2)}_{\ell 0}}{\partial \omega} \right)^{-1} \Bigg \vert_{\substack{\omega\rightarrow\omega^{(2)}_{\ell 0} }} \;,
\end{equation}
where the constants $A^{\pm (2)}_{\ell 0}$ are defined as
\begin{align}
A^{+ (2)}_{\ell 0} = \frac{\beta^{(2)}_1 \beta^{(2)}_2}{\alpha^{(2)}_1 \alpha^{(2)}_2} \frac{B^{(2)}_{out}}{C^{(2)}_{in}} \;, \hspace{1.3cm}
A^{- (2)}_{\ell 0} =  \frac{\gamma^{(2)}_1 \gamma^{(2)}_2}{\alpha^{(2)}_1 \alpha^{(2)}_2} \frac{B^{(2)}_{in}}{C^{(2)}_{in}} \;,
\end{align}
with the \enquote{phase factors} \cite{refDolanOttewill2011} extended to higher order as,
\begin{align}
\alpha_1^{(2)} & =  \exp \Bigg \{ +i \omega \int^{r=2}_{r=3} \left(1 + \frac{6}{r} \right)^{1/2} \left(1 - \frac{3}{r} \right) \frac{dr}{f} \Bigg \} \exp \{ + i \omega r_{\star} \}                                                   \nonumber \\
& =\exp \{ i \omega [6 - \sqrt{27} + 8 \ln 2 - 3 \ln \xi ] \}  \;, \\
\beta_1^{(2)} & =  \exp \Bigg \{ +i \omega \int^{r=\infty}_{r=3} \left(1 + \frac{6}{r} \right)^{1/2} \left(1 - \frac{3}{r} \right) \frac{dr}{f} \Bigg \} \exp \{ - i \omega r_{\star} \}                                                   \nonumber \\
& =\exp \{ i \omega [3 - \sqrt{27} + 4 \ln 2 - 3 \ln \xi ] \}  \\
\gamma_1^{(2)} & =  \exp \Bigg \{ -i \omega \int^{r=\infty}_{r=3} \left(1 + \frac{6}{r} \right)^{1/2} \left(1 - \frac{3}{r} \right) \frac{dr}{f} \Bigg \} \exp \{ + i \omega r_{\star} \}                                                    = 1/\beta_1^{(2)} \;, \\
\alpha_2^{(2)} & =  \lim_{r \rightarrow 2\;} \;\; \exp \Bigg \{ \sum_{k=0}^{2} \widetilde{S}^{+}_{k 0}(r)L^{-k} \Bigg \} = 1 \;, \\
\beta_2^{(2)} & =  \lim_{r \rightarrow \infty} \;\; \exp \Bigg \{ \sum_{k=0}^{2} \widetilde{S}^{+}_{k 0}(r)L^{-k} \Bigg \} = \left(2-\sqrt{3} \right)^{-1/2}\exp \bigg \{+\frac{47 i}{144 \sqrt{3} L} - \frac{145}{2592 \sqrt{3} L^2} \bigg \}
\;, \\
\gamma_2^{(2)} & =  \lim_{r \rightarrow \infty} \;\; \exp \Bigg \{ \sum_{k=0}^{2} \widetilde{S}^{-}_{k 0} (r)L^{-k} \Bigg \} = \left(2-\sqrt{3} \right)^{+1/2}\exp \bigg \{-\frac{47 i}{144 \sqrt{3} L} + \frac{145}{2592 \sqrt{3} L^2} \bigg \}
\;.
\end{align}

\par With the use of these definitions, we find that the out-going coefficient precisely matches Eq. (\ref{eq:AplusLO}) for $n=0$, with corrections in higher orders of $L^{-k}$. In the second line, we substitute the QNF up to order $\mathcal{O}(L^{-2}),$
\begin{align}
    A^{+ (2)}_{\ell 0} \Big \vert_{\omega \rightarrow \widetilde{\omega}^{(2)}_{\ell 0}} &\approx  \xi^{1/4} \exp \bigg \{- i \omega [3 + 4 \ln 2] +\frac{47 i}{144 \sqrt{3}} \frac{1}{L} -\frac{145}{2592 \sqrt{3}} \frac{1}{L^2} + \mathcal{O} (\epsilon)\bigg \} \Bigg \vert_{\omega \rightarrow {\omega^{(2)}_{\ell 0}}} \nonumber \\
    & \approx  \xi^{1/4} \exp \Bigg \{-\frac{i (3 + 4 \ln 2)}{3 \sqrt{3}}L
    -\frac{(3 + 4 \ln 2)}{6 \sqrt{3}} + \left[\frac{127 i}{432 \sqrt{3}}-\frac{7 i \ln 2}{162 \sqrt{3}} \right] \frac{1}{L} \nonumber \\
    & \hspace{2cm} + \left[ -\frac{143}{1944 \sqrt{3}}-\frac{137 \ln 2}{5832 \sqrt{3}}\right] \frac{1}{L^2} +{\mathcal{O}(\epsilon) }\Bigg \}\; . \label{eq:AplusL2}
\end{align}
\noindent Similarly, the exponent of the in-going coefficient is precisely that of Eq. (\ref{eq:AminusLO}), albeit augmented here by higher-order $L^{-k}$ terms. Note that this correction in $L^{-k}$ within the exponential is the additive inverse of that of Eq. (\ref{eq:AplusL2}). With the higher-order terms in the polynomial, we find that the $z$ terms cancel.
\begin{align}
    A^{- (2)}_{\ell 0} \Big \vert_{\omega \rightarrow \widetilde{\omega}^{(2)}_{\ell 0}} &\approx \epsilon \left(\frac{\left(26 +5 \sqrt{27}\right) \pi }{i L} \right)^{1/2} \frac{1619197 i+316420 i \sqrt{3}+\left(1982016+231552 \sqrt{3}\right) L-1492992 i L^2}{2614202 i-625336 i \sqrt{3}+\left(3915648-463104 \sqrt{3}\right) L-2985984 i L^2}  \nonumber \\
    & \quad \times \exp \bigg \{ -i \omega [9 - 2\sqrt{27} + 12 \ln 2  - 6 \ln \xi] -\frac{47 i}{144 \sqrt{3}} \frac{1}{L} + \frac{145}{2592 \sqrt{3} } \frac{1}{L^2} + \mathcal{O} (\epsilon) \bigg \}\Bigg \vert_{\omega \rightarrow {\omega^{(2)}_{\ell 0}}} \nonumber \\
    &\approx \epsilon \left(\frac{\left(26 +5 \sqrt{27}\right) \pi }{i L} \right)^{1/2} \frac{1619197 i+316420 i \sqrt{3}+\left(1982016+231552 \sqrt{3}\right) L-1492992 i L^2}{2614202 i-625336 i \sqrt{3}+\left(3915648-463104 \sqrt{3}\right) L-2985984 i L^2}  \nonumber \\
    & \quad \quad \times \exp \bigg \{ \left[2 i-i \sqrt{3}-\frac{4 i \ln 2}{\sqrt{3}}+\frac{4 i \ln \{ 2+ \sqrt{3} \}}{\sqrt{3}} \right] L +\left[ 1 -\frac{\sqrt{3}}{2} +\frac{2 \ln  \{ 2+ \sqrt{3} \}}{\sqrt{3}} -\frac{2 \ln 2}{\sqrt{3}} \right] \nonumber \\
    & \hspace{1cm}+ \left[ \frac{7 i}{108}-\frac{61 i}{144 \sqrt{3}}-\frac{7 i \ln  2}{54 \sqrt{3}}+\frac{7 i \ln \{2+ \sqrt{3} \}}{54 \sqrt{3}} \right] \frac{1}{L} + \Bigg [ \frac{137}{3888}+\frac{1}{324 \sqrt{3}}-\frac{137 \ln 2}{1944 \sqrt{3}}\nonumber \\
   & \hspace{1.6cm} +\frac{137 \ln \{2+ \sqrt{3} \})}{1944 \sqrt{3}}\Bigg ] \frac{1}{L^2} +{\mathcal{O}(\epsilon)}\bigg \}\;.
\end{align}
\noindent With all the necessary components of the QNEF computed explicitly to higher order, we present the final expressions for the QNEF at orders $\mathcal{O}(L^{(-1)})$ and $\mathcal{O}(L^{(-2)})$, respectively as:
\begin{align}
\mathcal{B}_{\ell 0}^{(1)}  
&\approx \left(\frac{-1133+134\sqrt{3}+864 i L}{-1147-134\sqrt{3}+864 i L}\right)\left(\frac{\exp \big \{ 2 i \omega_{\ell 0}^{(1)}\zeta+\frac{47 i}{72\sqrt{3}} \frac{1}{L}  \big \}}{\omega_{\ell 0}^{(1)}}\right) \left(\frac{\sqrt{L}}{\left(2+\sqrt{3}\right)\sqrt{-i\pi}}\right) \;, \label{eq:QNEFnloPenultimate} \\
\mathcal{B}_{\ell 0}^{(2)}  
&\approx \left(\frac{-1307101+312688\sqrt{3}-1728 i L \left(-1133+134\sqrt{3}+864i L\right)}{-1619197-316420\sqrt{3}-1728 i L\left(-1147-134\sqrt{3}+864 i L\right)}\right)\nonumber\\
& \qquad \times\left(\frac{ \exp \big \{ 2 i \omega_{\ell 0}^{(2)}\zeta+\frac{47 i}{72\sqrt{3}} \frac{1}{L} -\frac{145}{1296\sqrt{3}} \frac{1}{L^{2}} \big \}
 }{\omega_{\ell 0}^{(2)}}\right) \left(\frac{\sqrt{L}}{\left(2+\sqrt{3}\right)\sqrt{-i\pi}}\right),\label{eq:QNEFnnloPenultimate}
\end{align}
\noindent where the constant $\zeta$ is defined in Eq. (\ref{eq:zita}).
\par
\begin{figure}[t]
    \centering
    \includegraphics[width=0.8\textwidth]{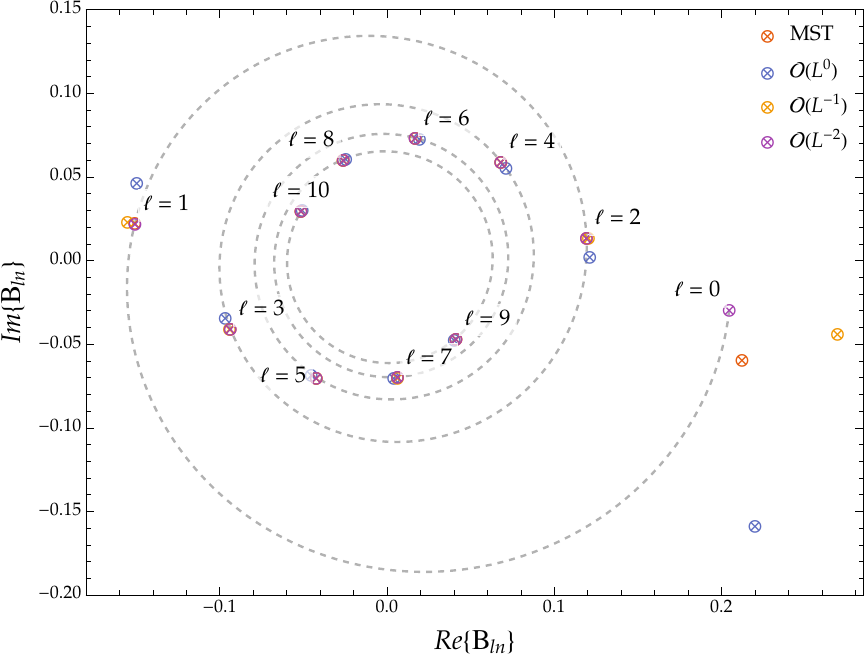}
    \vskip 0.3cm
    \caption{\textit{The real and imaginary parts of the QNEF at orders $L^0$, $L^{-1}$, and $L^{-2}$ for the fundamental mode. We compare against the numerical results generated using the MST formalism \cite{MST1996_AnalyticQNMsolsITeukolsky,MST1996_AnalyticQNMsolsIIRW}, listed in Table II of Ref. \cite{refDolanOttewill2013}. }}
    \label{fig:QNEFNNLO}
\end{figure}

 We plot the real and imaginary contributions in Fig. \ref{fig:QNEFNNLO} at leading order, and at the next two orders in $L^{-k}$. At order $k$, we evaluate the QNEF at the $k$-th order QNF, i.e. at order $L^{-1}$ we evaluate at $\omega^{(1)}_{\ell 0} = \sqrt{27}L - i\sqrt{27}/2 + 7\sqrt{27}/216L^{-1}$. To improve the accuracy of our QNEFs, we subject Eq. (\ref{eq:QNEF}) to an additional power series expansion in $L^{-k}$, where $k$ refers to the order at which we compute the QNEF. In so doing, we can compute the elusive $\ell=0$, $\ell=1$ terms that are otherwise difficult to capture accurately. We discuss this process and the manner in which we quantify accuracy in further detail in Section \ref{sec:Discuss}.

\par In order to validate our results, we also compare our QNEF values against those generated in Ref. \cite{refDolanOttewill2013} through the method introduced by Mano, Suzuki, and Takasugi (MST) in Refs. \cite{MST1996_AnalyticQNMsolsITeukolsky,MST1996_AnalyticQNMsolsIIRW}. This technique produces solutions to the radial component of the inhomogeneous QNM problem, as well as derived quantities such as the Wronskian, using Gaussian hypergeometric functions and Coloumb wave functions. In order to do this, two complementary infinite series are employed, whose radius of convergence extends towards spatial infinity and the event horizon, respectively. The approach is demonstrably accurate, particularly for small frequenices/late times. 

\par We note with interest that, reminiscent of photon orbits in the $r \rightarrow r_+$ limit, the QNEFs tend towards zero in the large-$\ell$ regime. This reinforces the connection between unstable null geodesics and the behaviour of QNMs in the eikonal regime \cite{refGoebel1972}.

\begin{table}[h]
\caption{\textit{The real and imaginary components of the QNEF at orders $L^0$, $L^{-1}$, and $L^{-2}$ for the fundamental mode. \label{tab:QNEF}}}
\begin{center}
\def\arraystretch{1.5}
\begin{tabular}{@{}CCCCCCC@{}}
\hline\noalign{\smallskip}
\ell & \hspace{0.5cm} \mathfrak{R}e \{ \mathcal{B}^{(0)}_{\ell 0} \} \hspace{0.5cm} \textcolor{white}{.} & \mathfrak{I}m \{ \mathcal{B}^{(0)}_{\ell 0} \} & \hspace{0.5cm} \mathfrak{R}e \{ \mathcal{B}^{(1)}_{\ell 0} \} \hspace{0.5cm} \textcolor{white}{.} & \mathfrak{I}m \{ \mathcal{B}^{(1)}_{\ell 0} \} & \hspace{0.5cm} \mathfrak{R}e \{ \mathcal{B}^{(2)}_{\ell 0} \}\hspace{0.5cm} \textcolor{white}{.}  & \mathfrak{I}m \{ \mathcal{B}^{(2)}_{\ell 0} \} \\
\noalign{\smallskip}\hline\noalign{\smallskip}
  0 & +0.22010 & -0.15859 & +0.26935 & -0.04343 & +0.20467 & -0.02932 \\
 1 & -0.14952 & +0.04663 & -0.15502 & +0.02304 & -0.15062 & +0.02227 \\
 2 & +0.12130 & +0.00240 & +0.12058 & +0.01364 & +0.11934 & +0.01352 \\
 3 & -0.09660 & -0.03438 & -0.09412 & -0.04071 & -0.09363 & -0.04050 \\
 4 & +0.07122 & +0.05573 & +0.06826 & +0.05933 & +0.06804 & +0.05914 \\
 5 & -0.04515 & -0.06821 & -0.04224 & -0.07005 & -0.04215 & -0.06990 \\
 6 & +0.01954 & +0.07266 & +0.01693 & +0.07331 & +0.01691 & +0.07320 \\
 7 & +0.00416 & -0.06992 & +0.00632 & -0.06976 & +0.00631 & -0.06968 \\
 8 & -0.02450 & +0.06107 & -0.02615 & +0.06038 & -0.02613 & +0.06032 \\
 9 & +0.04026 & -0.04746 & +0.04141 & -0.04646 & +0.04138 & -0.04643 \\
 10 & -0.05062 & 0.03070 & -0.05129 & +0.02957 & -0.05126 & +0.02956 \\
 \hline\noalign{\smallskip}
\end{tabular}
\end{center}
\end{table}


\subsection{Improvement on relative uncertainty in the quasinormal excitation factor \label{sec:Discuss}}

\par In this section, we shall determine how we can maximise and quantify the accuracy of our results. Our first step is to confirm the most precise way in which to evaluate the leading-order QNEF. As discussed in Section \ref{sss:QNEF}, there is a discrepancy between the original results reported in Eq. (31) and Eq. (A43) of Ref. \cite{refDolanOttewill2011}. We reiterate that this discrepancy is largely a consequence of the order to which the QNF is evaluated, with $\omega \rightarrow \omega^{(-1)}_{\ell 0} = \overline{\omega}_{-1} L$ in the denominator of the QNEF definition Eq. (\ref{eq:QNEF}) and $\omega  \rightarrow \omega^{(0)}_{\ell 0} = \overline{\omega}_{-1}L + \overline{\omega}_0$ in the exponent.

\par Let us assume that the $k^{th}$-order QNEF $\mathcal{B}^{(k)}_{\ell 0}$ must be evaluated at the $k^{th}$-order QNF $\omega_{\ell 0}^{(k)}$. At leading order, $k=0$ and the QNF is linear in $L$. If we substitute $\omega_{\ell 0}^{(0)}= \overline{\omega}_{-1}L + \overline{\omega}_0$ into Eq. (\ref{eq:QNEFloPenultimate}), then we find terms of order $\mathcal{O} (L^{-1/2})$:   

\begin{align}
\mathcal{B}_{\ell 0}^{(0)} & = \exp \Bigg \{ \frac{2i\zeta L}{\sqrt{27}}+\frac{\zeta}{\sqrt{27}} \Bigg \} \left(\frac{6\sqrt{3}}{\sqrt{-i\pi} \left(2+\sqrt{3}\right)}\right) \left(\frac{\sqrt{L}}{2 L-i}\right) \nonumber\\
&\approx \exp \Bigg \{ (2iL+1)\frac{\zeta}{\sqrt{27}} \Bigg \}\left(\frac{3\sqrt{3}}{\sqrt{-i\pi} \left(2+\sqrt{3}\right)}\right)\left(\frac{1}{\sqrt{L}}\left(1+\frac{i}{2L}+...\right)\right)\nonumber\\
&\sim \exp \Bigg \{ (2iL+1)\frac{\zeta}{\sqrt{27}} \Bigg \} \left(\frac{\sqrt{27i}}{\sqrt{\pi L} \left(2+\sqrt{3}\right)}\right) \;.\label{eq:QNEFloFinal} 
\end{align}
\noindent The leading term of the non-exponential part is of order $\mathcal{O}(L^{-1/2})$. We find this same behaviour for $\mathcal{B}_{\ell 0}^{(1)}$ and  $\mathcal{B}_{\ell 0}^{(2)}$: higher-order $L^{-k}$ terms beyond $k$ remain in the non-exponential components of the $k^{th}$-order QNEF. To contend with this, we introduce an additional power series expansion in $L^{-k}$ after we input the corresponding $k^{th}$-order QNFs, in order to retain the terms that contribute without modification to the $(k+1)$-order QNEFs.

\par We now have two expressions for the leading-order QNEF, $viz.$ Eq. (\ref{eq:QNEFloPenultimate}) and Eq. (\ref{eq:QNEFloFinal}). To quantify the accuracy of the leading-order QNEF, we compare it against the MST results \cite{MST1996_AnalyticQNMsolsIIRW,MST1996_AnalyticQNMsolsITeukolsky} quoted in Ref. \cite{refDolanOttewill2013}. We introduce a measure for the relative uncertainty, 
\begin{equation} \label{eq:RelErr}
|\delta \mathcal{B}^{(k)}_{\ell 0}|=\sqrt{  \mathfrak{R} e \Big  \{  \left(\mathcal{B}_{\ell 0}^{(k)} - \mathcal{B}_{\ell 0}^{MST}   \right)^{2} \Big \}  + \mathfrak{I} m  \Big \{  \left(\mathcal{B}_{\ell 0}^{(k)} - \mathcal{B}_{\ell 0}^{MST}   \right)^{2} \Big \}} \; .
\end{equation} 
\noindent We consider QNEFs for which $|\delta \mathcal{B}^{(k)}_{\ell 0}|$ is minimised to be more accurate.

\begin{table}[h]
\caption{\textit{Comparison of the relative error for the leading-order QNEF expressions Eqs. (\ref{eq:QNEFloPenultimate}) and (\ref{eq:QNEFloFinal}). }}
\begin{center}
\def\arraystretch{1.5}
\begin{tabular}{@{}CCCCC@{}}
\hline\noalign{\smallskip}
\ell & & |\delta \mathcal{B}^{(0)}_{\ell 0}| \; \mathrm{of}\; Eq.  \; (\ref{eq:QNEFloPenultimate}) & & |\delta \mathcal{B}^{(0)}_{\ell 0}| \; \mathrm{of}\;  Eq. \; (\ref{eq:QNEFloFinal})\\
\noalign{\smallskip}\hline\noalign{\smallskip}
1 & &  0.02579 & & 0.02384 \\
2 & &  0.01261 & & 0.01120 \\
3 & &  0.00773 & & 0.00678 \\
4 & &  0.00533 & & 0.00466 \\
5 & &  0.00396 & & 0.00345 \\
6 & &  0.00309 & & 0.00268 \\
7 & &  0.00249 & & 0.00217 \\
8 & &  0.00207 & & 0.00179 \\
9 & &  0.00175 & & 0.00152 \\
10& &  0.00151 & & 0.00131 \\ 
 \hline\noalign{\smallskip}
\end{tabular}
\end{center}
\label{tab:LOrelerr}
\end{table} 
\noindent  From Table \ref{tab:LOrelerr}, we find that the Eq.~(\ref{eq:QNEFloFinal}) results behave better than those of Eq.~(\ref{eq:QNEFloPenultimate}). We then surmise that the introduction of an additional power series expansion in $1/L$ is necessary {{to remove terms that become suppressed as we progress to higher orders and retain only the terms that remain unmodified}}. In the following tables, we demonstrate that this is also the case for higher-order QNEFs. 

\par Following the example of Eqs. (\ref{eq:QNEFloPenultimate}) and (\ref{eq:QNEFloFinal}), we subject Eqs. (\ref{eq:QNEFnloPenultimate}) and (\ref{eq:QNEFnnloPenultimate}) to the additional power series expansions to produce, respectively,
\begin{align}
\mathcal{B}_{\ell 0}^{(1)}&\sim \exp \Bigg \{ (2iL + 1)\frac{\zeta}{\sqrt{27}} +\frac{7i}{108}\left[\frac{141}{14\sqrt{3}}+\frac{\zeta}{\sqrt{27}}\right]\frac{1}{L} \Bigg \} \nonumber\\
&\quad \quad \quad \times \frac{\sqrt{27 i}}{\sqrt{\pi L}\left(2+\sqrt{3}\right)} \left(1+i\left[\frac{209}{432} -\frac{67\sqrt{3}}{216}\right] \frac{1}{L} \right) \;,\label{eq:QNEFnloFinal}
\end{align}
\begin{align}
\mathcal{B}_{\ell 0}^{(2)}&\sim 
\exp \Bigg \{ (2iL + 1)\frac{\zeta}{\sqrt{27}} +\frac{7i}{108}\left[\frac{141}{14\sqrt{3}}+\frac{\zeta}{\sqrt{27}}\right]\frac{1}{L} +\frac{137}{3888}\left[-\frac{435}{137\sqrt{3}}+\frac{\zeta}{\sqrt{27}}\right]\frac{1}{L^{2}}\Bigg \} \nonumber\\
& \quad \quad \quad \times \frac{\sqrt{27 i}}{\sqrt{\pi L}\left(2+\sqrt{3}\right)} \left(1+i\left[\frac{209}{432} -\frac{67\sqrt{3}}{216}\right] \frac{1}{L} +\left[-\frac{86257}{373248} +\frac{15131\sqrt{3}}{93312}\right]\frac{1}{L^2} \right) \;,\label{eq:QNEFnnloFinal}
\end{align}

\noindent {where the results in Table \ref{tab:QNEF} and Fig. \ref{fig:QNEFNNLO} were collected exactly from Eqs. (\ref{eq:QNEFloFinal}), (\ref{eq:QNEFnloFinal}), and (\ref{eq:QNEFnnloFinal}).} In Table \ref{tab:QNEFsRelErr}, we quantify the accuracy of the higher-order results of Eqs. (\ref{eq:QNEFnloFinal}) and (\ref{eq:QNEFnnloFinal}).

\begin{table}[h]
\caption{\textit{Comparison of the relative error for the higher-order QNEFs of Eqs. (\ref{eq:QNEFloFinal}), (\ref{eq:QNEFnloFinal}), and (\ref{eq:QNEFnnloFinal}). \label{tab:QNEFlo-relerr}}}
\begin{center}
\def\arraystretch{1.5}
\begin{tabular}{@{}CCCCCCC@{}}
\hline\noalign{\smallskip}
\ell & & |\delta \mathcal{B}^{(0)}_{\ell 0}| \; \mathrm{of}\; Eq.  \; (\ref{eq:QNEFloFinal}) & & |\delta \mathcal{B}^{(1)}_{\ell 0}| \; \mathrm{of}\;  Eq. \; (\ref{eq:QNEFnloFinal}) & & |\delta \mathcal{B}^{(2)}_{\ell 0}| \; \mathrm{of}\;  Eq. \; (\ref{eq:QNEFnnloFinal})\\
\noalign{\smallskip}\hline\noalign{\smallskip}
1 & &  0.02384 & & 0.00436 & & 0.00055\\
2 & &  0.01120 & & 0.00125 & & 8.94738\times 10^{-5}\\
3 & &  0.00678 & & 0.00054 & & 2.87952\times 10^{-5}\\
4 & &  0.00466 & & 0.00029 & & 1.13336\times 10^{-5} \\
5 & &  0.00345 & & 0.00017 & & 5.96197\times 10^{-6} \\
6 & &  0.00268 & & 0.00011 & & 3.12746\times 10^{-6} \\
7 & &  0.00217 & & 7.99649\times 10^{-5} & & 2.12480\times 10^{-6} \\
8 & &  0.00179 & & 5.86569\times 10^{-5} & & 1.79861\times 10^{-6} \\
9 & &  0.00152 & & 4.43349\times 10^{-5} & & 8.01453\times 10^{-7}\\
10& &  0.00131 & & 3.51013\times 10^{-5} & & 5.95535\times 10^{-7} \\
 \hline\noalign{\smallskip}
\end{tabular}
\end{center}
\label{tab:QNEFsRelErr}
\end{table} 

\section{Conclusion}

\par In this work, we have extended the Dolan-Ottewill method to calculate QNEFs to higher orders for the Schwarzschild black hole. By incorporating higher-order corrections using the WKB method, we achieved significantly greater accuracy in our QNEF calculations compared to leading-order results (see Table \ref{tab:QNEFsRelErr}). Specifically, our higher-order treatment reduced the relative error, particularly for lower multipolar numbers, as evidenced by the improved accuracy shown in Fig. \ref{fig:QNEFNNLO}.

\par Our results also demonstrate excellent agreement with those obtained using the MST method, validating the precision of our approach. The ability to extract QNEFs with higher accuracy has several important implications for future research and applications. These include enhanced GW modelling, where accurate QNEF calculations are crucial for modelling the post-merger phase of GWs from binary black hole mergers. Also, higher accuracy QNEF values will aid in better parameter estimation of post-merger black holes, facilitating more precise tests of GR in the strong-field regime.

\par Note that since the Dolan-Ottewill technique can be applied to a variety of spherically-symmetric black holes and perturbing fields of different spins, our higher-order extension enhances its utility across different contexts, including extremal black holes where traditional numerical methods may fail. In this way, we can perform more precise calculations while avoiding the use of computationally intensive approaches. Therefore, by providing a more accurate method for calculating QNEFs, our work supports the growing field of GW astronomy, offering a valuable resources for interpreting and analysing signals detected by the LVK collaboration.

\par As a final thought, we note that while gravitational perturbations of a Kerr black hole would be more closely aligned to the astrophysical reality, the comparatively simple setup produced here provides a suitable testing ground for our extension of this technique. We consider this work on static, spherically-symmetric black holes then a necessary precursor to the rotating case, which we will consider in a future work.

\section*{Acknowledgements}
HTC is supported in part by the National Science and Technology Council, Taiwan, under the Grants No. NSTC112-2112-M-032-006. ASC is supported in part by the National Research Foundation (NRF) of South Africa. AC is supported by a Campus France Scholarship as well as by the NRF and Department of Science and Innovation through the SA-CERN programme. CHC is supported in part by the Naresuan University research fund No. R2566C048 and the National Science, Research and Innovation Fund (NSRF) of Thailand, via the Program Management Unit for Human Resources $\&$ Institutional Development, Research and Innovation, grant number B39G670016.

\appendix

\renewcommand{\theequation}{A.\arabic{equation}}
\section{The QNEF in the Green's function approach  \label{sec:scattering}}


\par When studying the QNM contribution to the full black hole response in the wake of a perturbation, we may consider the Green's function solution to the inhomogeneous wave equation \cite{NollertSchmidt1992_QNMsInhomogeneous,Andersson1995_SchWavefunction}.
We consider here a retarded Green function for a scalar field on a Schwarzschild background space-time,
\begin{equation} \label{eq:genGreen}
\Box_x G_{\text{ret}} (x, x^{\prime}) = \frac{1}{\sqrt{-g}} \partial_{\mu} \left( \sqrt{ - g} g^{\mu \nu} \partial_{\nu} \right) G_{\text{ret}}  =  \delta^{(4)} (x-x^{\prime} ) \;.
\end{equation}
Eq. (\ref{eq:genGreen}), the scalar wave equation, is separable in the frequency domain for spherically-symmetric black hole space-times. We may therefore express the Green function through a spectral decomposition in the Schwarzschild space-time
\begin{align} \label{eq:expandGreen}
G_{\text{ret}} (x, x^{\prime})& =  \frac{1}{2\pi r r^{\prime}} \int^{+\infty +ic}_{-\infty +ic} d \omega \; \tilde{G}_{\ell  \omega} (r, r^{\prime}) e^{-i \omega (t - t^{\prime})}  \times \sum^{\infty}_{\ell = 0} (2\ell +1)P_{\ell}(\cos \gamma)  \;.
\end{align}
Here, $x,x^{\prime}$ represent space-time points at radii $r,r^{\prime}$ that are separated by the coordinate time $t-t^{\prime}$ and the spatial angle $\gamma$ (for which $\cos \gamma = \cos \theta \cos \theta^{\prime} + \sin \theta \sin \theta^{\prime}$). The constant $c$ is positive. $\tilde{G}_{\ell  \omega} (r, r^{\prime})$ is the radial Green function that satisfies
\begin{equation} \label{eq:radialGreen}
\left[ \frac{d^2}{dr_{\star}^2} + \omega^2 - V(r) \right]\tilde{G}_{\ell  \omega} (r, r^{\prime}) = -\delta (r_{\star}-r_{\star}^{\prime}) \;.
\end{equation}
The Green function is constructed using the two homogenous solutions of this Eq. (\ref{eq:radialGreen}), such that at the horizon
\begin{equation} 
u^{in}_{\ell \omega} \sim
  \begin{cases}
    e^{-i \omega r_{\star}} \;, & r_{\star} \rightarrow - \infty \;,\\
    A_{\ell  \omega}^{out} e^{+i \omega r_{\star}} + A_{\ell  \omega}^{in} e^{-i \omega r_{\star}} \;, & r_{\star} \rightarrow + \infty \;,\label{eq:in}
  \end{cases} 
\end{equation} 
while at spatial infinity,
\begin{equation}
u^{up}_{\ell \omega} \sim
  \begin{cases}
  B_{\ell  \omega}^{out} e^{+i \omega r_{\star}} + B_{\ell  \omega}^{in} e^{-i \omega r_{\star}} \;, & r_{\star} \rightarrow - \infty \; , \\
    e^{+i \omega r_{\star}} \;, & r_{\star} \rightarrow + \infty \;.\label{eq:up}
  \end{cases} 
\end{equation} 
$A_{\ell  \omega}^{out}$, $A_{\ell  \omega}^{in}$, $B_{\ell  \omega}^{out} $, and $B_{\ell  \omega}^{in} $ are complex constants. 

\par Specifically, the radial Green function can be written as
\begin{equation} \label{eq:radialGreenW}
\tilde{G}_{\ell \omega} (r, r^{\prime}) = -\frac{u_{in} (r_{<}) u_{up} (r_{>})}{W_{\ell \omega}} \;,
\end{equation}
where we denote min($r,r^{\prime}$) and max($r,r^{\prime}$) by $r_{<}$ and $r_{>}$, respectively. The Wronskian is 
\begin{equation}
W_{\ell \omega} = u^{in}_{\ell \omega} \frac{d}{dr_{\star}}u_{\ell \omega}^{up} - u^{up}_{\ell \omega} \frac{d}{dr_{\star}}u^{in}_{\ell \omega} = 2i \omega A_{\ell \omega}^{in} \;.
\end{equation}
From the Wronskian theorem, the solutions are linearly independent. However, the poles of the Green's function are located at the zeroes of $A_{\ell \omega}^{in}$ in the complex $\omega$-plane, which we shall call $\{ \omega_p \}$. When $\omega=\omega_p$, $u^{in}_{\ell \omega}$ and $u^{up}_{\ell \omega}$ become equivalent. This follows naturally, with $W_{\ell \omega}=0$ implying linearly-dependent solutions. This yields the QNM boundary conditions,
\begin{align}
      \psi \sim e^{-i \omega r_{\star}} \;, & \hspace{1cm} r_{\star} \rightarrow - \infty \; \label{eq:QNMBCin} \\
    \psi \sim e^{+i \omega r_{\star}} \;, &\hspace{1cm} r_{\star} \rightarrow + \infty \; , \label{eq:QNMBCout}
\end{align}
and demonstrates explicitly that QNFs correspond to the poles of the Green's function. 
\begin{figure}[t!]
    \centering
    \begin{subfigure}[t]{0.5\textwidth}
        \centering
        \includegraphics[width=0.9\textwidth]{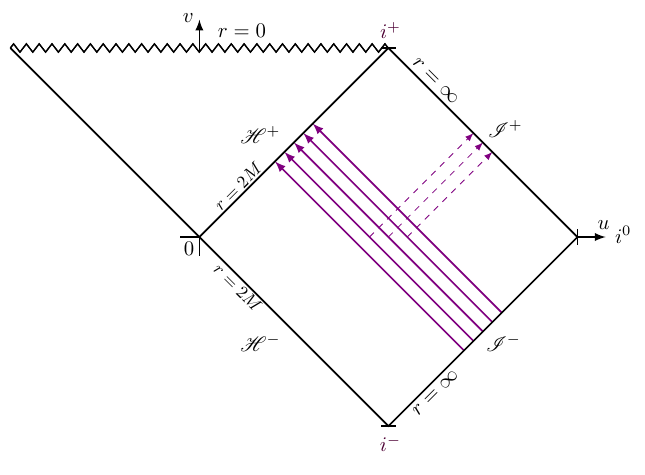}
        \caption{\textit{In-mode: purely in-going waves at $\mathscr{H^+}$.}}
    \end{subfigure}%
    ~ 
    \begin{subfigure}[t]{0.5\textwidth}
        \centering
        \includegraphics[width=0.9\textwidth]{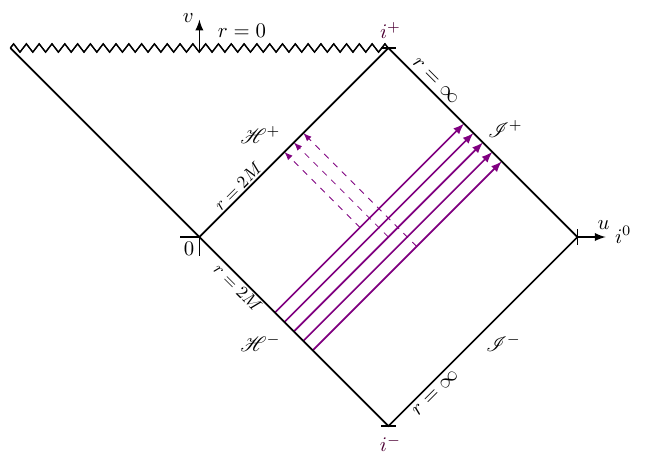}
        \caption{\textit{Up-mode: purely out-going waves at $\mathscr{I}^+$.}}
    \end{subfigure}
    \caption{\textit{Conformal diagrams of the Schwarzschild black hole space-time depicting the linearly-independent solutions to the scalar wave equation, Eqs. (\ref{eq:in}) and \ref{eq:up}) (adapted from Ref. \cite{Andersson2000_BHscattering}).   \label{fig:Penrose}}}
\end{figure}

\par Let us briefly elaborate on how these boundary conditions for the scattering problem may be refined to extract those of the QNFs. For some complex frequencies $\omega_{\ell  n}$, the in-going wave solution $u^{in}$ (Eq. (\ref{eq:in})) 
also satisfies an out-going boundary condition at infinity, namely $A^{in}_{\ell  \omega}=0$. This implies that the in-going solution is a multiple of the out-going solution, which in turn tells us that $u^{in}$ and $u^{up}$ are degenerate at these frequencies. Consequently, $u^{up}$ (Eq. (\ref{eq:up})) 
must satisfy an in-going wave condition at the black hole horizon, namely $B_{\ell \omega}^{out} = 0$. At these QNFs, we observe that $B^{in}_{\ell \omega_{\ell  n}}$, $A^{out}_{\ell \omega_{\ell  n}} = 1.$ 

\par As a visual aide, we plot the boundary conditions Eq. (\ref{eq:in}) and Eq. (\ref{eq:up}) in Fig. \ref{fig:Penrose}, which correspond to the future horizon $\mathscr{H^+}$ $(t \rightarrow +\infty, \;r_{\star} \rightarrow -\infty)$ and the future null infinity $\mathscr{I^+}$ $(t \rightarrow +\infty, \; r_{\star} \rightarrow +\infty)$, respectively. In each of these figures, we have used solid lines to denote those corresponding exclusively to the QNM boundary conditions of Eqs. (\ref{eq:QNMBCin}) and Eq. (\ref{eq:QNMBCout}), respectively.

We evaluate the Green function by substituting Eq. (\ref{eq:radialGreenW}) into the spectral decomposition, Eq. (\ref{eq:expandGreen}). To evaluate the integral, we apply the residue theorem and divide into:
\begin{itemize}
\item[$(i)$] the \enquote{direct part} i.e. the integral over the high frequency component;
\item[$(ii)$] the integral on the branch cut along the $\mathfrak{I}m \{ \omega \}$ axis;
\item[$(iii)$] the QNM contribution.
\end{itemize}
Part $(i)$ quickly approaches zero after the initial pulse \cite{refLeaver1986_SchwarzSpectralDecomp}, whereas part $(ii)$ influences the power-law spectrum at late times while also contributing to intermediate and late times.  Here, the focus is on part $(iii)$, where the QNMs affect intermediate and early times.

\par With this in mind, let us return to the problem at hand. If we exploit the fact that $u^{up}$ and $u^{in}$ are degenerate at the QNFs, we can write the QNM contribution to the Green function within the Schwarzschild space-time using Eq. (\ref{eq:expandGreen}),
\begin{align}  \label{eq:QNMGreen}
G_{\text{QNM}} (x, x^{\prime}) & =  \frac{2}{rr^{\prime}} \mathfrak{R}e \Bigg \{ \sum_{\ell}^{\infty}\sum_{n}^{\infty} (2 \ell + 1) P_{\ell} (\cos \gamma) \times \mathcal{B}_{\ell n} \tilde{u}_{in} (r) \tilde{u}_{in} (r^{\prime}) e^{-i \omega_{\ell  n} T} \Bigg \} \;,
\end{align}
where the normalised in-going wavefunction, 
\begin{equation} \label{eq:normalisedu}
\tilde{u}^{in} (r) \equiv u_{in} (r) \times \left[ A^{out}_{\ell  \omega} e^{i \omega_{\ell  n} r_{\star}} \right]^{-1} 
\end{equation}
behaves as $ \tilde{u}_{in} (r) \sim 1$ for $r \rightarrow +\infty.$ The \enquote{reflection time} \cite{refDolanOttewill2011},
\begin{equation}
    T= t-t^{\prime}-r_{\star}-r^{\prime}_{\star} \;,
\end{equation}
refers to the estimate of the time taken for the incident wave to travel from its initial position at $r^{\prime}_{\star}$ to the black hole, and for the reflected wave to reach an observer at $r_{\star}$ \cite{Andersson2000_BHscattering}. QNMs are expected to dominate the spectrum when $T>0$. In this regime, Eq. (\ref{eq:QNMGreen}) converges: higher overtones are suppressed in the sum over $n$. However, when $T<0$, Eq. (\ref{eq:QNMGreen}) becomes ill-defined due to the exponential divergence of the $e^{-i \omega_{\ell  n} T}$ factor. Within this black hole context, we assume that the initial data (i.e. the incident radiation) is only well defined far away from the black hole, and that the observer is also located in the large-$r_{\star}$ regime. In this way, we need only consider the asymptotic behaviour of the wavefunctions at large $r_{\star}$.

\par Finally, this brings us to the data-independent contribution to Eq. (\ref{eq:QNMGreen}), the QNEF, defined as the amount by which the QNMs are excited by the $\delta$-function in Eq. (\ref{eq:genGreen}) (see Eq. (\ref{eq:QNEF})). Recall that the QNEF is a quantity that indicates how and by how much a particular QNM is excited. With the increasing sensitivity of GW detectors to higher harmonics, the need to distinguish one mode from another within the superposition of observed QNMs becomes more pressing. For this reason, precise values of the QNEF are required. Furthermore, as a \enquote{no-hair} factor dependent exclusively on its associated black hole's parameters, the QNEF is especially useful in the development of waveform models and the testing of GR \cite{Berti2005_BHspectroscopy}. 

\par Through this description of the QNM as a scattering problem, we have derived the boundary conditions that must be imposed upon Eq. (\ref{eq:SWEpsi}) in order to perform the QNM computation. However, note carefully that Eq. (\ref{eq:QNMGreen}) conveys only the late-time component of the black hole signal for some specified initial data. Though there is a countable infinity of QNMs for each multipolar number $\ell$ for each overtone $n$, QNMs do not form a complete set. Recall that Eq. (\ref{eq:QNMGreen}) diverges for early times $T<0$ and fails to represent the power-law tail that dominates at very late times. 

\par Intuitively, we can appreciate this on the basis that the QNM system is intrinsically damped and therefore  not time-symmetric. As a consequence, the eigenvalue problem is non-Hermitian and the eigenvalues are complex. The corresponding eigenfunctions are then not normalisable and therefore do not form a complete set (see Refs \cite{Andersson2000_BHscattering,refNollert1999} for further discussion). 

\renewcommand{\theequation}{B.\arabic{equation}}
\section{Details on the Dolan-Ottewill method \label{sub:DO}}

\par In this section, we shall briefly present a derivation of Dolan and Ottewill's physically-motivated ansatz, presented in Eq. (\ref{eq:DOansatz}). To extract the parameters of interest, we follow Refs. \cite{Chandrasekhar1983,refCardosoLyapunov} in studying the equation of motion for a test particle near the spherically-symmetric black hole. The Lagrangian in the equatorial plane ($\theta = \pi/2$) is written as
\begin{equation} \label{eq:Lagrangian}
{\cal L} = \frac{1}{2} g_{\mu \nu} \dot{x}^{\mu} \dot{x}^{\nu} = \frac{1}{2} \left( -f(r) \dot{t}^2 + f(r)^{-1} \dot{r}^2 + r^2 \dot{\phi}^2 \right) \;,
\end{equation}
where the overdot represents a derivative with respect to an affine parameter. From the corresponding conjugate momenta, we obtain the angular momentum $\bar{L}$ and the photon energy $\bar{E}$,
\begin{equation}
\dot{\phi} = \frac{\bar{L}}{r^2} \;, \hspace{1.5cm} \dot{t} = \frac{\bar{E}}{f(r)} \;.
\end{equation}
These expressions allow us to write the Hamiltonian as
\begin{eqnarray}
{\cal H} & = & \left( p_t \dot{t} + p_{\phi} \dot{\phi} + p_r \dot{r} - {\cal L} \right) \nonumber \\
& \Rightarrow & 2 {\cal H} = \bar{E} \dot{t} - \bar{L}\dot{\phi} - f(r)^{-1} \dot{r}^2 = \delta_1 \;,
\end{eqnarray}
for which $\delta_1 = 0$ for null geodesics. Combining $\dot{r} \equiv V_r$ \cite{refCardosoLyapunov} with the condition for circular orbits $V_r = V_r' = 0$ \cite{refBardeenPressTeukolsky} produces
\begin{equation}
0 = \frac{\bar{E}^2}{\bar{L}^2} - \frac{f(r)}{r^2} \;.
\end{equation}
From this, Dolan and Ottewill define the function 
\begin{equation}
    k(r)=\sqrt{\frac{1}{b^2}-\frac{f(r)}{r^2}} \;,
\end{equation}
where $b=\bar{L}/\bar{E}$ is the impact parameter. From these considerations, the Dolan-Ottewill ansatz is established (c.f. Eq. (\ref{eq:DOansatz})).

\par The ansatz must satisfy the boundary conditions,
\begin{align}
f(r) \rightarrow 0 \;, \hspace{0.2cm} b_c k_c(r) \rightarrow -1 & \quad \text{as} \quad x \rightarrow - \infty\;, \\
f(r)/r^2 \rightarrow 0\;, \hspace{0.2cm} b_c k_c(r) \rightarrow +1 & \quad  \text{as} \quad x \rightarrow + \infty\;.
\end{align}
These convey that the event horizon is encountered as $x \rightarrow - \infty$ and an asymptotically-flat region is approached as $x \rightarrow + \infty$. 

\par As demonstrated in Section \ref{sec:Schwarz}, the components of the ansatz the Schwarzschild black hole space-time are given by $ r_c = 3 \;, \; b_c = \sqrt{27}$ for $M=1$; thus $ \rho (r) = \left(1-3/r \right) \left(1+6/r \right)^{1/2}$. For a scalar field, Eq. (\ref{eq:QNFseries}) to order $\mathcal{O} (L^{-6})$ yields
\begin{gather}
\begin{aligned} 
\sqrt{27} \omega_{\ell 0} &= L -\frac{i}{2} + \frac{7}{216}L^{-1} - \frac{137i}{7776}L^{-2} + \frac{2615}{1259712}L^{-3} \nonumber \\
& \hspace{1cm} + \frac{590983i}{362797056L}L^{-4} -\frac{42573661}{39182082048}L^{-5} + \frac{11084613257i}{8463329722368} L^{-6} 
    \end{aligned}
\end{gather}
for $n=0$, as demonstrated in Ref. \cite{refDolanOttewill2009} and confirmed in Ref. \cite{refOurLargeL}. Note that odd (even) values of $k$ correspond to real (imaginary)  expansion terms.

\vfill
\renewcommand{\theequation}{C.\arabic{equation}}
\section{Explicit perturbed $ {\widetilde{S}^{\pm}_{k 0}}$ functions for the fundamental QNM \label{app:ext}}
\begin{align}
    {\widetilde{S}^{+}_{0 0}} (r) &= \frac{1}{4} \ln \left(\frac{4 r}{6+r}\right)+\frac{1}{2} \left(\ln \left(\frac{\xi  r}{2}\right)-\ln \left(3+2 r+\sqrt{3 r (r+6)}\right)\right) \nonumber \\
    & \quad+i \sqrt{27} \epsilon  \left(\ln (r-3)-\ln \left(3+2 r+\sqrt{3 r
   (r+6)}\right)+\ln (\xi )\right)\;,\\
    {\widetilde{S}^{-}_{0 0}}(r) &= \frac{1}{4} \ln \left(\frac{4 r}{6+r}\right)+\frac{1}{2} \left(\ln \left(\frac{\xi  r}{2}\right)-\ln \left(3+2 r+\sqrt{3
   r (r+6)}\right)\right) \nonumber \\
   &\quad - (1 + i \sqrt{27} \epsilon)\left(\ln (r-3)-\ln \left(3+2 r+\sqrt{3 r (r+6)}\right)+\ln (\xi )\right)\;;\\
    {\widetilde{S}^{+}_{1 0}}(r) &= \left(\frac{67 i}{144 \sqrt{3}}+\frac{27 i (r-2)}{4 (r-3)^2 (r+6)}-\frac{i \sqrt{\frac{r+6}{3 r}} \left(1458-945 r-90 r^2+213 r^3+10 r^4\right)}{72 \left((r-3)^2
   (r+6)^2\right)}\right) \nonumber \\
   &\quad +\epsilon  \Bigg(1-\frac{5 \ln (\xi )}{4 \sqrt{3}}-\frac{81 \sqrt{3} (r-2)}{2 (r-3)^2 (r+6)}-\frac{\sqrt{r (r+6)} \left(90-45 r+2 r^2\right)}{4 \left((r-3)^2 (r+6)\right)} \nonumber \\
   &\quad \quad -\frac{1}{12} \left(5 \sqrt{3}\right)
   \ln (r-3)+\frac{1}{12} \left(5 \sqrt{3}\right) \ln \left(3+2 r+\sqrt{3 r (r+6)}\right)\Bigg)\;,\\
     {\widetilde{S}^{-}_{1 0}} (r) &=
    \left(-\frac{67 i}{144 \sqrt{3}}+\frac{27 i (r-2)}{4 (r-3)^2 (r+6)}+\frac{i \sqrt{\frac{r+6}{3 r}} \left(1458-945 r-90 r^2+213 r^3+10 r^4\right)}{72 \left((r-3)^2 (r+6)^2\right)}\right) \nonumber \\
    & \quad-\epsilon 
   \Bigg( 1-\frac{5 \ln (\xi )}{4 \sqrt{3}}+\frac{81 \sqrt{3} (r-2)}{2 (r-3)^2 (r+6)}-\frac{\sqrt{r (r+6)} \left(90-45 r+2 r^2\right)}{4 \left((r-3)^2 (r+6)\right)} \nonumber \\
   &\quad \quad -\frac{1}{12} \left(5 \sqrt{3}\right) \ln
   (r-3)+\frac{1}{12} \left(5 \sqrt{3}\right) \ln \left(3+2 r+\sqrt{3 r (r+6)}\right)\Bigg)
    \;;\\
     {\widetilde{S}^{+}_{2 0}} (r) &= -\frac{47}{2592 \sqrt{3}}+\frac{6237}{4 (-3+r)^4 (6+r)^3}-\frac{2187}{4 (-3+r)^4 r (6+r)^3}-\frac{17631 r}{16 (-3+r)^4 (6+r)^3}+\frac{3267 r^2}{8 (-3+r)^4 (6+r)^3} \nonumber \\
    &\quad -\frac{483 r^3}{16 (-3+r)^4
   (6+r)^3}-\frac{61 r^4}{4 (-3+r)^4 (6+r)^3}-\frac{115 r^5}{16 (-3+r)^4 (6+r)^3}-\frac{411 \sqrt{3} r \sqrt{\frac{6+r}{r}}}{4 (-3+r)^4 (6+r)^3} \nonumber \\
   & \quad \quad +\frac{685 \sqrt{3} r^2 \sqrt{\frac{6+r}{r}}}{8 (-3+r)^4
   (6+r)^3}-\frac{2497 r^3 \sqrt{\frac{6+r}{r}}}{16 \sqrt{3} (-3+r)^4 (6+r)^3}+\frac{1355 r^4 \sqrt{\frac{6+r}{r}}}{32 \sqrt{3} (-3+r)^4 (6+r)^3} \nonumber \\
   & \quad \quad \quad +\frac{475 r^5 \sqrt{\frac{6+r}{r}}}{144 \sqrt{3} (-3+r)^4
   (6+r)^3}+\frac{863 r^6 \sqrt{\frac{6+r}{r}}}{864 \sqrt{3} (-3+r)^4 (6+r)^3}-\frac{49 r^7 \sqrt{\frac{6+r}{r}}}{1296 \sqrt{3} (-3+r)^4 (6+r)^3}+\mathcal{O}(\epsilon) \;,\\
     {\widetilde{S}^{-}_{2 0}} (r) &= \frac{59}{864}+\frac{47}{2592 \sqrt{3}}+\frac{729}{2
   (-3+r)^4 (6+r)^3}-\frac{2187}{4 (-3+r)^4 r (6+r)^3}-\frac{1701 r}{16 (-3+r)^4 (6+r)^3} \nonumber \\
   &\quad +\frac{4941 r^2}{16 (-3+r)^4 (6+r)^3} -\frac{3975 r^3}{32 (-3+r)^4 (6+r)^3}-\frac{r^4}{2 (-3+r)^4 (6+r)^3}-\frac{7 r^5}{2
   (-3+r)^4 (6+r)^3}\nonumber \\ 
   & \quad \quad -\frac{59 r^6}{144 (-3+r)^4 (6+r)^3}  -\frac{59 r^7}{864 (-3+r)^4 (6+r)^3}+\frac{411 \sqrt{3} r \sqrt{\frac{6+r}{r}}}{4 (-3+r)^4 (6+r)^3} -\frac{685 \sqrt{3} r^2 \sqrt{\frac{6+r}{r}}}{8
   (-3+r)^4 (6+r)^3} \nonumber \\
   & \quad \quad \quad +\frac{2497 r^3 \sqrt{\frac{6+r}{r}}}{16 \sqrt{3} (-3+r)^4 (6+r)^3}-\frac{1355 r^4 \sqrt{\frac{6+r}{r}}}{32 \sqrt{3} (-3+r)^4 (6+r)^3}-\frac{475 r^5 \sqrt{\frac{6+r}{r}}}{144 \sqrt{3}
   (-3+r)^4 (6+r)^3} \nonumber \\
   &\quad \quad \quad \quad -\frac{863 r^6 \sqrt{\frac{6+r}{r}}}{864 \sqrt{3} (-3+r)^4 (6+r)^3}+\frac{49 r^7 \sqrt{\frac{6+r}{r}}}{1296 \sqrt{3} (-3+r)^4 (6+r)^3}+\mathcal{O}(\epsilon) \;.
\end{align}
\noindent Upon imposing $r \rightarrow r_p+ \alpha^{-1} z$ and the limit $\epsilon \rightarrow 0$, we can derive the asymptotics thereof, $viz.$
\begin{eqnarray}
 {\widetilde{S}^{+}_{0 0}}(z)&\sim &\ln{\left[\left(\frac{4}{3}\right)^{\frac{1}{4}}\left(\frac{\xi}{12}\right)^{\frac{1}{2}}\right]} +\left(\frac{\sqrt{3}}{9}z\right) \frac{1}{\sqrt{L}}-\left(\frac{2}{9}i+\frac{17}{216}z^{2}\right)\frac{1}{L}\nonumber\\
&&+\left(\frac{29 z^3}{486 \sqrt{3}}-\frac{19 i z}{54 \sqrt{3}}\right) \frac{1}{L^{3/2}}+\left(-\frac{247 z^4}{15552}+\frac{95 i z^2}{486}-\frac{7}{243}\right) \frac{1}{L^2} \;,\\
 {\widetilde{S}^{+}_{1 0}}(z) &\sim & \left(\frac{67 i}{144 \sqrt{3}}-\frac{79 i}{324}\right)+\left(\frac{5iz}{81\sqrt{3}}\right)\left(\frac{1}{\sqrt{L}}\right)+\left( \frac{10}{243}-\frac{229iz^{2}}{23328}\right)\frac{1}{L}\nonumber\\
&&+\left(\frac{491 z}{5832 \sqrt{3}}+\frac{41 i z^3}{6561 \sqrt{3}}\right) \frac{1}{L^{3/2}}-\left(\frac{43 i z^4}{26244}+\frac{1405 z^2}{52488}+\frac{613 i}{52488}\right)\frac{1}{L^2} \;,\\
{\widetilde{S}^{+}_{2 0}}(z)&\sim & \left(\frac{985}{93312}-\frac{47}{2592 \sqrt{3}} \right) -\frac{11 z}{5832 \sqrt{3}}\frac{1}{\sqrt{L}} +\left(-\frac{947 z^2}{839808}+\frac{11 i}{8748}\right)\frac{1}{L}\nonumber\\
&&+\left(\frac{107 z^3}{59049 \sqrt{3}}+\frac{1739 i z}{209952 \sqrt{3}}\right)\frac{1}{L^{3/2}}-\left(\frac{845645 z^4}{1088391168}-\frac{1675 i z^2}{1889568}-\frac{4271}{1889568}\right) \frac{1}{L^2} \;,\\
 {\widetilde{S}^{-}_{0 0}}(z) &\sim &\ln{\left[\left(\frac{4}{3}\right)^{\frac{1}{4}}\left(\frac{\xi}{12}\right)^{\frac{1}{2}}\right]} -\ln\left(\frac{\xi\left|z\right|}{2\sqrt{27L}}\right) +\left(\frac{2i}{\sqrt{3}z}+\frac{z}{\sqrt{3}}\right)\frac{1}{\sqrt{L}} \nonumber\\
&&-\left(\frac{2}{3z^{2}}-\frac{4i}{3}+\frac{35}{216}z^{2}\right) \frac{1}{L}+\left(\frac{17 z^3}{162 \sqrt{3}}-\frac{8 i}{9 \sqrt{3} z^3}-\frac{73 i z}{54 \sqrt{3}}-\frac{151}{54 \sqrt{3} z}\right) \frac{1}{L^{3/2}}\nonumber\\
&&+\left(-\frac{1195 z^4}{46656}+\frac{4}{9 z^4}+\frac{71 i z^2}{162}-\frac{151 i}{81 z^2}-\frac{43}{432}\right) \frac{1}{L^{2}} \;, \\
{\widetilde{S}^{-}_{1 0}}(z) &\sim &\left(\frac{i}{2z^{2}}\right)L +\left(\frac{2i}{3\sqrt{3}}\right)\left(\frac{2}{z}+\frac{3i}{z^{3}}\right)\sqrt{L} +\left(\frac{31i}{324}-\frac{67i}{144\sqrt{3}}-\frac{26}{9z^{2}}-\frac{2i}{z^{4}}\right)\nonumber\\
&&+\left(\frac{16}{3 \sqrt{3} z^5}-\frac{679 i}{54 \sqrt{3} z^3}-\frac{8}{3 \sqrt{3} z}-\frac{i z}{81 \sqrt{3}}\right) \frac{1}{\sqrt{L}}\nonumber\\
&&+\left(\frac{40 i}{9 z^6}+\frac{133}{9 z^4}+\frac{101 i z^2}{23328}-\frac{26963 i}{3888 z^2}-\frac{2}{243}\right) \frac{1}{L} \;,\\
{\widetilde{S}^{-}_{2 0}}(z) &=&-\frac{5}{8 z^4} L^{2}+\left(-\frac{\sqrt{3}}{z^3}-\frac{5 i}{\sqrt{3} z^5}\right) L^{3/2}+\left(\frac{25}{3 z^6}-\frac{11 i}{z^4}-\frac{73}{54 z^2}\right) L\nonumber\\
&&+\left(\frac{100 i}{3 \sqrt{3} z^7}+\frac{7667}{108 \sqrt{3} z^5}-\frac{632 i}{27 \sqrt{3} z^3}-\frac{35}{54 \sqrt{3} z}\right) L^{1/2}\nonumber\\
&&+\left(-\frac{350}{9 z^8}+\frac{18895 i}{162 z^6}+\frac{60119}{864 z^4}-\frac{473 i}{81 z^2}+\frac{2263}{93312}+\frac{47}{2592 \sqrt{3}}\right) \;.
\end{eqnarray}


\bibliography{zotero}

\begin{thebibliography}{56}%
\makeatletter
\providecommand \@ifxundefined [1]{%
 \@ifx{#1\undefined}
}%
\providecommand \@ifnum [1]{%
 \ifnum #1\expandafter \@firstoftwo
 \else \expandafter \@secondoftwo
 \fi
}%
\providecommand \@ifx [1]{%
 \ifx #1\expandafter \@firstoftwo
 \else \expandafter \@secondoftwo
 \fi
}%
\providecommand \natexlab [1]{#1}%
\providecommand \enquote  [1]{``#1''}%
\providecommand \bibnamefont  [1]{#1}%
\providecommand \bibfnamefont [1]{#1}%
\providecommand \citenamefont [1]{#1}%
\providecommand \href@noop [0]{\@secondoftwo}%
\providecommand \href [0]{\begingroup \@sanitize@url \@href}%
\providecommand \@href[1]{\@@startlink{#1}\@@href}%
\providecommand \@@href[1]{\endgroup#1\@@endlink}%
\providecommand \@sanitize@url [0]{\catcode `\\12\catcode `\$12\catcode
  `\&12\catcode `\#12\catcode `\^12\catcode `\_12\catcode `\%12\relax}%
\providecommand \@@startlink[1]{}%
\providecommand \@@endlink[0]{}%
\providecommand \url  [0]{\begingroup\@sanitize@url \@url }%
\providecommand \@url [1]{\endgroup\@href {#1}{\urlprefix }}%
\providecommand \urlprefix  [0]{URL }%
\providecommand \Eprint [0]{\href }%
\providecommand \doibase [0]{https://doi.org/}%
\providecommand \selectlanguage [0]{\@gobble}%
\providecommand \bibinfo  [0]{\@secondoftwo}%
\providecommand \bibfield  [0]{\@secondoftwo}%
\providecommand \translation [1]{[#1]}%
\providecommand \BibitemOpen [0]{}%
\providecommand \bibitemStop [0]{}%
\providecommand \bibitemNoStop [0]{.\EOS\space}%
\providecommand \EOS [0]{\spacefactor3000\relax}%
\providecommand \BibitemShut  [1]{\csname bibitem#1\endcsname}%
\let\auto@bib@innerbib\@empty
\bibitem [{\citenamefont {Regge}\ and\ \citenamefont {Wheeler}(1957)}]{refRW}%
  \BibitemOpen
  \bibfield  {author} {\bibinfo {author} {\bibfnamefont {T.}~\bibnamefont
  {Regge}}\ and\ \bibinfo {author} {\bibfnamefont {J.~A.}\ \bibnamefont
  {Wheeler}},\ }\bibfield  {title} {\bibinfo {title} {Stability of a
  schwarzschild singularity},\ }\href
  {https://doi.org/10.1103/PhysRev.108.1063} {\bibfield  {journal} {\bibinfo
  {journal} {Phys. Rev.}\ }\textbf {\bibinfo {volume} {108}},\ \bibinfo {pages}
  {1063} (\bibinfo {year} {1957})}\BibitemShut {NoStop}%
\bibitem [{\citenamefont {Berti}\ \emph {et~al.}(2009)\citenamefont {Berti},
  \citenamefont {Cardoso},\ and\ \citenamefont {Starinets}}]{refBertiCardoso}%
  \BibitemOpen
  \bibfield  {author} {\bibinfo {author} {\bibfnamefont {E.}~\bibnamefont
  {Berti}}, \bibinfo {author} {\bibfnamefont {V.}~\bibnamefont {Cardoso}},\
  and\ \bibinfo {author} {\bibfnamefont {A.~O.}\ \bibnamefont {Starinets}},\
  }\bibfield  {title} {\bibinfo {title} {Quasinormal modes of black holes and
  black branes},\ }\href {https://doi.org/10.1088/0264-9381/26/16/163001}
  {\bibfield  {journal} {\bibinfo  {journal} {Class. Quant. Grav.}\ }\textbf
  {\bibinfo {volume} {26}},\ \bibinfo {pages} {163001} (\bibinfo {year}
  {2009})},\ \Eprint {https://arxiv.org/abs/0905.2975} {arxiv:0905.2975
  [gr-qc]} \BibitemShut {NoStop}%
\bibitem [{\citenamefont {Konoplya}\ and\ \citenamefont
  {Zhidenko}(2011)}]{refKonoplyaZhidenkoReview}%
  \BibitemOpen
  \bibfield  {author} {\bibinfo {author} {\bibfnamefont {R.~A.}\ \bibnamefont
  {Konoplya}}\ and\ \bibinfo {author} {\bibfnamefont {A.}~\bibnamefont
  {Zhidenko}},\ }\bibfield  {title} {\bibinfo {title} {Quasinormal modes of
  black holes: {{From}} astrophysics to string theory},\ }\href
  {https://doi.org/10.1103/RevModPhys.83.793} {\bibfield  {journal} {\bibinfo
  {journal} {Rev. Mod. Phys.}\ }\textbf {\bibinfo {volume} {83}},\ \bibinfo
  {pages} {793} (\bibinfo {year} {2011})},\ \Eprint
  {https://arxiv.org/abs/1102.4014} {arxiv:1102.4014 [gr-qc]} \BibitemShut
  {NoStop}%
\bibitem [{\citenamefont {Anderson}\ and\ \citenamefont
  {Jensen}(2002)}]{Andersson2000_BHscattering}%
  \BibitemOpen
  \bibfield  {author} {\bibinfo {author} {\bibfnamefont {N.}~\bibnamefont
  {Anderson}}\ and\ \bibinfo {author} {\bibfnamefont {B.}~\bibnamefont
  {Jensen}},\ }\bibfield  {title} {\bibinfo {title} {Chapter 5.1.1 -
  {{Scattering}} by black holes},\ }in\ \href
  {https://doi.org/10.1016/B978-012613760-6/50085-1} {\emph {\bibinfo
  {booktitle} {Scattering}}},\ \bibinfo {editor} {edited by\ \bibinfo {editor}
  {\bibfnamefont {R.}~\bibnamefont {Pike}}\ and\ \bibinfo {editor}
  {\bibfnamefont {P.}~\bibnamefont {Sabatier}}}\ (\bibinfo  {publisher}
  {Academic Press},\ \bibinfo {address} {London},\ \bibinfo {year} {2002})\
  pp.\ \bibinfo {pages} {1607--1626},\ \Eprint
  {https://arxiv.org/abs/gr-qc/0011025} {arxiv:gr-qc/0011025} \BibitemShut
  {NoStop}%
\bibitem [{\citenamefont {Futterman}\ \emph {et~al.}(1988)\citenamefont
  {Futterman}, \citenamefont {Handler},\ and\ \citenamefont
  {Matzner}}]{FuttermanHandlerMatzner1988_BHscattering}%
  \BibitemOpen
  \bibfield  {author} {\bibinfo {author} {\bibfnamefont {J.~A.~H.}\
  \bibnamefont {Futterman}}, \bibinfo {author} {\bibfnamefont {F.~A.}\
  \bibnamefont {Handler}},\ and\ \bibinfo {author} {\bibfnamefont {R.~A.}\
  \bibnamefont {Matzner}},\ }\href@noop {} {\emph {\bibinfo {title} {Scattering
  from Black Holes}}},\ Cambridge Monographs on Mathematical Physics\ (\bibinfo
   {publisher} {Cambridge University Press},\ \bibinfo {address} {Cambridge},\
  \bibinfo {year} {1988})\BibitemShut {NoStop}%
\bibitem [{\citenamefont {Chandrasekhar}(1983)}]{Chandrasekhar1983}%
  \BibitemOpen
  \bibfield  {author} {\bibinfo {author} {\bibfnamefont {S.}~\bibnamefont
  {Chandrasekhar}},\ }\href@noop {} {\emph {\bibinfo {title} {The Mathematical
  Theory of Black Holes}}}\ (\bibinfo  {publisher} {Oxford Univerity Press},\
  \bibinfo {address} {New York},\ \bibinfo {year} {1983})\BibitemShut {NoStop}%
\bibitem [{\citenamefont
  {Vishveshwara}(1970{\natexlab{a}})}]{Vishveshwara1970_scattering}%
  \BibitemOpen
  \bibfield  {author} {\bibinfo {author} {\bibfnamefont {C.~V.}\ \bibnamefont
  {Vishveshwara}},\ }\bibfield  {title} {\bibinfo {title} {Scattering of
  gravitational radiation by a {{Schwarzschild}} black-hole},\ }\href
  {https://doi.org/10.1038/227936a0} {\bibfield  {journal} {\bibinfo  {journal}
  {Nature}\ }\textbf {\bibinfo {volume} {227}},\ \bibinfo {pages} {936}
  (\bibinfo {year} {1970}{\natexlab{a}})}\BibitemShut {NoStop}%
\bibitem [{\citenamefont
  {Echeverria}(1989)}]{Echeverria1989_BHpropertiesestimate}%
  \BibitemOpen
  \bibfield  {author} {\bibinfo {author} {\bibfnamefont {F.}~\bibnamefont
  {Echeverria}},\ }\bibfield  {title} {\bibinfo {title} {Gravitational-wave
  measurements of the mass and angular momentum of a black hole},\ }\href
  {https://doi.org/10.1103/PhysRevD.40.3194} {\bibfield  {journal} {\bibinfo
  {journal} {Phys. Rev. D}\ }\textbf {\bibinfo {volume} {40}},\ \bibinfo
  {pages} {3194} (\bibinfo {year} {1989})}\BibitemShut {NoStop}%
\bibitem [{\citenamefont
  {Vishveshwara}(1970{\natexlab{b}})}]{Vishveshwara1970_stability}%
  \BibitemOpen
  \bibfield  {author} {\bibinfo {author} {\bibfnamefont {C.~V.}\ \bibnamefont
  {Vishveshwara}},\ }\bibfield  {title} {\bibinfo {title} {Stability of the
  {{Schwarzschild}} metric},\ }\href {https://doi.org/10.1103/PhysRevD.1.2870}
  {\bibfield  {journal} {\bibinfo  {journal} {Phys. Rev. D}\ }\textbf {\bibinfo
  {volume} {1}},\ \bibinfo {pages} {2870} (\bibinfo {year}
  {1970}{\natexlab{b}})}\BibitemShut {NoStop}%
\bibitem [{\citenamefont {Hintz}\ and\ \citenamefont
  {Vasy}(2017)}]{Hintz2015_SCC}%
  \BibitemOpen
  \bibfield  {author} {\bibinfo {author} {\bibfnamefont {P.}~\bibnamefont
  {Hintz}}\ and\ \bibinfo {author} {\bibfnamefont {A.}~\bibnamefont {Vasy}},\
  }\bibfield  {title} {\bibinfo {title} {Analysis of linear waves near the
  {{Cauchy}} horizon of cosmological black holes},\ }\href
  {https://doi.org/10.1063/1.4996575} {\bibfield  {journal} {\bibinfo
  {journal} {J. Math. Phys.}\ }\textbf {\bibinfo {volume} {58}},\ \bibinfo
  {pages} {081509} (\bibinfo {year} {2017})},\ \Eprint
  {https://arxiv.org/abs/1512.08004} {arxiv:1512.08004 [math.AP]} \BibitemShut
  {NoStop}%
\bibitem [{\citenamefont {Ferrari}\ and\ \citenamefont
  {Mashhoon}(1984{\natexlab{a}})}]{refFerrMashh1}%
  \BibitemOpen
  \bibfield  {author} {\bibinfo {author} {\bibfnamefont {V.}~\bibnamefont
  {Ferrari}}\ and\ \bibinfo {author} {\bibfnamefont {B.}~\bibnamefont
  {Mashhoon}},\ }\bibfield  {title} {\bibinfo {title} {New approach to the
  quasinormal modes of a black hole},\ }\href
  {https://doi.org/10.1103/PhysRevD.30.295} {\bibfield  {journal} {\bibinfo
  {journal} {Phys. Rev. D}\ }\textbf {\bibinfo {volume} {30}},\ \bibinfo
  {pages} {295} (\bibinfo {year} {1984}{\natexlab{a}})}\BibitemShut {NoStop}%
\bibitem [{\citenamefont {Ferrari}\ and\ \citenamefont
  {Mashhoon}(1984{\natexlab{b}})}]{refFerrMashh2}%
  \BibitemOpen
  \bibfield  {author} {\bibinfo {author} {\bibfnamefont {V.}~\bibnamefont
  {Ferrari}}\ and\ \bibinfo {author} {\bibfnamefont {B.}~\bibnamefont
  {Mashhoon}},\ }\bibfield  {title} {\bibinfo {title} {Oscillations of a black
  hole},\ }\href {https://doi.org/10.1103/PhysRevLett.52.1361} {\bibfield
  {journal} {\bibinfo  {journal} {Phys. Rev. Lett.}\ }\textbf {\bibinfo
  {volume} {52}},\ \bibinfo {pages} {1361} (\bibinfo {year}
  {1984}{\natexlab{b}})}\BibitemShut {NoStop}%
\bibitem [{\citenamefont {Cho}\ \emph {et~al.}(2010)\citenamefont {Cho},
  \citenamefont {Cornell}, \citenamefont {Doukas},\ and\ \citenamefont
  {Naylor}}]{refAIM}%
  \BibitemOpen
  \bibfield  {author} {\bibinfo {author} {\bibfnamefont {H.~T.}\ \bibnamefont
  {Cho}}, \bibinfo {author} {\bibfnamefont {A.~S.}\ \bibnamefont {Cornell}},
  \bibinfo {author} {\bibfnamefont {J.}~\bibnamefont {Doukas}},\ and\ \bibinfo
  {author} {\bibfnamefont {W.}~\bibnamefont {Naylor}},\ }\bibfield  {title}
  {\bibinfo {title} {Black hole quasinormal modes using the asymptotic
  iteration method},\ }\href {https://doi.org/10.1088/0264-9381/27/15/155004}
  {\bibfield  {journal} {\bibinfo  {journal} {Class. Quant. Grav.}\ }\textbf
  {\bibinfo {volume} {27}},\ \bibinfo {pages} {155004} (\bibinfo {year}
  {2010})},\ \Eprint {https://arxiv.org/abs/0912.2740} {arxiv:0912.2740
  [gr-qc]} \BibitemShut {NoStop}%
\bibitem [{\citenamefont {Cho}\ \emph {et~al.}(2012)\citenamefont {Cho},
  \citenamefont {Cornell}, \citenamefont {Doukas}, \citenamefont {Huang},\ and\
  \citenamefont {Naylor}}]{Cho:2011sf}%
  \BibitemOpen
  \bibfield  {author} {\bibinfo {author} {\bibfnamefont {H.~T.}\ \bibnamefont
  {Cho}}, \bibinfo {author} {\bibfnamefont {A.~S.}\ \bibnamefont {Cornell}},
  \bibinfo {author} {\bibfnamefont {J.}~\bibnamefont {Doukas}}, \bibinfo
  {author} {\bibfnamefont {T.~R.}\ \bibnamefont {Huang}},\ and\ \bibinfo
  {author} {\bibfnamefont {W.}~\bibnamefont {Naylor}},\ }\bibfield  {title}
  {\bibinfo {title} {{A New Approach to Black Hole Quasinormal Modes: A Review
  of the Asymptotic Iteration Method}},\ }\href
  {https://doi.org/10.1155/2012/281705} {\bibfield  {journal} {\bibinfo
  {journal} {Adv. Math. Phys.}\ }\textbf {\bibinfo {volume} {2012}},\ \bibinfo
  {pages} {281705} (\bibinfo {year} {2012})},\ \Eprint
  {https://arxiv.org/abs/1111.5024} {arXiv:1111.5024 [gr-qc]} \BibitemShut
  {NoStop}%
\bibitem [{\citenamefont {Cornell}\ \emph {et~al.}(2022)\citenamefont
  {Cornell}, \citenamefont {Ncube},\ and\ \citenamefont
  {Harmsen}}]{Cornell:2022enn}%
  \BibitemOpen
  \bibfield  {author} {\bibinfo {author} {\bibfnamefont {A.~S.}\ \bibnamefont
  {Cornell}}, \bibinfo {author} {\bibfnamefont {A.}~\bibnamefont {Ncube}},\
  and\ \bibinfo {author} {\bibfnamefont {G.}~\bibnamefont {Harmsen}},\
  }\bibfield  {title} {\bibinfo {title} {{Using physics-informed neural
  networks to compute quasinormal modes}},\ }\href
  {https://doi.org/10.1103/PhysRevD.106.124047} {\bibfield  {journal} {\bibinfo
   {journal} {Phys. Rev. D}\ }\textbf {\bibinfo {volume} {106}},\ \bibinfo
  {pages} {124047} (\bibinfo {year} {2022})},\ \Eprint
  {https://arxiv.org/abs/2205.08284} {arXiv:2205.08284 [physics.comp-ph]}
  \BibitemShut {NoStop}%
\bibitem [{\citenamefont {Berti}\ \emph {et~al.}(2006)\citenamefont {Berti},
  \citenamefont {Cardoso},\ and\ \citenamefont
  {Will}}]{Berti2005_BHspectroscopy}%
  \BibitemOpen
  \bibfield  {author} {\bibinfo {author} {\bibfnamefont {E.}~\bibnamefont
  {Berti}}, \bibinfo {author} {\bibfnamefont {V.}~\bibnamefont {Cardoso}},\
  and\ \bibinfo {author} {\bibfnamefont {C.~M.}\ \bibnamefont {Will}},\
  }\bibfield  {title} {\bibinfo {title} {On gravitational-wave spectroscopy of
  massive black holes with the space interferometer {{LISA}}},\ }\href
  {https://doi.org/10.1103/PhysRevD.73.064030} {\bibfield  {journal} {\bibinfo
  {journal} {Phys. Rev. D}\ }\textbf {\bibinfo {volume} {73}},\ \bibinfo
  {pages} {064030} (\bibinfo {year} {2006})},\ \Eprint
  {https://arxiv.org/abs/gr-qc/0512160} {arxiv:gr-qc/0512160} \BibitemShut
  {NoStop}%
\bibitem [{\citenamefont {Dolan}\ and\ \citenamefont
  {Ottewill}(2011)}]{refDolanOttewill2011}%
  \BibitemOpen
  \bibfield  {author} {\bibinfo {author} {\bibfnamefont {S.~R.}\ \bibnamefont
  {Dolan}}\ and\ \bibinfo {author} {\bibfnamefont {A.~C.}\ \bibnamefont
  {Ottewill}},\ }\bibfield  {title} {\bibinfo {title} {Wave propagation and
  quasinormal mode excitation on schwarzschild spacetime},\ }\href
  {https://doi.org/10.1103/PhysRevD.84.104002} {\bibfield  {journal} {\bibinfo
  {journal} {Phys. Rev. D}\ }\textbf {\bibinfo {volume} {84}},\ \bibinfo
  {pages} {104002} (\bibinfo {year} {2011})},\ \Eprint
  {https://arxiv.org/abs/1106.4318} {arxiv:1106.4318 [gr-qc]} \BibitemShut
  {NoStop}%
\bibitem [{\citenamefont {Andersson}(1995)}]{Andersson1995_SchWavefunction}%
  \BibitemOpen
  \bibfield  {author} {\bibinfo {author} {\bibfnamefont {N.}~\bibnamefont
  {Andersson}},\ }\bibfield  {title} {\bibinfo {title} {Excitation of
  {{Schwarzschild}} black hole quasinormal modes},\ }\href
  {https://doi.org/10.1103/PhysRevD.51.353} {\bibfield  {journal} {\bibinfo
  {journal} {Phys. Rev. D}\ }\textbf {\bibinfo {volume} {51}},\ \bibinfo
  {pages} {353} (\bibinfo {year} {1995})}\BibitemShut {NoStop}%
\bibitem [{\citenamefont {Nollert}\ and\ \citenamefont
  {Schmidt}(1992)}]{NollertSchmidt1992_QNMsInhomogeneous}%
  \BibitemOpen
  \bibfield  {author} {\bibinfo {author} {\bibfnamefont {H.-P.}\ \bibnamefont
  {Nollert}}\ and\ \bibinfo {author} {\bibfnamefont {B.~G.}\ \bibnamefont
  {Schmidt}},\ }\bibfield  {title} {\bibinfo {title} {Quasinormal modes of
  {{Schwarzschild}} black holes: {{Defined}} and calculated via {{Laplace}}
  transformation},\ }\href {https://doi.org/10.1103/PhysRevD.45.2617}
  {\bibfield  {journal} {\bibinfo  {journal} {Phys. Rev. D}\ }\textbf {\bibinfo
  {volume} {45}},\ \bibinfo {pages} {2617} (\bibinfo {year}
  {1992})}\BibitemShut {NoStop}%
\bibitem [{\citenamefont {Yang}\ \emph {et~al.}(2014)\citenamefont {Yang},
  \citenamefont {Zhang}, \citenamefont {Zimmerman},\ and\ \citenamefont
  {Chen}}]{Yang2014_KerrGreen}%
  \BibitemOpen
  \bibfield  {author} {\bibinfo {author} {\bibfnamefont {H.}~\bibnamefont
  {Yang}}, \bibinfo {author} {\bibfnamefont {F.}~\bibnamefont {Zhang}},
  \bibinfo {author} {\bibfnamefont {A.}~\bibnamefont {Zimmerman}},\ and\
  \bibinfo {author} {\bibfnamefont {Y.}~\bibnamefont {Chen}},\ }\bibfield
  {title} {\bibinfo {title} {Scalar {{Green}} function of the {{Kerr}}
  spacetime},\ }\href {https://doi.org/10.1103/PhysRevD.89.064014} {\bibfield
  {journal} {\bibinfo  {journal} {Phys. Rev. D}\ }\textbf {\bibinfo {volume}
  {89}},\ \bibinfo {pages} {064014} (\bibinfo {year} {2014})},\ \Eprint
  {https://arxiv.org/abs/1311.3380} {arxiv:1311.3380 [gr-qc]} \BibitemShut
  {NoStop}%
\bibitem [{\citenamefont {Berti}\ and\ \citenamefont
  {Cardoso}(2006)}]{BertiCardoso2006_KerrExcitation}%
  \BibitemOpen
  \bibfield  {author} {\bibinfo {author} {\bibfnamefont {E.}~\bibnamefont
  {Berti}}\ and\ \bibinfo {author} {\bibfnamefont {V.}~\bibnamefont
  {Cardoso}},\ }\bibfield  {title} {\bibinfo {title} {Quasinormal ringing of
  kerr black holes. {{I}}. {{The}} excitation factors},\ }\href
  {https://doi.org/10.1103/PhysRevD.74.104020} {\bibfield  {journal} {\bibinfo
  {journal} {Phys. Rev. D}\ }\textbf {\bibinfo {volume} {74}},\ \bibinfo
  {pages} {104020} (\bibinfo {year} {2006})},\ \Eprint
  {https://arxiv.org/abs/gr-qc/0605118} {arxiv:gr-qc/0605118} \BibitemShut
  {NoStop}%
\bibitem [{\citenamefont {Oshita}(2021)}]{Oshita2021_ExcitationOvertones}%
  \BibitemOpen
  \bibfield  {author} {\bibinfo {author} {\bibfnamefont {N.}~\bibnamefont
  {Oshita}},\ }\bibfield  {title} {\bibinfo {title} {Ease of excitation of
  black hole ringing: {{Quantifying}} the importance of overtones by the
  excitation factors},\ }\href {https://doi.org/10.1103/PhysRevD.104.124032}
  {\bibfield  {journal} {\bibinfo  {journal} {Phys. Rev. D}\ }\textbf {\bibinfo
  {volume} {104}},\ \bibinfo {pages} {124032} (\bibinfo {year} {2021})},\
  \Eprint {https://arxiv.org/abs/2109.09757} {arxiv:2109.09757 [gr-qc]}
  \BibitemShut {NoStop}%
\bibitem [{\citenamefont {Abbott}\ \emph {et~al.}(2016)\citenamefont {Abbott}
  \emph {et~al.}}]{refLIGO}%
  \BibitemOpen
  \bibfield  {author} {\bibinfo {author} {\bibfnamefont {B.~P.}\ \bibnamefont
  {Abbott}} \emph {et~al.} (\bibinfo {collaboration} {LIGO Scientific,
  Virgo}),\ }\bibfield  {title} {\bibinfo {title} {Observation of gravitational
  waves from a binary black hole merger},\ }\href
  {https://doi.org/10.1103/PhysRevLett.116.061102} {\bibfield  {journal}
  {\bibinfo  {journal} {Phys. Rev. Lett.}\ }\textbf {\bibinfo {volume} {116}},\
  \bibinfo {pages} {061102} (\bibinfo {year} {2016})},\ \Eprint
  {https://arxiv.org/abs/1602.03837} {arxiv:1602.03837 [gr-qc]} \BibitemShut
  {NoStop}%
\bibitem [{\citenamefont {Abbott}\ \emph
  {et~al.}(2019{\natexlab{a}})\citenamefont {Abbott} \emph
  {et~al.}}]{refLIGO2018Run1_GWTC1}%
  \BibitemOpen
  \bibfield  {author} {\bibinfo {author} {\bibfnamefont {B.~P.}\ \bibnamefont
  {Abbott}} \emph {et~al.} (\bibinfo {collaboration} {LIGO Scientific,
  Virgo}),\ }\bibfield  {title} {\bibinfo {title} {{{GWTC-1}}: {{A}}
  gravitational-wave transient catalog of compact binary mergers observed by
  {{LIGO}} and virgo during the first and second observing runs},\ }\href
  {https://doi.org/10.1103/PhysRevX.9.031040} {\bibfield  {journal} {\bibinfo
  {journal} {Phys. Rev. X}\ }\textbf {\bibinfo {volume} {9}},\ \bibinfo {pages}
  {031040} (\bibinfo {year} {2019}{\natexlab{a}})},\ \Eprint
  {https://arxiv.org/abs/1811.12907} {arxiv:1811.12907 [astro-ph.HE]}
  \BibitemShut {NoStop}%
\bibitem [{\citenamefont {Abbott}\ \emph
  {et~al.}(2021{\natexlab{a}})\citenamefont {Abbott} \emph
  {et~al.}}]{refLIGO2020Run2_GWTC2}%
  \BibitemOpen
  \bibfield  {author} {\bibinfo {author} {\bibfnamefont {R.}~\bibnamefont
  {Abbott}} \emph {et~al.} (\bibinfo {collaboration} {LIGO Scientific,
  Virgo}),\ }\bibfield  {title} {\bibinfo {title} {{{GWTC-2}}: {{Compact}}
  binary coalescences observed by {{LIGO}} and virgo during the first half of
  the third observing run},\ }\href
  {https://doi.org/10.1103/PhysRevX.11.021053} {\bibfield  {journal} {\bibinfo
  {journal} {Phys. Rev. X}\ }\textbf {\bibinfo {volume} {11}},\ \bibinfo
  {pages} {021053} (\bibinfo {year} {2021}{\natexlab{a}})},\ \Eprint
  {https://arxiv.org/abs/2010.14527} {arxiv:2010.14527 [gr-qc]} \BibitemShut
  {NoStop}%
\bibitem [{\citenamefont {Abbott}\ \emph {et~al.}(2023)\citenamefont {Abbott}
  \emph {et~al.}}]{refLIGOrecent_GWTC3}%
  \BibitemOpen
  \bibfield  {author} {\bibinfo {author} {\bibfnamefont {R.}~\bibnamefont
  {Abbott}} \emph {et~al.} (\bibinfo {collaboration} {KAGRA, VIRGO, LIGO
  Scientific}),\ }\bibfield  {title} {\bibinfo {title} {{{GWTC-3}}: {{Compact}}
  binary coalescences observed by {{LIGO}} and {{Virgo}} during the second part
  of the third observing run},\ }\href
  {https://doi.org/10.1103/PhysRevX.13.041039} {\bibfield  {journal} {\bibinfo
  {journal} {Phys. Rev. X}\ }\textbf {\bibinfo {volume} {13}},\ \bibinfo
  {pages} {041039} (\bibinfo {year} {2023})},\ \Eprint
  {https://arxiv.org/abs/2111.03606} {arxiv:2111.03606 [gr-qc]} \BibitemShut
  {NoStop}%
\bibitem [{\citenamefont {Abbott}\ \emph
  {et~al.}(2019{\natexlab{b}})\citenamefont {Abbott} \emph
  {et~al.}}]{LIGO2019_GWTC1-GRtest}%
  \BibitemOpen
  \bibfield  {author} {\bibinfo {author} {\bibfnamefont {B.~P.}\ \bibnamefont
  {Abbott}} \emph {et~al.} (\bibinfo {collaboration} {LIGO Scientific,
  Virgo}),\ }\bibfield  {title} {\bibinfo {title} {Tests of general relativity
  with the binary black hole signals from the {{LIGO-Virgo}} catalog
  {{GWTC-1}}},\ }\href {https://doi.org/10.1103/PhysRevD.100.104036} {\bibfield
   {journal} {\bibinfo  {journal} {Phys. Rev. D}\ }\textbf {\bibinfo {volume}
  {100}},\ \bibinfo {pages} {104036} (\bibinfo {year} {2019}{\natexlab{b}})},\
  \Eprint {https://arxiv.org/abs/1903.04467} {arxiv:1903.04467 [gr-qc]}
  \BibitemShut {NoStop}%
\bibitem [{\citenamefont {Abbott}\ \emph
  {et~al.}(2021{\natexlab{b}})\citenamefont {Abbott} \emph
  {et~al.}}]{LIGO2020_GWTC2-GRtest_pyRing3}%
  \BibitemOpen
  \bibfield  {author} {\bibinfo {author} {\bibfnamefont {R.}~\bibnamefont
  {Abbott}} \emph {et~al.} (\bibinfo {collaboration} {LIGO Scientific,
  Virgo}),\ }\bibfield  {title} {\bibinfo {title} {Tests of general relativity
  with binary black holes from the second {{LIGO-Virgo}} gravitational-wave
  transient catalog},\ }\href {https://doi.org/10.1103/PhysRevD.103.122002}
  {\bibfield  {journal} {\bibinfo  {journal} {Phys. Rev. D}\ }\textbf {\bibinfo
  {volume} {103}},\ \bibinfo {pages} {122002} (\bibinfo {year}
  {2021}{\natexlab{b}})},\ \Eprint {https://arxiv.org/abs/2010.14529}
  {arxiv:2010.14529 [gr-qc]} \BibitemShut {NoStop}%
\bibitem [{\citenamefont {Abbott}\ \emph
  {et~al.}(2021{\natexlab{c}})\citenamefont {Abbott} \emph
  {et~al.}}]{LIGO2021_GWTC3-GRtest}%
  \BibitemOpen
  \bibfield  {author} {\bibinfo {author} {\bibfnamefont {R.}~\bibnamefont
  {Abbott}} \emph {et~al.} (\bibinfo {collaboration} {LIGO Scientific, VIRGO,
  KAGRA}),\ }\href@noop {} {\bibinfo {title} {Tests of general relativity with
  {{GWTC-3}}}} (\bibinfo {year} {2021}{\natexlab{c}}),\ \Eprint
  {https://arxiv.org/abs/2112.06861} {arxiv:2112.06861 [gr-qc]} \BibitemShut
  {NoStop}%
\bibitem [{\citenamefont {Chrysostomou}\ \emph {et~al.}(2023)\citenamefont
  {Chrysostomou}, \citenamefont {Cornell}, \citenamefont {Deandrea},
  \citenamefont {Ligout},\ and\ \citenamefont
  {Tsimpis}}]{Chrysostomou2023_EPJC}%
  \BibitemOpen
  \bibfield  {author} {\bibinfo {author} {\bibfnamefont {A.}~\bibnamefont
  {Chrysostomou}}, \bibinfo {author} {\bibfnamefont {A.}~\bibnamefont
  {Cornell}}, \bibinfo {author} {\bibfnamefont {A.}~\bibnamefont {Deandrea}},
  \bibinfo {author} {\bibfnamefont {{\'E}.}~\bibnamefont {Ligout}},\ and\
  \bibinfo {author} {\bibfnamefont {D.}~\bibnamefont {Tsimpis}},\ }\bibfield
  {title} {\bibinfo {title} {Black holes and nilmanifolds: Quasinormal modes as
  the fingerprints of extra dimensions?},\ }\href
  {https://doi.org/10.1140/epjc/s10052-023-11496-w} {\bibfield  {journal}
  {\bibinfo  {journal} {Eur. Phys. J. C}\ }\textbf {\bibinfo {volume} {83}},\
  \bibinfo {pages} {325} (\bibinfo {year} {2023})},\ \Eprint
  {https://arxiv.org/abs/2211.08489} {arxiv:2211.08489 [gr-qc]} \BibitemShut
  {NoStop}%
\bibitem [{\citenamefont {Abbott}\ \emph {et~al.}(2020)\citenamefont {Abbott}
  \emph {et~al.}}]{refLIGOguide}%
  \BibitemOpen
  \bibfield  {author} {\bibinfo {author} {\bibfnamefont {B.~P.}\ \bibnamefont
  {Abbott}} \emph {et~al.} (\bibinfo {collaboration} {LIGO Scientific,
  Virgo}),\ }\bibfield  {title} {\bibinfo {title} {A guide to
  {{LIGO}}--{{Virgo}} detector noise and extraction of transient
  gravitational-wave signals},\ }\href
  {https://doi.org/10.1088/1361-6382/ab685e} {\bibfield  {journal} {\bibinfo
  {journal} {Class. Quant. Grav.}\ }\textbf {\bibinfo {volume} {37}},\ \bibinfo
  {pages} {055002} (\bibinfo {year} {2020})},\ \Eprint
  {https://arxiv.org/abs/1908.11170} {arxiv:1908.11170 [gr-qc]} \BibitemShut
  {NoStop}%
\bibitem [{\citenamefont {Baibhav}\ \emph {et~al.}(2018)\citenamefont
  {Baibhav}, \citenamefont {Berti}, \citenamefont {Cardoso},\ and\
  \citenamefont {Khanna}}]{Baibhav2017_BHspectroscopyGW}%
  \BibitemOpen
  \bibfield  {author} {\bibinfo {author} {\bibfnamefont {V.}~\bibnamefont
  {Baibhav}}, \bibinfo {author} {\bibfnamefont {E.}~\bibnamefont {Berti}},
  \bibinfo {author} {\bibfnamefont {V.}~\bibnamefont {Cardoso}},\ and\ \bibinfo
  {author} {\bibfnamefont {G.}~\bibnamefont {Khanna}},\ }\bibfield  {title}
  {\bibinfo {title} {Black hole spectroscopy: {{Systematic}} errors and
  ringdown energy estimates},\ }\href
  {https://doi.org/10.1103/PhysRevD.97.044048} {\bibfield  {journal} {\bibinfo
  {journal} {Phys. Rev. D}\ }\textbf {\bibinfo {volume} {97}},\ \bibinfo
  {pages} {044048} (\bibinfo {year} {2018})},\ \Eprint
  {https://arxiv.org/abs/1710.02156} {arxiv:1710.02156 [gr-qc]} \BibitemShut
  {NoStop}%
\bibitem [{\citenamefont {Ota}\ and\ \citenamefont
  {Chirenti}(2020)}]{OtaChirenti2019_Harmonics}%
  \BibitemOpen
  \bibfield  {author} {\bibinfo {author} {\bibfnamefont {I.}~\bibnamefont
  {Ota}}\ and\ \bibinfo {author} {\bibfnamefont {C.}~\bibnamefont {Chirenti}},\
  }\bibfield  {title} {\bibinfo {title} {Overtones or higher harmonics?
  {{Prospects}} for testing the no-hair theorem with gravitational wave
  detections},\ }\href {https://doi.org/10.1103/PhysRevD.101.104005} {\bibfield
   {journal} {\bibinfo  {journal} {Phys. Rev. D}\ }\textbf {\bibinfo {volume}
  {101}},\ \bibinfo {pages} {104005} (\bibinfo {year} {2020})},\ \Eprint
  {https://arxiv.org/abs/1911.00440} {arxiv:1911.00440 [gr-qc]} \BibitemShut
  {NoStop}%
\bibitem [{\citenamefont {Baibhav}\ and\ \citenamefont
  {Berti}(2019)}]{Baibhav2018_BHspectroscopyL}%
  \BibitemOpen
  \bibfield  {author} {\bibinfo {author} {\bibfnamefont {V.}~\bibnamefont
  {Baibhav}}\ and\ \bibinfo {author} {\bibfnamefont {E.}~\bibnamefont
  {Berti}},\ }\bibfield  {title} {\bibinfo {title} {Multimode black hole
  spectroscopy},\ }\href {https://doi.org/10.1103/PhysRevD.99.024005}
  {\bibfield  {journal} {\bibinfo  {journal} {Phys. Rev. D}\ }\textbf {\bibinfo
  {volume} {99}},\ \bibinfo {pages} {024005} (\bibinfo {year} {2019})},\
  \Eprint {https://arxiv.org/abs/1809.03500} {arxiv:1809.03500 [gr-qc]}
  \BibitemShut {NoStop}%
\bibitem [{\citenamefont {Dolan}\ and\ \citenamefont
  {Ottewill}(2009)}]{refDolanOttewill2009}%
  \BibitemOpen
  \bibfield  {author} {\bibinfo {author} {\bibfnamefont {S.~R.}\ \bibnamefont
  {Dolan}}\ and\ \bibinfo {author} {\bibfnamefont {A.~C.}\ \bibnamefont
  {Ottewill}},\ }\bibfield  {title} {\bibinfo {title} {On an expansion method
  for black hole quasinormal modes and {{Regge}} poles},\ }\href
  {https://doi.org/10.1088/0264-9381/26/22/225003} {\bibfield  {journal}
  {\bibinfo  {journal} {Class. Quant. Grav.}\ }\textbf {\bibinfo {volume}
  {26}},\ \bibinfo {pages} {225003} (\bibinfo {year} {2009})},\ \Eprint
  {https://arxiv.org/abs/0908.0329} {arxiv:0908.0329 [gr-qc]} \BibitemShut
  {NoStop}%
\bibitem [{\citenamefont {Goebel}(1972)}]{refGoebel1972}%
  \BibitemOpen
  \bibfield  {author} {\bibinfo {author} {\bibfnamefont {C.~J.}\ \bibnamefont
  {Goebel}},\ }\bibfield  {title} {\bibinfo {title} {Comments on the
  {{``vibrations''}} of a black hole},\ }\href@noop {} {\bibfield  {journal}
  {\bibinfo  {journal} {Astrophys. J.}\ }\textbf {\bibinfo {volume} {172}}
  (\bibinfo {year} {1972})}\BibitemShut {NoStop}%
\bibitem [{\citenamefont {Cardoso}\ \emph {et~al.}(2009)\citenamefont
  {Cardoso}, \citenamefont {Miranda}, \citenamefont {Berti}, \citenamefont
  {Witek},\ and\ \citenamefont {Zanchin}}]{refCardosoLyapunov}%
  \BibitemOpen
  \bibfield  {author} {\bibinfo {author} {\bibfnamefont {V.}~\bibnamefont
  {Cardoso}}, \bibinfo {author} {\bibfnamefont {A.~S.}\ \bibnamefont
  {Miranda}}, \bibinfo {author} {\bibfnamefont {E.}~\bibnamefont {Berti}},
  \bibinfo {author} {\bibfnamefont {H.}~\bibnamefont {Witek}},\ and\ \bibinfo
  {author} {\bibfnamefont {V.~T.}\ \bibnamefont {Zanchin}},\ }\bibfield
  {title} {\bibinfo {title} {Geodesic stability, {{Lyapunov}} exponents, and
  quasinormal modes},\ }\href {https://doi.org/10.1103/PhysRevD.79.064016}
  {\bibfield  {journal} {\bibinfo  {journal} {Phys. Rev. D}\ }\textbf {\bibinfo
  {volume} {79}},\ \bibinfo {pages} {4016} (\bibinfo {year}
  {2009})}\BibitemShut {NoStop}%
\bibitem [{\citenamefont {Konoplya}\ and\ \citenamefont
  {Zhidenko}(2023)}]{Konoplya2023_BeyondEikonal}%
  \BibitemOpen
  \bibfield  {author} {\bibinfo {author} {\bibfnamefont {R.~A.}\ \bibnamefont
  {Konoplya}}\ and\ \bibinfo {author} {\bibfnamefont {A.}~\bibnamefont
  {Zhidenko}},\ }\bibfield  {title} {\bibinfo {title} {{Analytic expressions
  for quasinormal modes and grey-body factors in the eikonal limit and
  beyond}},\ }\href {https://doi.org/10.1088/1361-6382/ad0a52} {\bibfield
  {journal} {\bibinfo  {journal} {Class. Quant. Grav.}\ }\textbf {\bibinfo
  {volume} {40}},\ \bibinfo {pages} {245005} (\bibinfo {year} {2023})},\
  \Eprint {https://arxiv.org/abs/2309.02560} {arXiv:2309.02560 [gr-qc]}
  \BibitemShut {NoStop}%
\bibitem [{\citenamefont {Fernando}\ and\ \citenamefont
  {Correa}(2012)}]{refFernandoCorrea}%
  \BibitemOpen
  \bibfield  {author} {\bibinfo {author} {\bibfnamefont {S.}~\bibnamefont
  {Fernando}}\ and\ \bibinfo {author} {\bibfnamefont {J.}~\bibnamefont
  {Correa}},\ }\bibfield  {title} {\bibinfo {title} {Quasinormal modes of
  bardeen black hole: {{Scalar}} perturbations},\ }\href
  {https://doi.org/10.1103/PhysRevD.86.064039} {\bibfield  {journal} {\bibinfo
  {journal} {Phys. Rev. D}\ }\textbf {\bibinfo {volume} {86}},\ \bibinfo
  {pages} {064039} (\bibinfo {year} {2012})}\BibitemShut {NoStop}%
\bibitem [{\citenamefont {Li}\ \emph {et~al.}(2015)\citenamefont {Li},
  \citenamefont {Lin},\ and\ \citenamefont {Yang}}]{refLiLinYang}%
  \BibitemOpen
  \bibfield  {author} {\bibinfo {author} {\bibfnamefont {J.}~\bibnamefont
  {Li}}, \bibinfo {author} {\bibfnamefont {K.}~\bibnamefont {Lin}},\ and\
  \bibinfo {author} {\bibfnamefont {N.}~\bibnamefont {Yang}},\ }\bibfield
  {title} {\bibinfo {title} {Nonlinear electromagnetic quasinormal modes and
  {{Hawking}} radiation of a regular black hole with magnetic charge},\ }\href
  {https://doi.org/10.1140/epjc/s10052-015-3347-3} {\bibfield  {journal}
  {\bibinfo  {journal} {Eur. Phys. J. C}\ }\textbf {\bibinfo {volume} {75}},\
  \bibinfo {pages} {131} (\bibinfo {year} {2015})},\ \Eprint
  {https://arxiv.org/abs/1409.5988} {arxiv:1409.5988 [gr-qc]} \BibitemShut
  {NoStop}%
\bibitem [{\citenamefont {Chen}\ \emph {et~al.}(2021)\citenamefont {Chen},
  \citenamefont {Cho}, \citenamefont {Chrysostomou},\ and\ \citenamefont
  {Cornell}}]{refOurLargeL}%
  \BibitemOpen
  \bibfield  {author} {\bibinfo {author} {\bibfnamefont {C.-H.}\ \bibnamefont
  {Chen}}, \bibinfo {author} {\bibfnamefont {H.-T.}\ \bibnamefont {Cho}},
  \bibinfo {author} {\bibfnamefont {A.}~\bibnamefont {Chrysostomou}},\ and\
  \bibinfo {author} {\bibfnamefont {A.~S.}\ \bibnamefont {Cornell}},\
  }\bibfield  {title} {\bibinfo {title} {Quasinormal modes for integer and
  half-integer spins within the large angular momentum limit},\ }\href
  {https://doi.org/10.1103/PhysRevD.104.024009} {\bibfield  {journal} {\bibinfo
   {journal} {Phys. Rev. D}\ }\textbf {\bibinfo {volume} {104}},\ \bibinfo
  {pages} {024009} (\bibinfo {year} {2021})},\ \Eprint
  {https://arxiv.org/abs/2103.07777} {arxiv:2103.07777 [gr-qc]} \BibitemShut
  {NoStop}%
\bibitem [{\citenamefont {Berti}(2004)}]{Berti2004}%
  \BibitemOpen
  \bibfield  {author} {\bibinfo {author} {\bibfnamefont {E.}~\bibnamefont
  {Berti}},\ }\bibfield  {title} {\bibinfo {title} {Black hole quasinormal
  modes: {{Hints}} of quantum gravity?},\ }\href@noop {} {\bibfield  {journal}
  {\bibinfo  {journal} {Conf. Proc. C}\ }\textbf {\bibinfo {volume}
  {0405132}},\ \bibinfo {pages} {145} (\bibinfo {year} {2004})},\ \Eprint
  {https://arxiv.org/abs/gr-qc/0411025} {arxiv:gr-qc/0411025} \BibitemShut
  {NoStop}%
\bibitem [{\citenamefont {Iyer}\ and\ \citenamefont {Will}(1987)}]{refBHWKB1}%
  \BibitemOpen
  \bibfield  {author} {\bibinfo {author} {\bibfnamefont {S.}~\bibnamefont
  {Iyer}}\ and\ \bibinfo {author} {\bibfnamefont {C.~M.}\ \bibnamefont
  {Will}},\ }\bibfield  {title} {\bibinfo {title} {Black-hole normal modes: {{A
  WKB}} approach. {{I}}. {{Foundations}} and application of a higher-order
  {{WKB}} analysis of potential-barrier scattering},\ }\href
  {https://doi.org/10.1103/PhysRevD.35.3621} {\bibfield  {journal} {\bibinfo
  {journal} {Phys. Rev. D}\ }\textbf {\bibinfo {volume} {35}},\ \bibinfo
  {pages} {3621} (\bibinfo {year} {1987})}\BibitemShut {NoStop}%
\bibitem [{\citenamefont {Pan}\ and\ \citenamefont {Jing}(2006)}]{refPanJing}%
  \BibitemOpen
  \bibfield  {author} {\bibinfo {author} {\bibfnamefont {Q.-Y.}\ \bibnamefont
  {Pan}}\ and\ \bibinfo {author} {\bibfnamefont {J.-L.}\ \bibnamefont {Jing}},\
  }\bibfield  {title} {\bibinfo {title} {Quasinormal modes of the
  {{Schwarzschild}} black hole with arbitrary spin fields: {{Numerical}}
  analysis},\ }\href {https://doi.org/10.1142/S0217732306020287} {\bibfield
  {journal} {\bibinfo  {journal} {Mod. Phys. Lett. A}\ }\textbf {\bibinfo
  {volume} {21}},\ \bibinfo {pages} {2671} (\bibinfo {year}
  {2006})}\BibitemShut {NoStop}%
\bibitem [{\citenamefont {Shu}\ and\ \citenamefont {Shen}(2005)}]{refShuShen}%
  \BibitemOpen
  \bibfield  {author} {\bibinfo {author} {\bibfnamefont {F.-W.}\ \bibnamefont
  {Shu}}\ and\ \bibinfo {author} {\bibfnamefont {Y.-G.}\ \bibnamefont {Shen}},\
  }\bibfield  {title} {\bibinfo {title} {Quasinormal modes in {{Schwarschild}}
  black holes due to arbitrary spin fields},\ }\href
  {https://doi.org/10.1016/j.physletb.2005.05.077} {\bibfield  {journal}
  {\bibinfo  {journal} {Phys. Lett. B}\ }\textbf {\bibinfo {volume} {619}},\
  \bibinfo {pages} {340} (\bibinfo {year} {2005})},\ \Eprint
  {https://arxiv.org/abs/gr-qc/0501098} {arxiv:gr-qc/0501098} \BibitemShut
  {NoStop}%
\bibitem [{\citenamefont {Konoplya}(2003)}]{Konoplya2003}%
  \BibitemOpen
  \bibfield  {author} {\bibinfo {author} {\bibfnamefont {R.~A.}\ \bibnamefont
  {Konoplya}},\ }\bibfield  {title} {\bibinfo {title} {Quasinormal behavior of
  the d-dimensional {{Schwarzschild}} black hole and higher order {{WKB}}
  approach},\ }\href {https://doi.org/10.1103/PhysRevD.68.024018} {\bibfield
  {journal} {\bibinfo  {journal} {Phys. Rev. D}\ }\textbf {\bibinfo {volume}
  {68}},\ \bibinfo {pages} {024018} (\bibinfo {year} {2003})},\ \Eprint
  {https://arxiv.org/abs/gr-qc/0303052} {arxiv:gr-qc/0303052} \BibitemShut
  {NoStop}%
\bibitem [{\citenamefont {Zhidenko}(2004)}]{refZhidenko2004}%
  \BibitemOpen
  \bibfield  {author} {\bibinfo {author} {\bibfnamefont {A.}~\bibnamefont
  {Zhidenko}},\ }\bibfield  {title} {\bibinfo {title} {Quasi-normal modes of
  {{Schwarzschild-de Sitter}} black holes},\ }\href
  {https://doi.org/10.1088/0264-9381/21/1/019} {\bibfield  {journal} {\bibinfo
  {journal} {Class. Quantum Gravity}\ }\textbf {\bibinfo {volume} {21}},\
  \bibinfo {pages} {273} (\bibinfo {year} {2004})}\BibitemShut {NoStop}%
\bibitem [{\citenamefont {Schutz}\ and\ \citenamefont
  {Will}(1985)}]{refBHWKB0}%
  \BibitemOpen
  \bibfield  {author} {\bibinfo {author} {\bibfnamefont {B.~F.}\ \bibnamefont
  {Schutz}}\ and\ \bibinfo {author} {\bibfnamefont {C.~M.}\ \bibnamefont
  {Will}},\ }\bibfield  {title} {\bibinfo {title} {Black hole normal modes -
  {{A}} semianalytic approach},\ }\href {https://doi.org/10.1086/184453}
  {\bibfield  {journal} {\bibinfo  {journal} {Astrophys. J.}\ }\textbf
  {\bibinfo {volume} {291}},\ \bibinfo {pages} {L33} (\bibinfo {year}
  {1985})}\BibitemShut {NoStop}%
\bibitem [{\citenamefont {Will}(1986)}]{refBHWKB0.5}%
  \BibitemOpen
  \bibfield  {author} {\bibinfo {author} {\bibfnamefont {C.~M.}\ \bibnamefont
  {Will}},\ }\bibfield  {title} {\bibinfo {title} {Approximation methods in
  gravitational-radiation theory},\ }\href {https://doi.org/10.1139/p86-023}
  {\bibfield  {journal} {\bibinfo  {journal} {Can. J. Phys.}\ }\textbf
  {\bibinfo {volume} {64}},\ \bibinfo {pages} {140} (\bibinfo {year}
  {1986})}\BibitemShut {NoStop}%
\bibitem [{\citenamefont {Abramowitz}\ and\ \citenamefont
  {Stegun}(1964)}]{refmathbible}%
  \BibitemOpen
  \bibfield  {author} {\bibinfo {author} {\bibfnamefont {M.}~\bibnamefont
  {Abramowitz}}\ and\ \bibinfo {author} {\bibfnamefont {I.~A.}\ \bibnamefont
  {Stegun}},\ }\href@noop {} {\emph {\bibinfo {title} {Handbook of Mathematical
  Functions with National Bureau of Standards Applied Mathematics Series}}},\
  \bibinfo {edition} {6th}\ ed.\ (\bibinfo  {publisher} {U.S. Government
  Printing Office},\ \bibinfo {address} {Washington D.C.},\ \bibinfo {year}
  {1964})\BibitemShut {NoStop}%
\bibitem [{\citenamefont {Casals}\ \emph {et~al.}(2013)\citenamefont {Casals},
  \citenamefont {Dolan}, \citenamefont {Ottewill},\ and\ \citenamefont
  {Wardell}}]{refDolanOttewill2013}%
  \BibitemOpen
  \bibfield  {author} {\bibinfo {author} {\bibfnamefont {M.}~\bibnamefont
  {Casals}}, \bibinfo {author} {\bibfnamefont {S.}~\bibnamefont {Dolan}},
  \bibinfo {author} {\bibfnamefont {A.~C.}\ \bibnamefont {Ottewill}},\ and\
  \bibinfo {author} {\bibfnamefont {B.}~\bibnamefont {Wardell}},\ }\bibfield
  {title} {\bibinfo {title} {Self-force and green function in schwarzschild
  spacetime via quasinormal modes and branch cut},\ }\href
  {https://doi.org/10.1103/PhysRevD.88.044022} {\bibfield  {journal} {\bibinfo
  {journal} {Phys. Rev. D}\ }\textbf {\bibinfo {volume} {88}},\ \bibinfo
  {pages} {044022} (\bibinfo {year} {2013})},\ \Eprint
  {https://arxiv.org/abs/1306.0884} {arxiv:1306.0884 [gr-qc]} \BibitemShut
  {NoStop}%
\bibitem [{\citenamefont {Mano}\ \emph
  {et~al.}(1996{\natexlab{a}})\citenamefont {Mano}, \citenamefont {Suzuki},\
  and\ \citenamefont {Takasugi}}]{MST1996_AnalyticQNMsolsITeukolsky}%
  \BibitemOpen
  \bibfield  {author} {\bibinfo {author} {\bibfnamefont {S.}~\bibnamefont
  {Mano}}, \bibinfo {author} {\bibfnamefont {H.}~\bibnamefont {Suzuki}},\ and\
  \bibinfo {author} {\bibfnamefont {E.}~\bibnamefont {Takasugi}},\ }\bibfield
  {title} {\bibinfo {title} {{Analytic solutions of the Teukolsky equation and
  their low frequency expansions}},\ }\href
  {https://doi.org/10.1143/PTP.95.1079} {\bibfield  {journal} {\bibinfo
  {journal} {Prog. Theor. Phys.}\ }\textbf {\bibinfo {volume} {95}},\ \bibinfo
  {pages} {1079} (\bibinfo {year} {1996}{\natexlab{a}})},\ \Eprint
  {https://arxiv.org/abs/gr-qc/9603020} {arXiv:gr-qc/9603020} \BibitemShut
  {NoStop}%
\bibitem [{\citenamefont {Mano}\ \emph
  {et~al.}(1996{\natexlab{b}})\citenamefont {Mano}, \citenamefont {Suzuki},\
  and\ \citenamefont {Takasugi}}]{MST1996_AnalyticQNMsolsIIRW}%
  \BibitemOpen
  \bibfield  {author} {\bibinfo {author} {\bibfnamefont {S.}~\bibnamefont
  {Mano}}, \bibinfo {author} {\bibfnamefont {H.}~\bibnamefont {Suzuki}},\ and\
  \bibinfo {author} {\bibfnamefont {E.}~\bibnamefont {Takasugi}},\ }\bibfield
  {title} {\bibinfo {title} {{Analytic solutions of the Regge-Wheeler equation
  and the postMinkowskian expansion}},\ }\href
  {https://doi.org/10.1143/PTP.96.549} {\bibfield  {journal} {\bibinfo
  {journal} {Prog. Theor. Phys.}\ }\textbf {\bibinfo {volume} {96}},\ \bibinfo
  {pages} {549} (\bibinfo {year} {1996}{\natexlab{b}})},\ \Eprint
  {https://arxiv.org/abs/gr-qc/9605057} {arXiv:gr-qc/9605057} \BibitemShut
  {NoStop}%
\bibitem [{\citenamefont {Leaver}(1986)}]{refLeaver1986_SchwarzSpectralDecomp}%
  \BibitemOpen
  \bibfield  {author} {\bibinfo {author} {\bibfnamefont {E.~W.}\ \bibnamefont
  {Leaver}},\ }\bibfield  {title} {\bibinfo {title} {Spectral decomposition of
  the perturbation response of the {{Schwarzschild}} geometry},\ }\href
  {https://doi.org/10.1103/PhysRevD.34.384} {\bibfield  {journal} {\bibinfo
  {journal} {Phys. Rev. D}\ }\textbf {\bibinfo {volume} {34}},\ \bibinfo
  {pages} {384} (\bibinfo {year} {1986})}\BibitemShut {NoStop}%
\bibitem [{\citenamefont {Nollert}(1999)}]{refNollert1999}%
  \BibitemOpen
  \bibfield  {author} {\bibinfo {author} {\bibfnamefont {H.-P.}\ \bibnamefont
  {Nollert}},\ }\bibfield  {title} {\bibinfo {title} {Quasinormal modes: The
  characteristic `sound' of black holes and neutron stars},\ }\href
  {https://doi.org/10.1088/0264-9381/16/12/201} {\bibfield  {journal} {\bibinfo
   {journal} {Class. Quant. Grav.}\ }\textbf {\bibinfo {volume} {16}},\
  \bibinfo {pages} {R159} (\bibinfo {year} {1999})}\BibitemShut {NoStop}%
\bibitem [{\citenamefont {Bardeen}\ \emph {et~al.}(1972)\citenamefont
  {Bardeen}, \citenamefont {Press},\ and\ \citenamefont
  {Teukolsky}}]{refBardeenPressTeukolsky}%
  \BibitemOpen
  \bibfield  {author} {\bibinfo {author} {\bibfnamefont {J.~M.}\ \bibnamefont
  {Bardeen}}, \bibinfo {author} {\bibfnamefont {W.~H.}\ \bibnamefont {Press}},\
  and\ \bibinfo {author} {\bibfnamefont {S.~A.}\ \bibnamefont {Teukolsky}},\
  }\bibfield  {title} {\bibinfo {title} {Rotating black holes: {{Locally}}
  nonrotating frames, energy extraction, and scalar synchrotron radiation},\
  }\href {https://doi.org/10.1086/151796} {\bibfield  {journal} {\bibinfo
  {journal} {Astrophys. J.}\ }\textbf {\bibinfo {volume} {178}},\ \bibinfo
  {pages} {347} (\bibinfo {year} {1972})}\BibitemShut {NoStop}%
\end{thebibliography}%

\end{document}